\documentclass[a4paper,11pt,english,dvipsnames]{article}
\usepackage{enumerate}
\usepackage{latexsym}
\usepackage{amsmath, amsthm}
\usepackage{cases}
\usepackage{amssymb}
\usepackage{bm}
\usepackage{graphicx}
\graphicspath{{images/}}
\usepackage{array}
\usepackage{caption}
\usepackage{subcaption}
\usepackage{xcolor}
\usepackage{pifont}
\usepackage{accents}
\usepackage{soul}
\usepackage{hyperref}
\usepackage{anyfontsize}
\usepackage{algorithm}
\usepackage{algorithmic}
\usepackage[utf8]{inputenc}
\usepackage[]{natbib}
\usepackage{comment}
\usepackage{enumitem}
\allowdisplaybreaks[1]
\usepackage{multirow}
\usepackage{caption}
\usepackage{subcaption}
\usepackage{tikz}
\usetikzlibrary{positioning, shapes, arrows}
\usepackage{tkz-berge}
\usetikzlibrary{bayesnet}
\usetikzlibrary{positioning,arrows.meta,quotes}
\usetikzlibrary{shapes,decorations}
\usetikzlibrary{bayesnet}
\tikzset{>=latex}
\usepackage{tikz,tikz-3dplot}
\usepackage{amssymb}
\usepackage{amsfonts}
\usepackage{times}
\newtheorem{proposition}{Proposition}

\title{Markov Switching Multiple-equation Tensor Regressions}
\author{Roberto Casarin$^\dag$ Radu Craiu$^\ddag$ and Qing Wang$^\dag$\\\vspace{10pt}\\
		{\centering {\small{$^\dag$Ca' Foscari University of Venice, Italy}}}\\
		{\centering {\small{$^\ddag$University of Toronto, Canada}}}}

\begin{document}
\maketitle
\begin{abstract}
We propose a new flexible tensor model for multiple-equation regression that accounts for latent regime changes. The model allows for dynamic coefficients and multi-dimensional covariates that vary across equations. We assume the coefficients are driven by a common hidden Markov process that addresses structural breaks to enhance the model flexibility and preserve parsimony. We introduce a new Soft PARAFAC hierarchical prior to achieve dimensionality reduction while preserving the structural information of the covariate tensor. The proposed prior includes a new multi-way shrinking effect to address over-parametrization issues. We developed theoretical results to help hyperparameter choice. An efficient MCMC algorithm based on random scan Gibbs and back-fitting strategy is developed to achieve better computational scalability of the posterior sampling. The validity of the MCMC algorithm is demonstrated theoretically, and its computational efficiency is studied using numerical experiments in different parameter settings. The effectiveness of the model framework is illustrated using two original real data analyses. The proposed model exhibits superior performance when compared to the current benchmark, Lasso regression.

\end{abstract}

\section{Introduction} 
As data grow in volume and complexity, it is increasingly common to record them as high-dimensional arrays or tensors. Such structures appear in many applications and fields, such as neuroimaging \citep{spencer2022bayesian, guha2021bayesian}, biostatistics, financial networks \citep{billio2023MStensor}, or even more generally, in time series \citep{billio2023tensor}. People are often interested in characterizing the relationship between a tensor predictor and a scalar outcome \citep{guha2021bayesian} or tensor outcome \citep{wang2022bayesian}. Tensor regression has been studied extensively in this regard in a linear model framework. Nevertheless, a common challenge within the framework of regression is model misspecification. One of the sources of model misspecification is the presence of dynamic regimes, which naturally call for models with time-varying parameters. In this paper, we follow the parsimonious strategy for dynamic behaviour that is provided by Hidden Markov Models (HMM) \citep{Fruhwirth2006} also known as Markov Switching (MS) \citep{Kau11,Bil13,Cas18,CasarinSartoreTronzano18,Bian19,Cas22}. Efforts have been made to address abrupt structural changes in temporal networks by \cite{billio2023MStensor}. They proposed a tensor-on-tensor logistic regression model combining a low-rank decomposition and HMM to model the coefficient tensor. The contribution of our paper is multi-fold. First, we extend the soft tensor linear regression of \cite{guha2021bayesian,papadogeorgou2021soft,wang2022bayesian} to an HMM or MS framework to accommodate structural breaks. Second, we consider a multi-equation setting in which the latent process provides a time-varying structure to several tensor regression models involving different response variables and, possibly, different sets of covariates. Third, we propose a Bayesian inference procedure that relies on numerical exploration of the posterior via a new and efficient Gibbs Sampler, which reduces computational costs and improves scalability. Finally, using a common latent process is intended to address two goals: 1) it facilitates the integration of information about the latent process from multiple outcomes, and 2) it robustifies the estimation of regime changes, which must be supported by multiple outcome variables simultaneously.

The complex structure of multi-dimensional data naturally poses challenges such as over-parametrization and overfitting issues. In the tensor regression framework, a simple and straightforward approach is to vectorize the tensor predictor and regress the response variable on a large vector of tensor entries with some form of penalization and variable selection. However, this approach completely ignores the structural relationships embedded in the tensor predictor. Most research on tensor regression focuses on dimensionality reduction for tensor predictors or coefficients. Various dimensionality reduction strategies have been proposed to cope with these issues. For instance, \cite{zhang2019tensor} adopted a two-stage procedure to study the relationship between individuals' structural connectomes and human traits, using principal component analysis on the tensor predictors to achieve dimension reduction and then fitting a model using lower dimensional summaries of tensor predictors. Similarly, \cite{caffo2010two} carried out SVD on high dimensional fMRI data to study the relationship between functional connectivity and Alzheimer's disease risk. However, this approach suffers from the unsupervised nature of PCA, and the loss of structural information on the tensor predictors and interpretation of estimated coefficients could be difficult. Thus, we follow a different approach based on the reduction of dimensionality of tensor coefficients, which preserves the structural dependence of the predictor tensor.

Within the frequentist paradigm, \cite{pmlr-v48-yu16} applied Tucker decomposition on the tensor coefficients and proposed a fast algorithm (Tensor Projected Gradient) to minimize the empirical loss function. \cite{zhou2013tensor} used PARAFAC decomposition, a special case of Tucker decomposition, on the tensor coefficients and relied on maximum likelihood estimation to perform neuroimaging data analysis. \cite{kossaifi2020tensor} used a neural network combined with Tucker representation to address multi-way data analysis.

Within the Bayesian paradigm, \cite{yu2018tensor, zhao2014tensor} proposed non-parametric methods based on Gaussian Process priors. In a scalar on tensor regression framework, \cite{guhaniyogi2017bayesian} proposed a novel multi-way shrinkage prior on the Parallel Factor (PARAFAC) representation, also called CP or Poyadic decomposition (see \cite{kolda2009tensor}) of the coefficient tensor. Their work was extended by \cite{spencer2022bayesian}, who explored a more general Tucker decomposition on the tensor coefficients. In follow-up work, \cite{guha2021bayesian} proposed a {Bayesian network shrinkage prior} and used a spike-and-slab prior to determine which brain nodes are most influential to creativity. In the case of tensor on tensor regression, \cite{wang2022bayesian} proposed to use the Tucker decomposition of the coefficient tensor without assuming the dimension of the core tensor.  In this paper, we follow a soft PARAFAC framework \citep{papadogeorgou2021soft} where the hierarchical prior distribution of \cite{guhaniyogi2017bayesian} is modified to allow the coefficient tensor to deviate randomly from the rigid low-rank PARAFAC representation. We modified the multi-way shrinkage priors from \citep{papadogeorgou2021soft} and \cite{guhaniyogi2017bayesian} to improve the tractability of the conditional posterior distributions and developed an efficient MCMC algorithm to achieve better scalability. 
The latter relies on a random scan Gibbs sampler within the back-fitting strategy, usually employed in Bayesian tensor regression models \citep{levine2006optimizing, Rob2013, Hastie2000}.

The paper is organized as follows: in section \ref{MS}, we revisit the concept of soft PARAFAC decomposition for dimensionality reduction and introduce the  Markov-Switching Multiple-equation Tensor Regression (MSMETR) and the Bayesian framework for inference. In section \ref{posapp}, we propose a new MCMC algorithm based on Random Partial Scan Gibbs and back fitting strategy, prove its ergodicity and demonstrate its performance using numerical experiments (simulation resutls are shown in Appendix \ref{sec: sim_resu} of the Supplement). In section \ref{empapp}, we test our model with two applications that show the gain in performance in terms of in-sample fitting and out-of-sample forecasting. The paper ends with section \ref{concl}, which contains conclusions and future promising directions.

\section{A Markov-Switching Multiple-equation Tensor Regression Model}\label{MS}

To simplify the exposition in this section, we assume covariates are matrix-valued and common to all the equations. Furthermore, the error terms are assumed to be independent across equations, but the approach generalizes to equation-specific covariate tensors and dependent errors. In our model, we assume a system of $N$ equations has time-varying parameters 
\begin{numcases}{}
y_{1t}= \mu_1(s_t)+<B_1(s_t),X_t> + \sigma_1(s_t)\varepsilon_{1t} \notag\\
\quad \quad \vdots\label{eq.model}\\
y_{Nt}= \mu_N(s_t)+<B_N(s_t),X_t> + \sigma_N(s_t)\varepsilon_{Nt} \notag
\end{numcases}
$t=1,2,\ldots,T$ where $y_{\ell t}$ is a scalar response variable,  $X_t$ is a $n \times m$ matrix of covariates,  $B_{\ell}(s_t)$ are $n\times m$ coefficient matrices, $\varepsilon_{\ell t}$ are i.i.d from $\mathcal{N}(0,1)$, $\{s_t\}_{t=1}^{T}$ is a common latent process, and $<\cdot,\cdot>$ denotes the inner product for tensors (\cite{kolda2009tensor}). 

The latent process is a $K$-state homogeneous Markov chain process with transition probability $\mathbb{P}(s_{t}=j\vert s_{t-1}=i)=p_{ij}$, $i,j=1,\dots,K$ and the parametrization used is
\begin{equation}
\mu_{\ell}(s_t)=\sum_{k=1}^{K}\mu_{\ell k}\mathbb{I}(s_t=k),\quad B_{\ell}(s_{t})=\sum_{k=1}^{K}B_{\ell k}\mathbb{I}(s_t=k),\quad \sigma^2_{\ell}(s_{t})=\sum_{k=1}^{K}\sigma^2_{\ell k}\mathbb{I}(s_t=k)
\end{equation}
Alternative parameterizations for the coefficients can be used; see \cite{Fruhwirth2006} for conditionally linear single-equation models and \cite{CasarinSartoreTronzano18} for conditionally linear multiple-equation models.

Since, in many applications, the number of covariates in Eq. \ref{eq.model} is large, a dimensionality reduction strategy is needed. In this paper, we follow \cite{papadogeorgou2021soft} and \cite{billio2023tensor, billio2023MStensor} and consider a low-rank representation combined with a hierarchical prior distribution. The hierarchical prior allows for shrinking effects in the coefficient matrices $B_{\ell k}$, $k=1,\ldots, K$, and the low-rank representation induces further shrinking effects along different modes. We assume a PARAFAC representation and decompose the state-specific coefficient matrix as follows
\begin{equation}
B_{\ell k}=\sum_{d=1}^{D}B_{\ell, 1,k}^{(d)} \circ B_{\ell, 2,k}^{(d)}
\end{equation}
where $\circ$ is the {Hadamard product} and $B_{\ell, j,k}^{(d)}$ for $j=1,2$ are multiplicative factors (\cite{papadogeorgou2021soft}). To simplify exposition, we assume there are two modes, $m=2$, and the number of elements in each mode is $p_1$ and $p_2$, respectively. 

The hierarchical prior distribution includes three stages. At the first stage a matrix-variate normal distribution \citep{Gup99} is assumed
\begin{align}
B_{\ell, m,k}^{(d)} \sim \mathcal{MN}_{p_1, p_2}\left(G_{\ell, m,k}^{(d)}, \tau_{\ell,k}\sigma_{\ell,m,k}^2\zeta^{(d)}_{\ell,k} I_{p_1}, I_{p_2}\right)\label{eq.b.prior}
\end{align}
    where $\mathcal{MN}_{p_1,p_2}\left(M, U, V\right)$ denotes the matrix-variate normal distribution with $p_1 \times p_2$ mean matrix $M$, $p_1\times p_1$ row covariance matrix $U$ and $p_2 \times p_2$ column covariance matrix $V$.
The location matrix $G_{\ell,m, k}^{(d)}$ is parametrized as follows:
\begin{align}
    G_{\ell,m,k}^{(d)}=
    \begin{cases}
        \boldsymbol{\gamma}_{\ell,1}^{(d)}\otimes
        \boldsymbol{\iota}_{p_2},  \mbox{ if } m=1 \\
        \boldsymbol{\iota}_{p_1}\otimes\boldsymbol{\gamma}_{\ell,2}^{(d)}, \mbox{ if } m=2
    \end{cases}\notag
\end{align}
where $\otimes$ denotes the \textit{outer product}, $\boldsymbol{\iota}_n=(1,\ldots,1)'$ is the $n$-dimensional unit vector, $\boldsymbol{\gamma}_{\ell,1}^{(d)}$ and $\boldsymbol{\gamma}_{\ell,2}^{(d)}$ are the PARAFAC marginals in the conditional mean of the factors $B_{\ell,k}^{(d)}$, such that
\begin{align}
    \mathbb{E}\left(B_{\ell,k} \mid \boldsymbol{\gamma}_{\ell,1,k}^{(d)}, \boldsymbol{\gamma}_{\ell,2,k}^{(d)}\right) &= \sum_{d=1}^D \mathbb{E}\left(B_{\ell,1,k}^{(d)}\right) \circ \mathbb{E}\left(B_{\ell,2,k}^{(d)}\right)\notag = \sum_{d=1}^D \left(G_{\ell,1,k}^{(d)} \circ G_{\ell,2,k}^{(d)}\right)\notag  = \sum_{d=1}^D \boldsymbol{\gamma}_{\ell,1,k}^{(d)} \otimes \boldsymbol{\gamma}_{\ell,2,k}^{(d)} \notag
\end{align}
In the second stage, we assume that the marginals from the PARAFAC decomposition follow a multivariate normal distribution 
\begin{eqnarray}
\boldsymbol{\gamma}_{\ell,m,k}^{(d)}&\sim&\mathcal{N}_{p_{m,k}}(\boldsymbol{0},\tau_{\ell,k}\zeta^{(d)}_{\ell,k}W_{\ell,m,k}^{(d)})\label{eq.gamma.prior}
\end{eqnarray}
and assume the distributions are centred around the null vector and have random scales to allow for shrinkage at different levels. At the third stage, we borrow from \cite{guhaniyogi2017bayesian} and specify the priors for the scales to induce shrinkage across components and rows
\begin{eqnarray}
\tau_{\ell,k} & \sim & \mathcal{G}a(a_\tau , b_\tau)\\
\sigma_{\ell,m,k}^{2}&\sim&\mathcal{G}a(a_{\sigma},b_{\sigma})\\
w_{\ell,m,j_{m},k}^{(d)}&\sim &\mathcal{E}xp((\lambda_{\ell,m,k}^{(d)})^2/2)\\
\lambda_{\ell,m,k}^{(d)}&\sim &\mathcal{G}a(a_\lambda ,b_\lambda)\\
(\zeta^{(1)}_{\ell,k},\ldots,\zeta^{(D)}_{\ell,k})&\sim &\mathcal{D}ir(\alpha/D,\ldots,\alpha/D)\label{eq.zeta.prior}
\end{eqnarray}
where $\tau_{\ell,k}$ is a global scale which contributes to the variances of both $\boldsymbol{\gamma}^{(d)}_{\ell,m,k}$ and $B^{(d)}_{\ell,m,k}$. The matrices $W^{(d)}_{\ell,m,k} = \text{diag}(w^{(d)}_{\ell,1,1,k},\ldots,w^{(d)}_{\ell,m,j_m,k},\ldots,w^{(d)}_{\ell,M,p_{M},k})$ are the row-specific parameters that shrink the individual elements of the marginals and, together with the prior on $\lambda^{(d)}_{\ell,m,k}$, lead to an adaptive LASSO-type penalty on $\boldsymbol{\gamma}^{(d)}_{\ell,m,k}$ \citep{armagan2013generalized}. The parameter $\zeta^{(d)}_{\ell,k}$ is component-specific and allows a subset of the $D$ components to contribute substantially to the PARAFAC approximation while leaving the values of other components close to zero.

The choice of the prior hyperparameter value is crucial in Bayesian inference and can greatly affect the model's performance. We turn to study the induced prior for the $\ell$th coefficient tensor $\boldsymbol{B}_{\ell}$ to elicit the default choice of hyperparameters. In particular, the variance of the entry of coefficient tensor $\boldsymbol{B}_{\ell}$ for the soft PARAFAC can be written as a function of the hyperparameters:
\begin{align}
    \mathbb{V}(B_{\ell,ij}) &= \frac{a_{\tau}(a_{\tau}+1)}{b_{\tau}^2}C\left(\frac{a_{\sigma}}{b_{\sigma}}+\frac{2b_{\lambda}^2}{\left(a_{\lambda}-1\right)\left(a_{\lambda}-2\right)}\right)^2 \label{eq:hyp1}
\end{align}
where $C=\frac{\frac{\alpha}{D}+1}{\alpha+1}$. Moreover, the variance of the coefficient entries for the hard PARAFAC is:
\begin{align*}
\mathbb{V}^{\text{hard}}\left(B_{\ell,ij}\right) = \frac{a_{\tau}\left(a_{\tau}+1\right)}{b_{\tau}^2}C\left(\frac{2b_{\lambda}^2}{\left(a_{\lambda}-1\right)\left(a_{\lambda}-2\right)}\right)^2
\end{align*}

We define the \textit{relative additional variance} introduced by the softening of the PARAFAC as:
\begin{align*}
    AV = \frac{\mathbb{V}\left(B_{\ell,ij}\right)-\mathbb{V}^{\text{hard}}\left(B_{\ell,ij}\right)}{\mathbb{V}\left(B_{\ell,ij}\right)}
\end{align*}
which can be used to elicit the choice of the hyper-parameters.  
\begin{proposition}\label{prop:Hyper}
    For a matrix coefficient, target variance $V^* \in (0, \infty)$, target additional variance $AV^* \in [0,1)$, we have the following expression,
    \begin{align}
        \frac{a_{\sigma}}{b_{\sigma}} = \frac{b_{\tau}}{a_{\tau}}\sqrt{\frac{a_{\tau}V^*}{(a_{\tau}+1)C}}\left(1-\sqrt{1-AV^*}\right) \label{eq:hyp2}
    \end{align}
\end{proposition}
Equation \eqref{eq:hyp1} and Proposition 1 are used in the simulations and empirical applications to help choose hyperparameters. In particular, we impose restrictions on the induced prior variance such that $\mathbb{V}(B_{\ell,ij}) = 1$ and $AV=10\%$. Moreover, we set $\alpha=1, a_{\tau}=3, a_{\sigma}=0.5, a_{\lambda} = 3, b_{\lambda}=\sqrt[2M]{a_{\lambda}}$ following \cite{papadogeorgou2021soft} and we compute the values of $b_{\tau}$ and $b_{\sigma}$ from Equations \eqref{eq:hyp1} and \eqref{eq:hyp2} for which $V^*=1$ and $AV^*=10\%$.

The choice of rank $D$ for the soft PARAFAC decomposition of the tensor coefficient can lead to significant changes in computational costs, with a higher value of $D$ triggering drastic increases in computational time. However, the increase in $D$ doesn't necessarily guarantee a vast boost in inferential performance. Intuitively, the soft PARAFAC can expand away from the low-rank hard PARAFAC structure and achieve a higher-rank representation of the tensor coefficient.

We demonstrate through simulation studies (Appendix \ref{sec: sim_resu} in the Supplement) that our proposed MCMC procedure is robust to the selection of $D$, where true tensor coefficients with low to full ranks are used for data generation in different simulation settings. The inference performance are similar when $D \in \{3, 5, 7\}$. Moreover, additional simulations are carried out as a robustness check in which the true coefficients are contaminated with white noise such that the ranks are considered full for all different coefficients, and the MCMC procedure can recover the patterns of the true coefficients reasonably well for all values of  $D \in \{ 3, 5, 7\}$ (Figure \ref{fig:sim_main_noisy} in Appendix \ref{sec: sim_resu}).

Alternatively, one can resort to model selection methods to choose the best value for $D$, since the  Bayesian Information Criterion (BIC) or the Akaike Information Criterion (AIC) can be computed for models with different values of $D$. 

\section{Posterior Approximation} \label{posapp}
In this section, we assume tensor-valued covariates and denote with $B_{\ell, m,\tilde{j}_{m},k}^{(d)}$ the $j_{m}^{th}$ slice of tensor $B_{\ell, m,k}^{(d)}$ along mode $m$, where $\tilde{j}_{m} = (:,:,\cdots,j_{m},\cdots,:)$ and with $B_{\ell, m,\tilde{j}_{m},k}^{(d)}$ the $p_1\times \cdots \times p_{m-1} \times p_{m+1} \times \cdots \times p_{M}$ tensor with $m-1$ modes. Let $\boldsymbol{\beta}_{\ell, m,j_{m}, k}^{(d)}=\text{vec}(B_{\ell, m,\tilde{j}_{m}, k}^{(d)})$ be a $q_{m} \times 1$ vector, where $q_{m} = \prod_{l \neq m}p_l$. We further define the collections   $\boldsymbol{\beta}_k = (\boldsymbol{\beta}_{1k}, \ldots, \boldsymbol{\beta}_{Nk})$ and $\boldsymbol{\gamma}_k = (\boldsymbol{\gamma}_{1k}, \ldots,$ $ \boldsymbol{\gamma}_{Nk})$, with $\boldsymbol{\beta}_{\ell, k} =(\boldsymbol{\beta}_{\ell,1,1,k}^{(1)},\ldots, \boldsymbol{\beta}_{\ell,m,j_m,k}^{(d)},\ldots,\boldsymbol{\beta}_{\ell,M,p_M,k}^{(D)})'$ and $\boldsymbol{\gamma}_{\ell,k}=(\gamma_{\ell,1,1,k}^{(1)}$, $\ldots,\gamma^{(d)}_{\ell,m,j_m,k}$, $\ldots,\gamma_{\ell,M,p_M,k}^{(D)})'$, for $l=1,\ldots, N$, $k=1,\ldots,K$, $m=1,\ldots,M$,  $d = 1,\ldots,D$ and $j_{m} = 1,\ldots,p_{m}$.

We summarize our Bayesian model in the Directed Acyclic Graph (DAG) representation of Fig. \ref{DAG} where $\mathbf{y}_t = (y_{1t}, \ldots, y_{Nt})$ is the collection of response variables across equations, $\boldsymbol{p}=(p_{11},\ldots,p_{1k},\ldots,p_{k1},\ldots,p_{kk})$ is the collection of transition probabilities, and $\boldsymbol{\beta}=(\boldsymbol{\beta}_{1},\ldots,\boldsymbol{\beta}_{K})$ and $\boldsymbol{\gamma}=(\boldsymbol{\gamma}_{1},\ldots,\boldsymbol{\gamma}_{K})$ denotes the collections across equations and states of the regression coefficients and PARFAC factors, respectively. 

In the same diagram $\boldsymbol{\zeta}=(\boldsymbol{\zeta}_1,\ldots,\boldsymbol{\zeta}_K)$, $\boldsymbol{\tau}=(\tau_{1},\ldots,\tau_{K})'$, $\boldsymbol{w}=(\boldsymbol{w}_{1},\ldots,\boldsymbol{w}_{K})$ and $\boldsymbol{\lambda}=(\boldsymbol{\lambda}_{1},\ldots,\boldsymbol{\lambda}_{K})$ denote the collections of the hyper-parameters at the second and third stage of the hierarchical prior where $\boldsymbol{\zeta}_{k} = (\boldsymbol{\zeta}_{1,k}, \ldots, \boldsymbol{\zeta}_{N,k})$, $\boldsymbol{\zeta}_{\ell, k} = (\zeta_{\ell, k}^{(1)},\ldots$, $\zeta^{(d)}_{\ell,k},\ldots,\zeta_{\ell, k}^{(D)})'$, $\boldsymbol{\lambda}_{k} = (\lambda_{1k}, \ldots, \lambda_{Nk})$, $\boldsymbol{\lambda}_{\ell,k}=(\lambda_{\ell,1,k}^{(1)},\ldots,\lambda^{(d)}_{\ell,m,k},\ldots,\lambda_{\ell,M,k}^{(D)})'$ and $\boldsymbol{w}_{k} = (\boldsymbol{w}_{1k}, \ldots, \boldsymbol{w}_{Nk})$,  $\boldsymbol{w}_{\ell,k} = (w_{\ell,1,1,k}^{(1)}$, $\ldots, w^{(d)}_{\ell,m,j_m,k},\ldots,$ $w_{\ell,M,p_M,k}^{(D)})'$. 

\tikzset{obs/.style={rectangle, draw, minimum width = 5ex, minimum height = 5ex, inner sep = 0.1},
latent/.style={circle, draw, minimum size =5ex,fill=gray!10,inner sep = 0.1},
par/.style={circle, draw, minimum size =5ex, fill=white,inner sep = 0.1}}
	\begin{figure}[t]
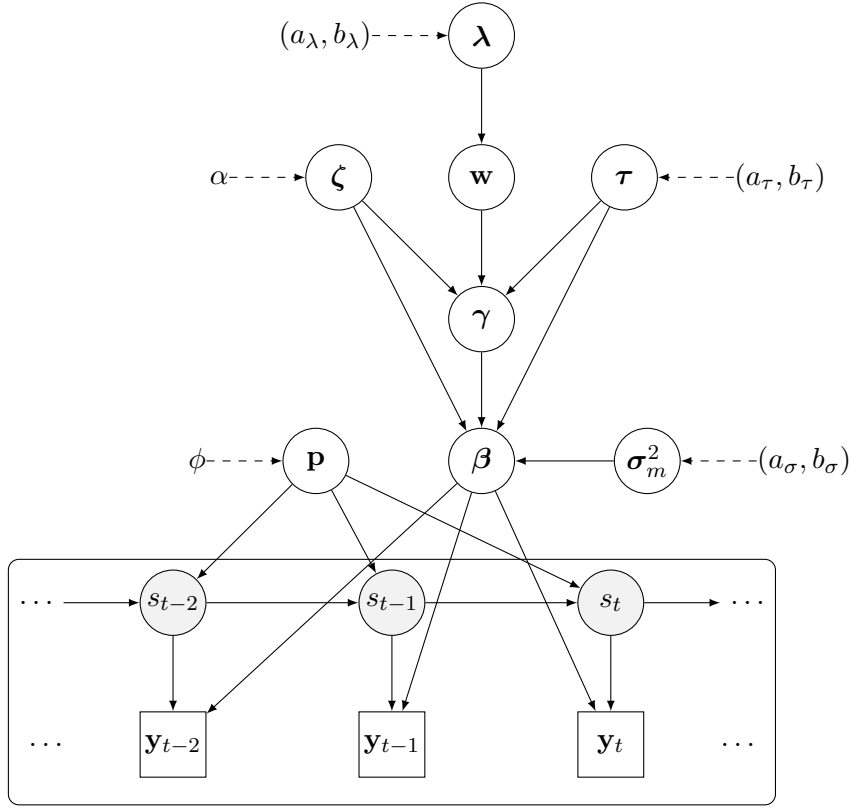

		\begin{center}
			\tikz[scale=1]{	
				\node[obs] (y1) {$\mathbf{y}_{t-1}$} ; %
				\node[obs, right = of y1,xshift = 1cm] (y2) {$\mathbf{y}_{t}$} ; %
				\node[const, right = of y2,xshift = 0cm] (y3) {$\ldots$} ; %
				\node[obs, left = of y1, xshift = -1cm] (y12) {$\mathbf{y}_{t-2}$} ;
				\node[const, left= of y12,xshift = 0cm] (y13) {$\ldots$} ; %
				\node[latent, above =of y1] (s1) {$s_{t-1}$} ; %
				\node[latent,right= of s1,xshift = 1cm] (s2) {$s_{t}$} ; %
				\node[const, right= of s2,xshift = 0cm] (s3) {$\,\,\ldots$} ; %
				\node[latent,left= of s1,xshift = -1cm] (s12) {$s_{t-2}$} ;
				\node[const, left= of s12,xshift = 0cm] (s13) {$\ldots\,\,$} ; %
				\edge {s1}{y1}
				\edge {s2}{y2}
				\edge {s12}{y12}
				\edge {s1}{s2}
				\edge {s2}{s3}
				\edge {s12}{s1}
				\edge {s13}{s12}
				\node[par,  above= of s1, xshift = -1cm, yshift = 0cm] (p) {$\mathbf{p}$};
				\node[const,  left= of p] (f) {$\phi$};
				\node[par,  right= of p, xshift = 0.3cm] (beta) {$\boldsymbol{\beta}$};
                    \node[par, right= of beta, xshift = 0.3cm] (sigma) {$\boldsymbol{\sigma}^2_{m}$};
				\node[par,  above= of beta, xshift = 0cm, yshift = 0cm] (gamma) {$\boldsymbol{\gamma}$};
    			\node[par,  above= of gamma, xshift = 0cm, yshift = 0cm] (w) {$\mathbf{w}$};
         		\node[par,  left= of w, xshift = 0cm, yshift = 0cm] (zeta) {$\boldsymbol{\zeta}$};
         		\node[par,  right= of w, xshift = 0cm, yshift = 0cm] (tau) {$\boldsymbol{\tau}$};
        		\node[par,  above= of w, xshift = 0cm, yshift = 0cm] (lambda) {$\boldsymbol{\lambda}$};
           		\node[const,  left= of lambda] (alambda) {$(a_{\lambda},b_{\lambda})$};
           		\node[const,  left= of zeta] (azeta) {$\alpha$};
                    \node[const,  right= of tau] (atau) {$(a_{\tau},b_{\tau})$};
                    \node[const, right= of sigma] (asig) {$(a_{\sigma}, b_{\sigma})$};
				\edge {beta}{y1}
				\edge {beta}{y2}
				\edge {beta}{y12}
				\edge {p}{s1}
				\edge {p}{s2}
				\edge {p}{s12}
				\edge{gamma}{beta}
				\edge{zeta}{gamma}
				\edge{zeta}{beta}
				\edge{tau}{gamma}
				\edge{tau}{beta}
                    \edge{w}{gamma}
                    \edge{lambda}{w}
                    \edge{sigma}{beta}
				\edge[style={dashed}]{alambda}{lambda}
				\edge[style={dashed}]{f}{p}
                    \edge[style={dashed}]{azeta}{zeta}
                    \edge[style={dashed}]{atau}{tau}
                    \edge[style={dashed}]{asig}{sigma}
				\plate {sec} {(y13) (y12) (y1) (y2) (y3) (s13) (s12) (s1) (s2) (s3)} {} ;
			}
		\end{center}
		\caption{
  DAG of the Bayesian Markov-switching Matrix Regression model. It exhibits the hierarchical structure of the observations $\mathbf{y}_t$ (boxes), the latent state variables $s_{t}$ (grey circles), the parameters $\beta_{\ell, m,j_m,k}^{(d)}$, $\sigma^2_{\ell,k}$, the hyper-parameters of the first stage $\gamma_{\ell,m,j_m,k}^{(d)}$, the second stage $\tau_{\ell,k}$, $\zeta^{(d)}_{\ell,m,k}$ and $w_{\ell,m,j_m,k}^{(d)}$ and the third stage $\lambda_{\ell,m,k}^{(d)}$ (white circles). The directed arrows show the conditional independence structure of the model.}\label{DAG}
	\end{figure}

\subsection{Sampling method}
The joint posterior distribution is not tractable, so we develop a MCMC algorithm to sample from it. Specifically, we use a Gibbs sampling procedure which combines two sampling strategies: i) back-fitting sampling  \citep{Hastie2000} for the coefficients and ii) forward filtering and backward sampling for the latent states \citep{Fruhwirth2006}. To cope with the computational cost of the Monte Carlo approximation, we implement a version of random scan Gibbs \cite{Rob2013}.

Let $\boldsymbol{\theta} = \left(\boldsymbol{\theta}_{1},\ldots, \boldsymbol{\theta}_{K}\right)$ be the collection of the state-specific parameters $\boldsymbol{\theta}_{k}=(\boldsymbol{\beta}_{k}$, $\boldsymbol{\gamma}_{k}, \boldsymbol{\zeta}_{k}$, $\tau_{k}, \boldsymbol{\lambda}_{k}, \boldsymbol{w}_{k})$ and denote with $\mathbf{y}=(\mathbf{y}_1,\ldots,\mathbf{y}_T)$, $\mathbf{X}=(X_1,\ldots, X_T)$ and $\boldsymbol{s} = (s_1, \ldots, s_T)$ the collection of response variables, covariates and state variables, respectively.

Since the joint posterior $p(\boldsymbol{\theta}\vert \mathbf{y},\mathbf{X})$ is not tractable, we follow a data augmentation strategy and introduce the joint posterior $p(\boldsymbol{\theta},\mathbf{s}\vert \mathbf{y},\mathbf{X})$. We sample groups of parameters and latent variables from their full conditional distributions, following  a block Gibbs scheme. Our sampling strategy deviates in three ways from the one in \cite{papadogeorgou2021soft}. First, we include the global shrinkage parameter $\tau$ not only in the prior for $\boldsymbol{\gamma}$ but also in the prior for $\boldsymbol{\beta}$. Second, we integrate out $\boldsymbol{\gamma}$ from the full conditional of $\boldsymbol{\beta}$ to allow $\boldsymbol{\beta}$ to depend directly on the observed data. The resulting collapsed Gibbs sampler allows us to achieve exact sampling for $\boldsymbol{\beta}$ and $\boldsymbol{\gamma}$ and to improve the sampler efficiency \citep{robert2007bayesian}. Third, we apply random scan Gibbs to increase the efficiency of the sampler \citep{Rob2013}.

At the first step of the Gibbs sampler, the $B_{\ell, k}$s and the PARAFAC margins are drawn from their full conditional distributions.  The back-fitting sampling strategy allows sampling from tractable distributions, i.e. conditionally normal distributions, and for splitting the parameter vector into blocks. Its implementation relies on the following equivalent representation of the regression model.
\begin{proposition}\label{prop:Backfitting}
The model in Eq. \eqref{eq.model} can be written as:
\begin{eqnarray}
&&y_{\ell t}=\sum_{k=1}^{K}\left({\boldsymbol{\beta}_{\ell, m,j_{m},k}^{(d)\top}} \Psi^{(d)}_{\ell, j_m,m, t,k}+R^{(d)}_{\ell, j_m,m,t,k}+R^{(d)}_{\ell, t,k}\right)\mathbb{I}(s_t=k)+ \sum_{k=1}^{K}\sigma^2_{\ell, k}\varepsilon_{\ell,t }\mathbb{I}(s_t=k)\notag
\end{eqnarray}
$\ell = 1, \ldots, N$, where the residual terms $R_{\ell, dt,k}$ and $R_{\ell, jmdt,k}$ and the auxiliary covariate vector $\Psi_{\ell, jmdt}$ are:
\begin{eqnarray}
R^{(d)}_{\ell, t}(s_t)&&=\sum_{d'\neq d}<B_{\ell, 1}(s_t)^{(d')}\circ \cdots \circ B_{\ell, M}^{(d')}(s_t),X_t> \notag\\
R^{(d)}_{\ell, m, j_m, t}(s_t)&&=<(B_{\ell, 1}^{(d)}(s_t)\circ \cdots \circ B_{\ell, M}^{(d)}(s_t))_{-j_{m}}, (X_t)_{-j_{m}}>\notag\\
\Psi^{(d)}_{\ell, m, j_m, t}(s_t) &&= \text{vec}((B_{\ell, 1}^{(d)}(s_t)\circ\cdots\circ B_{\ell, m-1}^{(d)}(s_t)\circ B_{\ell, m+1}^{(d)}(s_t) \circ \cdots \circ B_{\ell, M}^{(d)}(s_t)\circ X_t)_{\tilde{j}_{m}}) \notag
\end{eqnarray}
where $(A)_{-j_m}$ indicates the tensor after removing the $j_m$-th slice along mode $m$ and  $\tilde{j}_m=(:,\ldots,:,j_m,:,\ldots,:)$ is the collection of indexes along $m-1$ modes while keeping fix the index $j_m$ of the mode $m$.
\end{proposition}
At every iteration of the sampling algorithm, we randomly select the factor index $d$ from the proposal distribution $g(d)$ with support $\{1,2,\ldots, D\}$ and the model index $m$ from the distribution $h(m)$ with support $\{1,2, \ldots, M\}$. For $k = 1, \ldots, K $ and $ \ell = 1, \ldots, N$, all the elements of $B_{\ell, k}$ and the PARAFAC margins $\gamma^{(d)}_{\ell,m,j_m,k}$ are sampled from their full conditional distributions:
\begin{enumerate}
    \item Draw $\boldsymbol{\beta}_{\ell,m,j_m,k}^{(d)}$ from  $f(\boldsymbol{\beta}_{\ell,m,j_m,k}^{(d)}\vert \boldsymbol{y}, B, \beta_{\ell,m,-j_m,k}^{(d)},\boldsymbol{\zeta}^{(d)}_{\ell,k},\boldsymbol{\tau}_{\ell,k}, \boldsymbol{\lambda}^{(d)}_{\ell,m,k}, \boldsymbol{w}^{(d)}_{\ell,m,j_m,k})$ for $d \in \{1, \ldots, D\}$ which is a multivariate normal distribution, where the $d$ and $m$ have been randomly selected, i.e. $d\sim g(d)$ and $m\sim h(m)$.
    \item Draw $\gamma^{(d)}_{\ell,m,j_m,k}$ from $f(\gamma^{(d)}_{\ell,m,j_m,k}\vert \boldsymbol{\beta},\boldsymbol{\zeta}^{(d)}_{\ell,k},\boldsymbol{\tau}_{\ell,k}, \boldsymbol{\lambda}^{(d)}_{\ell,m,k}, \boldsymbol{w}^{(d)}_{\ell,m,j_m,k})$, which is a univariate normal distribution.
\end{enumerate}

The remaining parameters and hyper-parameters are sampled via the following Gibbs updates:

\begin{enumerate}[resume]
    \item Draw $\boldsymbol{\zeta}_{\ell, k}$ from the Generalized Inverse Gaussian distribution $f(\boldsymbol{\zeta}_{\ell, k} \vert \boldsymbol{\beta}_k, \boldsymbol{\gamma}_{\ell,k}, \boldsymbol{\lambda}^{(d)}_{\ell,m,k}, \boldsymbol{w}^{(d)}_{\ell,m,j_m,k})$.
    \item Draw $\boldsymbol{\tau}_{\ell,k}$ from the Generalized Inverse Gaussian distribution $f(\boldsymbol{\tau}_{\ell,k} \vert \boldsymbol{\beta}_{\ell,k}, \boldsymbol{\gamma}_{\ell,k}, \boldsymbol{\zeta}^{(d)}_{\ell,k}, \boldsymbol{w}^{(d)}_{\ell,k})$.
    \item Draw $\boldsymbol{\lambda}_{\ell,k}$ from $f(\boldsymbol{\lambda}_{\ell,k} \vert \boldsymbol{\gamma}_{\ell,k}, \boldsymbol{\tau}_{\ell,k}, \boldsymbol{\zeta}_{\ell,k} )$ which is a Gamma distribution.
    \item Draw $\boldsymbol{w}_{\ell,k}$ from the Generalized Inverse Gaussian distribution $f(\boldsymbol{w}_{\ell,k} \vert \boldsymbol{\gamma}_{\ell,k}, \boldsymbol{\lambda}_{\ell,k}, \boldsymbol{\tau}_{\ell,k}, \boldsymbol{\zeta}_{\ell,k})$.
    \item Draw transitional probabilities $(p_{k1},\ldots, p_{kK})$ from Dirichlet distribution $f(p_{k1},\ldots, p_{kK}\vert \boldsymbol{s})$.
\end{enumerate}
Regarding the hidden states, we apply a Forward-Filtering Backward-Sampling (FFBS) strategy.
\begin{enumerate}[resume]
    \item Compute iteratively the vector of smoothed probabilities $\xi_{t\vert T} = p(s_t \vert \boldsymbol{\theta}, \mathbf{y})$ by using the Hamilton filter recursions, and draw the state vector $s_t$ from the multinomial distribution $\mathcal{M}(1, \xi_{t\vert T})$.
\end{enumerate}
The derivation of the full conditional distributions for the parameters and the FFBS recursions can be found in Appendix \ref{app: deri}.

The current implementation is a variation of the usual Gibbs with a random scan. More concretely, consider that of interest is the posterior distribution $\pi(\boldsymbol{\theta})$ where $\boldsymbol{\theta} \in R^d$, but at each iteration, only a random subset of fixed size, say $k \le d$, of the parameter vector, is updated. Moreover, every set of indices of size $k$ has an equal chance of being selected. We describe in  Algorithm \ref{rpsg} the steps of the sampler,  which we call a random-partial-scan Gibbs (PRSG).
\begin{algorithm}
\caption{The steps in a Random-Partial-Scan Gibbs} \label{rpsg}
\begin{algorithmic}
     \STATE S1: Draw uniformly $I\subset \{1,\ldots,d\}$ a random set of indices of size $k\le d$ so that each subset has an equal chance of being selected.
    \STATE S2: If $I=(i_1,\ldots,i_k)$, update $\theta_I=(\theta_{i_1}, \theta_{i_2},\ldots, \theta_{i_k})$  using a random scan and leave the other components of $\theta$ unchanged.
    \end{algorithmic}
    \end{algorithm}

The transition kernel of the PRSG  satisfies the detailed balance condition, hence:
\begin{proposition}\label{prop:randGibbs}
 The chain generated by the PRSG sampler described in Algorithm \ref{rpsg} is an ergodic Markov chain with stationary distribution $\pi$.
\end{proposition}

To illustrate the performance of our proposed algorithm for tensor regression, we carried out an extensive simulation study for both simple and Markov Switching tensor regression. We study the proposed MCMC algorithm's efficiency by examining the MCMC chain empirical autocorrelation function (ACF) and the mean square error (MSE) of the true and sampled coefficient values. The regression model and the MCMC algorithm provide reasonably accurate estimates of the coefficients, and the regimes are successfully identified in the different experimental settings (see Appendix \ref{sec: sim_resu} in the Supplementary for further details).

\section{Empirical Application} \label{empapp}
We test the validity of our tensor regression model using two real-world applications. In particular, we would like to show through the applications that our tensor regression model with Markov Switching: i) outperforms the competing models (ordinary least square and linear LASSO) in terms of both in-sample and out-of-sample fitting; ii)  captures the structural / regime changes in the data by taking advantage of the Markov Switching Model.

\subsection{Volatility Index of US market}

In the first application, we study the relationship between the daily volatility index of the US market, also known as VIX and the crude oil ETF oil volatility index (OVX) with a number of other financial indicators. This is motivated by the fact that VIX has been recognized as the key benchmark index for measuring the market's expectations and sentiments, and predicting VIX is a crucial step for traders and investors when developing their trading strategies. In this regard, \cite{fernandes2014modeling} studied the long-range dependence in the VIX data by including a vector of the average of the logarithm of VIX for the last $k \in \left\{1, 5, 10, 22, 66\right\}$ days (to mirror daily, weekly, bi-weekly, monthly and quarterly component) in a family of heterogeneous autoregressive (HAR) processes.

We follow similar strategy as \cite{fernandes2014modeling} in forecasting VIX, but adapt it into a multiple-equation tensor regression framework, where we regress VIX on OVX (Eq. \eqref{eq:fin}) and vice versa (Eq. \eqref{eq:fin2}) together with other covariates: the $k$ day log return for S\&P 500, exchange rate (proxy by US dollar index), spot price of WTI crude oil for $k \in \{1, \ldots, 44\}$. To take advantage of the tensor structure, we construct the covariates for each response variable as a $4 \times 44$ matrix, which implies that the coefficient to be estimated is also a $4 \times 44$ matrix. The model specification is shown in Equation \eqref{eq:fin} and Equation (\ref{eq:fin2}).
    \begin{align}
    \text{VIX}_t &= \mu_{1}(s_t) + \left<B_{1}(s_t), \begin{pmatrix}
        \text{SP}_{t-1} & \ldots & \text{SP}_{t-h} & \ldots & \text{SP}_{t-44}\\
        \text{ER}_{t-1} & \ldots & \text{ER}_{t-h} & \ldots & \text{ER}_{t-44}\\
        \text{Oil}_{t-1} & \ldots &\text{Oil}_{t-h} & \ldots & \text{Oil}_{t-44}\\
        \text{OVX}_{t-1} &\ldots & \text{OVX}_{t-h} &\ldots & \text{OVX}_{t-44}
    \end{pmatrix}\right> + \sigma_{1}(s_t)\epsilon_{1t}
    \label{eq:fin}\\
    \text{OVX}_t &= \mu_{2}(s_t) + \left<B_{2}(s_t), \begin{pmatrix}
        \text{SP}_{t-1} & \ldots & \text{SP}_{t-h} & \ldots & \text{SP}_{t-44}\\
        \text{ER}_{t-1} & \ldots & \text{ER}_{t-h} & \ldots & \text{ER}_{t-44}\\
        \text{Oil}_{t-1} & \ldots &\text{Oil}_{t-h} & \ldots & \text{Oil}_{t-44}\\
        \text{VIX}_{t-1} &\ldots & \text{VIX}_{t-h} &\ldots & \text{VIX}_{t-44} \label{eq:fin2}
    \end{pmatrix}\right> + \sigma_{2}(s_t)\epsilon_{2t}
    \end{align}

\begin{figure}[p]
    \centering
    \begin{tabular}{cc}
        \includegraphics[width=.43\linewidth]{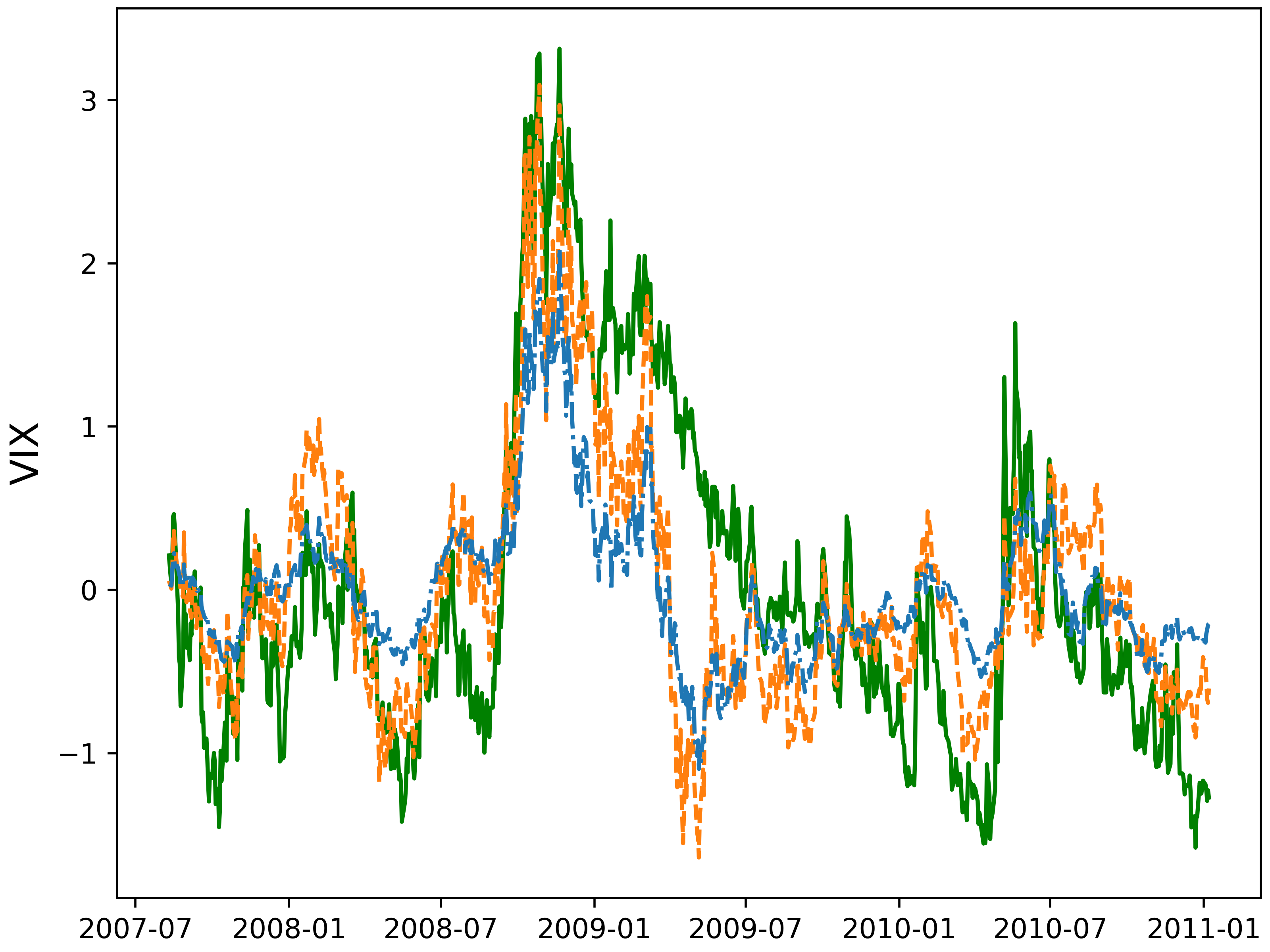} &\includegraphics[width=.43\linewidth]{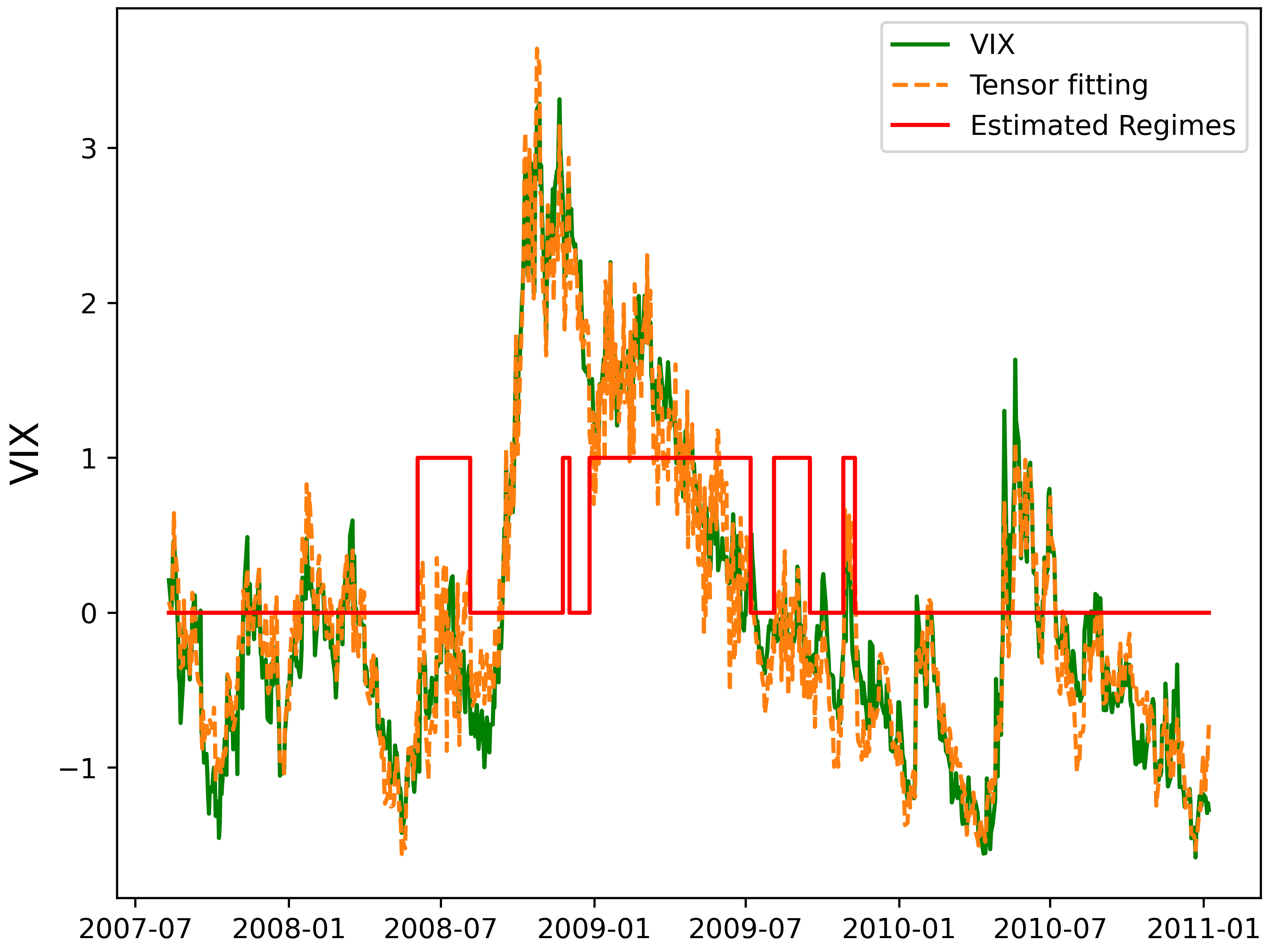}  \\
        \includegraphics[width=.43\linewidth]{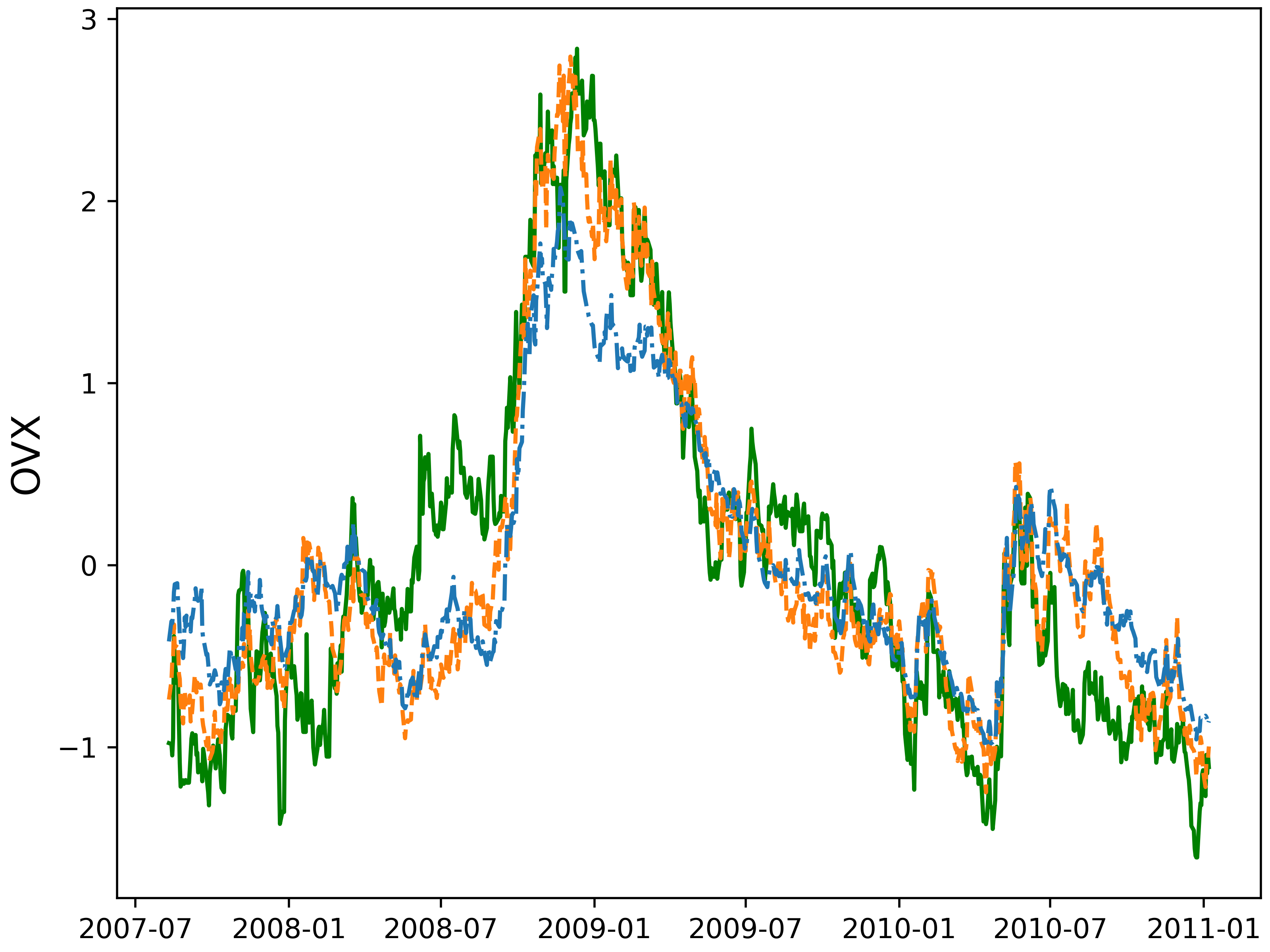} & \includegraphics[width=.43\linewidth]{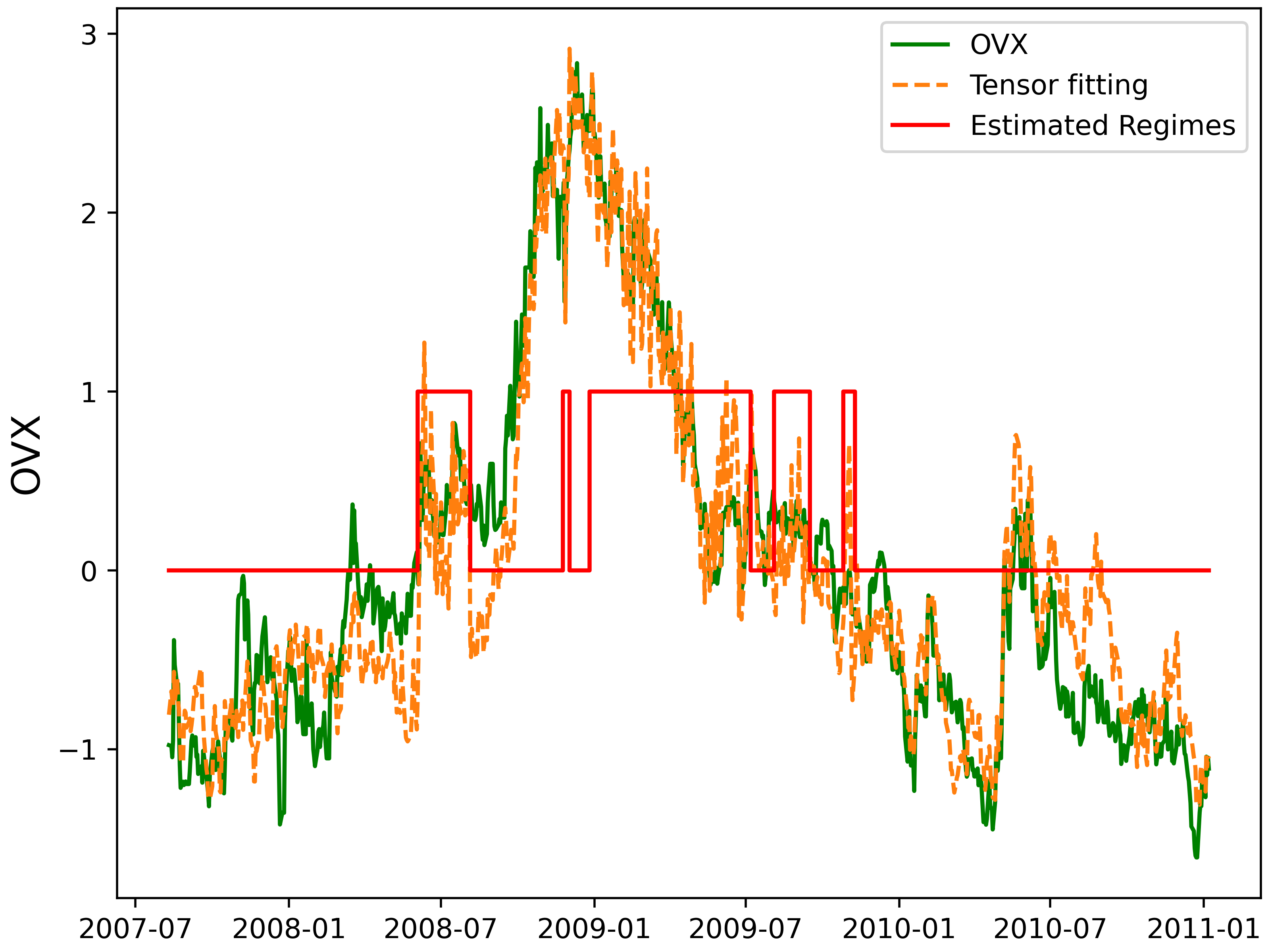} 
    \end{tabular}
    \caption{\small Left: In-sample fitting for Least Square (orange dashed) and LASSO (blue dashed). Right: In-sample fitting of Tensor Regression with Markov Switching (orange dashed) and estimated hidden states (red solid). The green solid line represents the VIX and VOX indexes (top and bottom).}
    \label{fig:insamp_fin}
\end{figure}

We show the in-sample fitting results of the MSMETR against the competing models in Figure \ref{fig:insamp_fin}. The in-sample fitting of the OLS and Linear LASSO regression fails to capture the structural changes in the series of VIX and OVX. However, these structural changes are successfully captured by an MSMETR, for which we assumed there are two possible regimes representing high and low levels of volatility. Indeed, the MSMETR identified two different regimes with distinctive regime-specific intercepts. Regime 2, representing a high level of volatility, has a higher intercept value than Regime 1, which represents a low level of volatility. 

\begin{figure}[p]
    \centering
    \begin{tabular}{cc}
        \includegraphics[width=.43\linewidth]{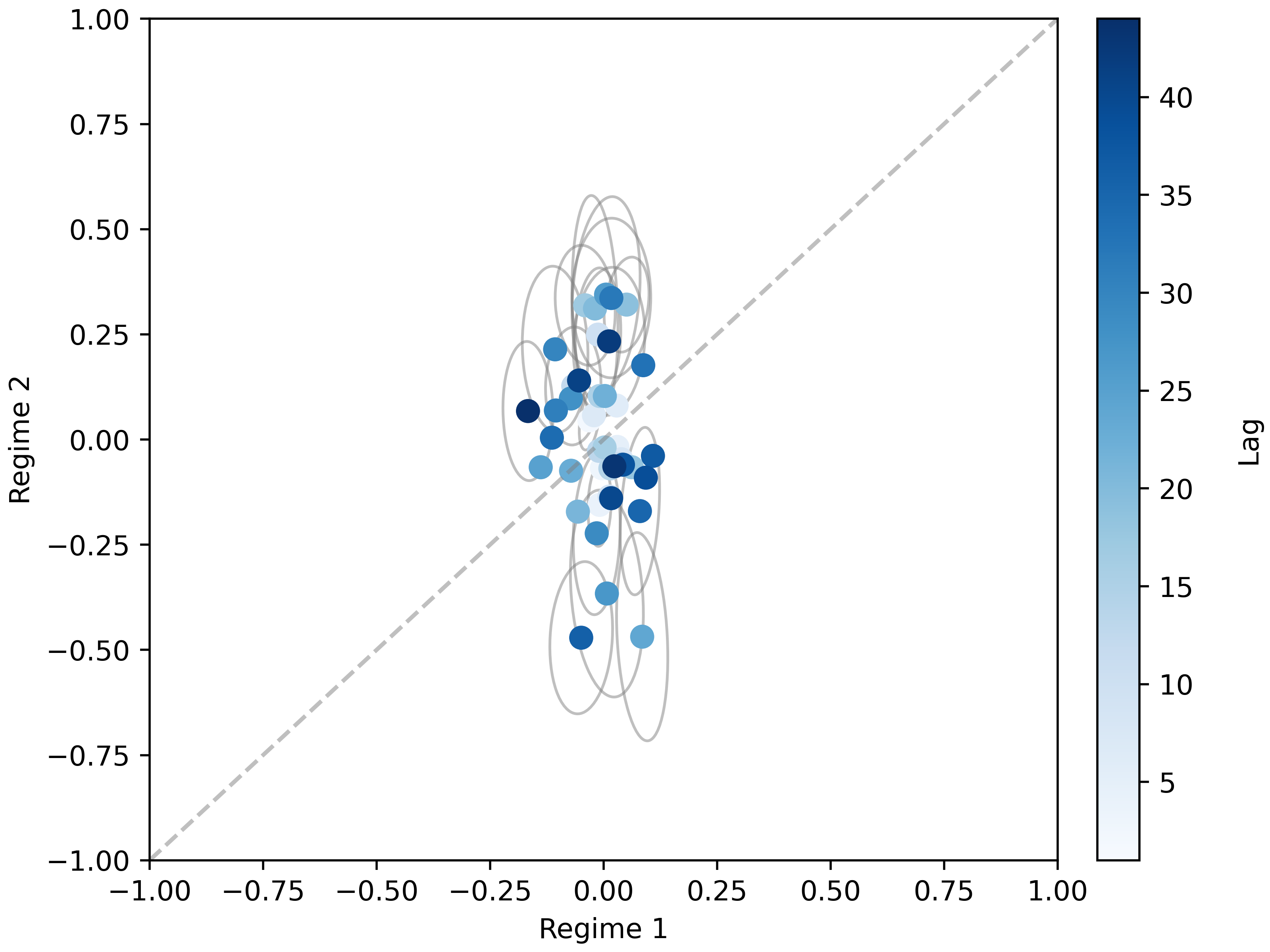} & \includegraphics[width=.43\linewidth]{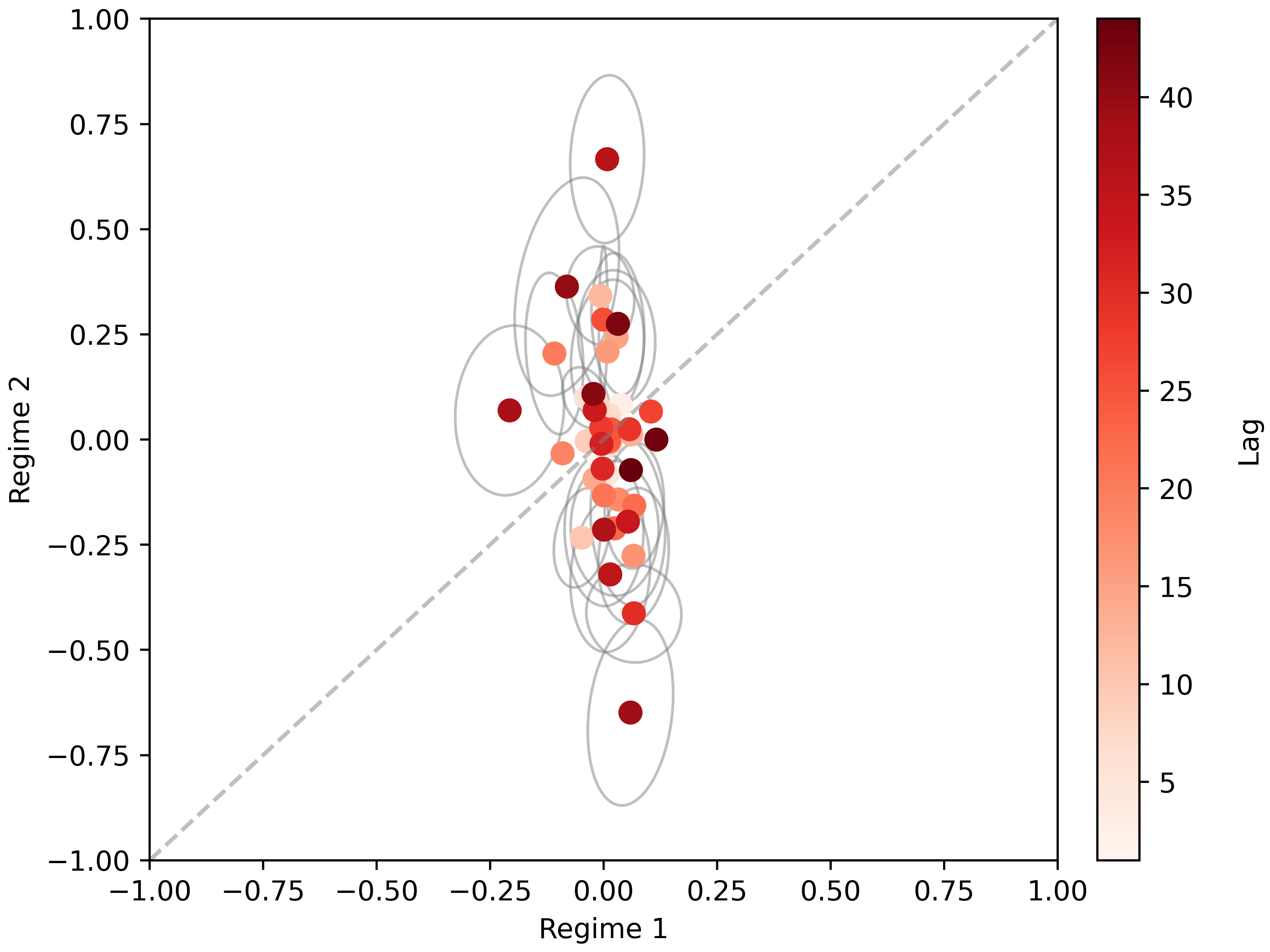}\\
        \includegraphics[width=.43\linewidth]{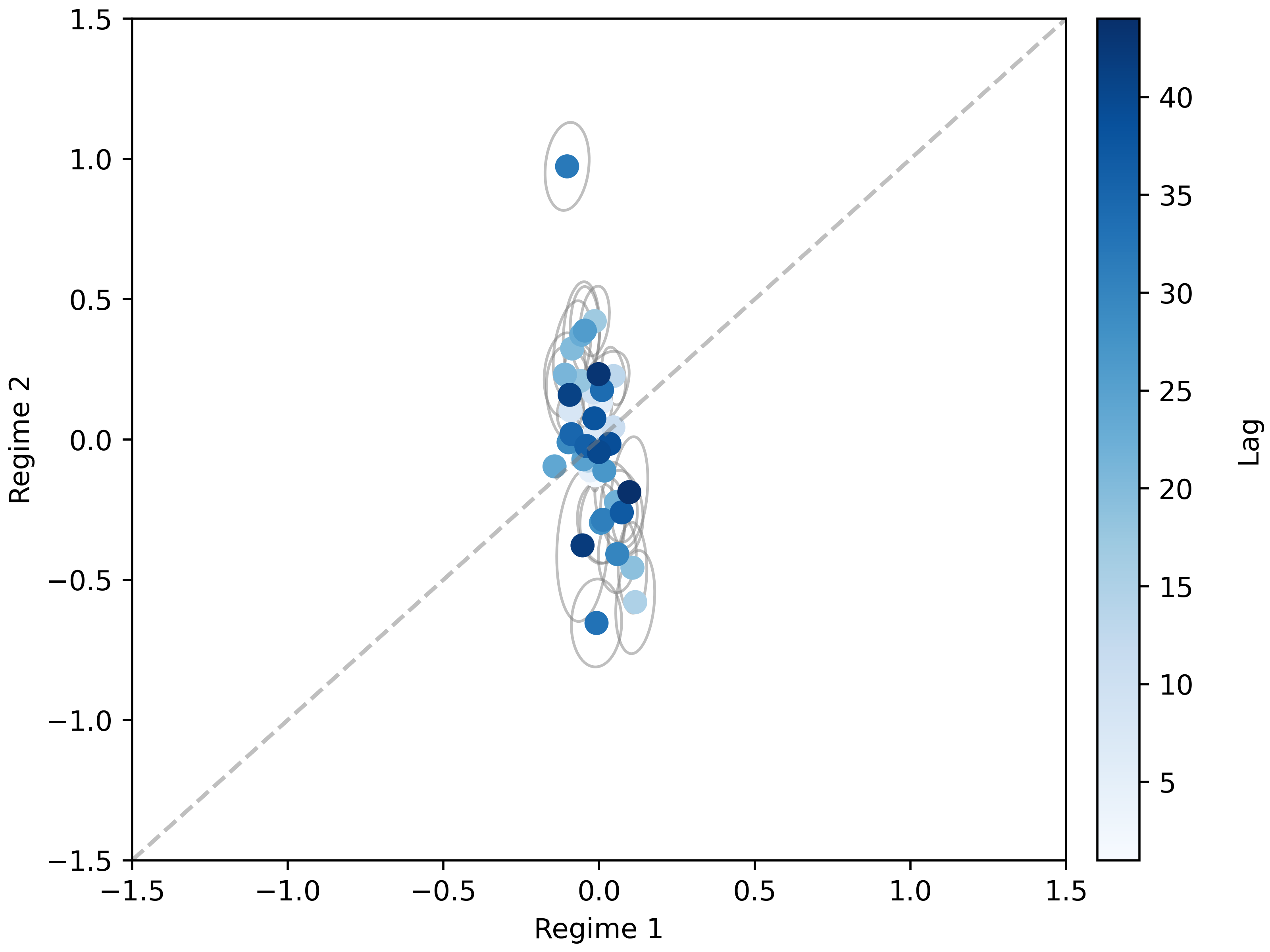} & \includegraphics[width=.43\linewidth]{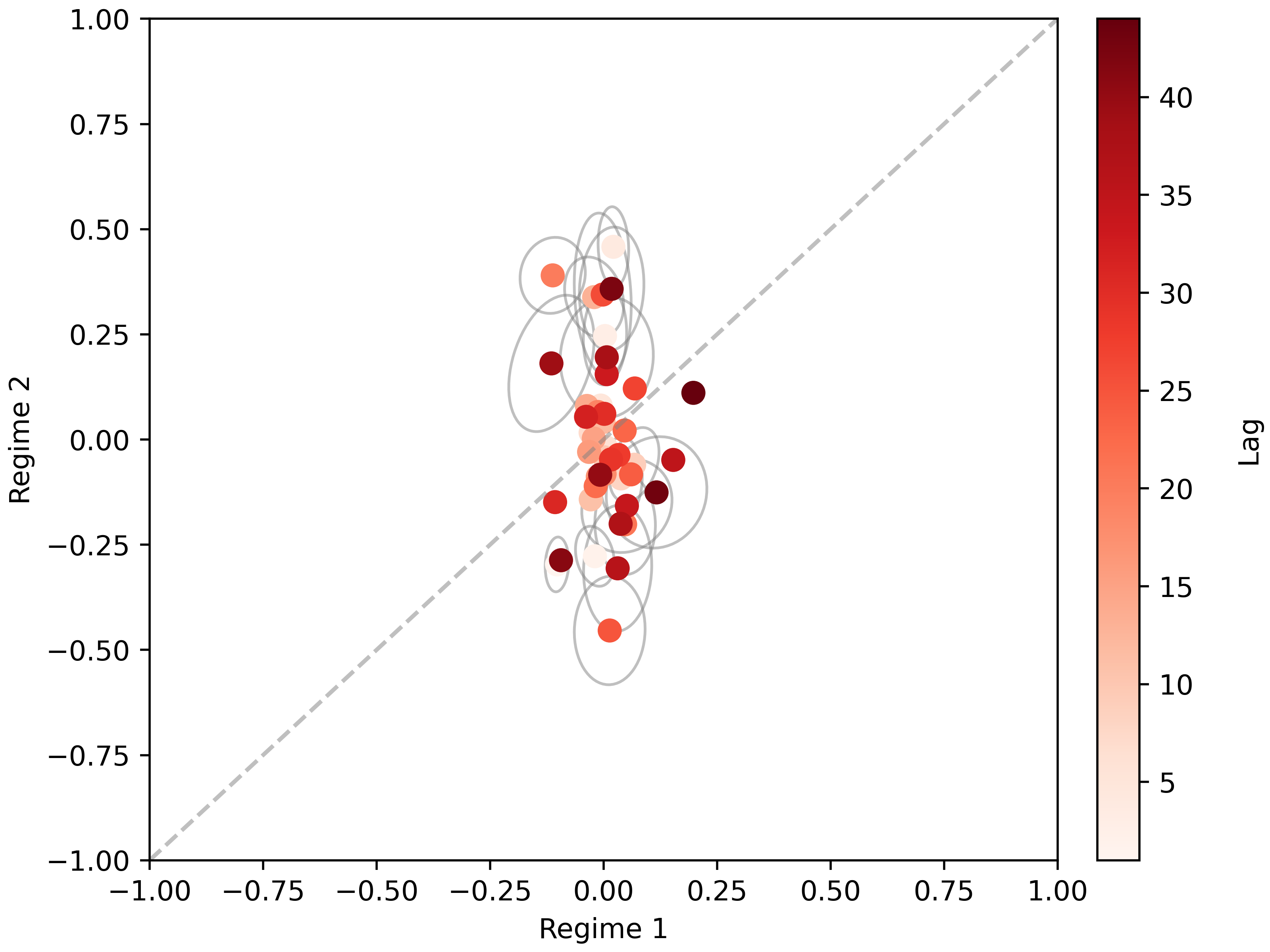}
    \end{tabular}
    \caption{Effects of $k$-day Oil (top) and S\&P 500 (bottom) $\log$ returns on VIX (left) and OVX (right) for $k \in \{1, \ldots, 44\}$. Lighter and darker colors represent smaller and larger $k$, respectively. $90\%$ Highest Posterior Density regions (gray ellipses) are plotted for the coefficients which exhibit asymmetric effects across regimes (HPD ellipse does not  intersect the $45^{\circ}$ line). }
    \label{fig:scatter}
\end{figure}

We show the effects of $k$-day $\log$ return of oil prices and S\&P 500 on VIX (blue dots) and OVX (red dots), respectively, in Figure \ref{fig:scatter}. The dots in the plots correspond to the values of parameters in the low-volatility ($s_t=1)$ and in the high-volatility ($s_t=2$) regimes. The $90\%$ HPD regions (grey ellipses) provide evidence of coefficient heterogeneity across regimes (asymmetric effects), equations (market asymmetry) and lags (long-term effects). 

Regarding the asymmetric effects, we found evidence of the limited impact of the $k$-day oil and S\&P 500 $\log$ returns on both VIX and OVX in the low-volatility regime.  The values of coefficients are mostly centred around zero in this regime. As for the market asymmetry, the returns on oil have a larger effect on OVX than on VIX in the high-volatility regime. Oil returns have a small impact on VIX and a more pronounced impact on OVX.

There is also strong evidence of non-negligible long-term effects of oil prices (dark red points) on oil volatility, consistently with previous findings \citep{bandi2006long, corsi2009simple}. The returns on the S\&P 500 have a very limited effect on VIX in the low-volatility regime. In contrast, the effects of returns at all lags are more substantial during high-volatility periods. The effects of S\&P 500 on OVX follow a similar pattern; the coefficients with medium lags tend to have a larger effect than lower and higher lags. From the shape of the ellipses, we can tell that the coefficients are mostly uncorrelated across regimes, with fewer coefficients showing small positive or negative correlations. The coefficient posterior variance in the low-volatility regime is generally larger than in the high-volatility.

In Table \ref{tab:stats fin} of the Supplement, we report the mean square error (MSE) and mean absolute error (MAE) for the in-sample fitting and out-of-sample forecasting with prediction horizons of $1$-day and $5$-day. Our proposed MSMETR consistently outperforms the Least Squares and linear LASSO across almost all the measures.

\subsection{Oil Prices on Stock Return}
For the second application, we extend our matrix-variate tensor regression model to a 3-mode tensor regression model by constructing the covariates $\mathcal{X}_t$ as a three-dimensional array for each observation. Therefore, the coefficients $\mathcal{B}_{\ell}$ also form a three-dimensional array with the same size as the covariates. In this application, we contribute to the debate on the interdependence between financial and oil markets \citep[e.g., see][]{xiao2022good,XIAO2023106969} and examine the impact of oil price volatility on the stock market returns (S\&P 500) at an aggregate level and on the financial sector, energy sector and other sectors of S\&P 500 at the disaggregate level. In particular, we classified the oil price volatility into Good Oil Volatility (GV), where the realized volatility is positive, and Bad Oil Volatility (BV), where the realized volatility is negative.

Our approach is different from the one in \cite{xiao2022good} in that we consider a Mixed Data Sampling (MIDAS) \citep{ghysels2004midas} framework by taking advantage of the tensor structure of the covariates. Our tensor regression setting naturally accommodates the multi-array structure of the covariates when data are sampled at different frequencies with different lags and reduces the number of parameters to be estimated by shrinking the unimportant parameters to small values. In particular,  the response variable $R_{\ell, t}$ is the 4-week log-return of market $\ell$ at time $t$, where $\ell =$ \{S\&P 500, financial sector, energy sector, other sectors in S\&P 500\}. The covariates are sampled weekly at the 1st week, 2nd week, 3rd week, 4th week before time $t$, indexed by $t-\frac{1}{4}, t-\frac{2}{4}, t-\frac{3}{4}, t-\frac{4}{4}$. Together with the GV and BV, the other covariates are the exchange rate volatility (ER), TED spread volatility (IR) and VIX index volatility (VI), following a similar specification as in \cite{xiao2022good}. 

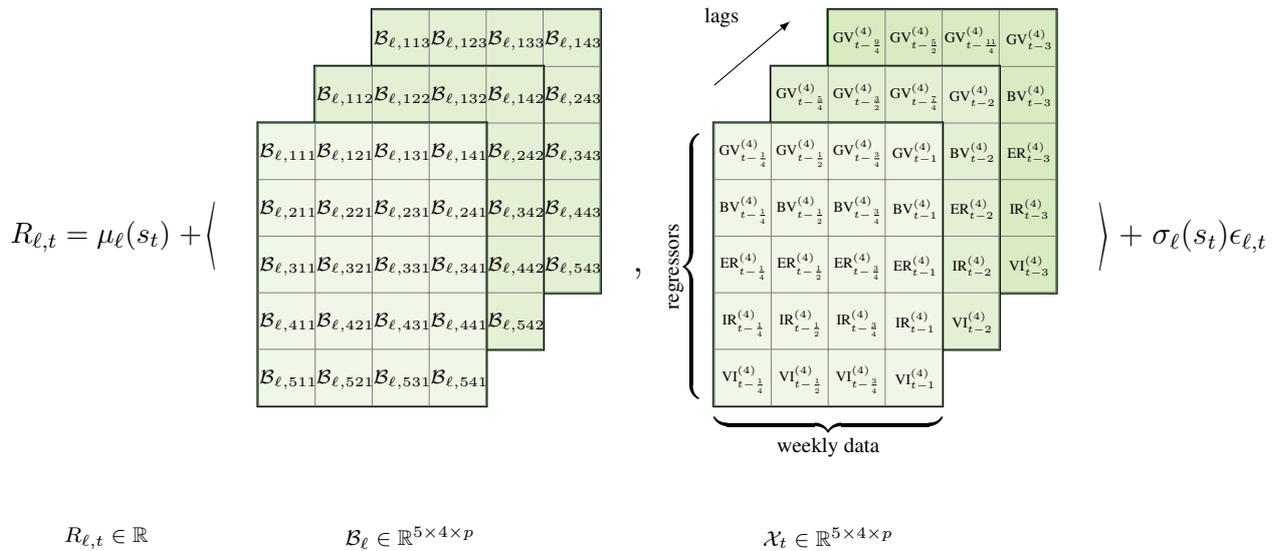
\begin{figure}[h!]
    \centering


\begin{tikzpicture}
    
    \draw [very thick] (1.5+1.5,-5.25+1.5) rectangle (4.5+1.5,-1.5+1.5);
    \filldraw [fill=LimeGreen!30!white,draw=green!40!black] (1.5+1.5,-5.25+1.5) rectangle (4.5+1.5,-1.5+1.5);
    \draw [step=0.75, very thin, color=gray] (1.5+1.5,-5.25+1.5) grid (4.5+1.5,-1.5+1.5);
    
    \draw [very thick] (1.5+.75,-5.25+.75) rectangle (4.5+.75,-1.5+.75);
    \filldraw [fill=LimeGreen!20!white,draw=green!40!black] (1.5+.75,-5.25+.75) rectangle (4.5+.75,-1.5+.75);
    \draw [step=0.75, very thin, color=gray] (1.5+.75,-5.25+.75) grid (4.5+.75,-1.5+.75);
    
    \draw [very thick] (1.5,-5.25) rectangle (4.5,-1.5);
    \filldraw [fill=LimeGreen!10!white,draw=green!40!black] (1.5,-5.25) rectangle (4.5,-1.5);
    \draw [step=0.75, very thin, color=gray] (1.5,-5.25) grid (4.5,-1.5);
    
    \draw (3,-7) node
    {{\color{black}\scriptsize{$\mathcal{X}_{t}\in\mathbb{R}^{5 \times 4 \times p}$}}};
    
    \draw [very thick] (-4.5+1.5,-5.25+1.5) rectangle (-1.5+1.5,-1.5+1.5);
    \filldraw [fill=LimeGreen!20!white,draw=green!40!black] (-4.5+1.5,-5.25+1.5) rectangle (-1.5+1.5,-1.5+1.5);
    \draw [step=0.75, very thin, color=gray] (-4.5+1.5,-5.25+1.5) grid (-1.5+1.5,-1.5+1.5);
    
    \draw [very thick] (-4.5+.75,-5.25+.75) rectangle (-1.5+.75,-1.5+.75);
    \filldraw [fill=LimeGreen!20!white,draw=green!40!black] (-4.5+.75,-5.25+.75) rectangle (-1.5+.75,-1.5+.75);
    \draw [step=0.75, very thin, color=gray] (-4.5+.75,-5.25+.75) grid (-1.5+.75,-1.5+.75);
    
    \draw [very thick] (-4.5,-5.25) rectangle (-1.5,-1.5);
    \filldraw [fill=LimeGreen!10!white,draw=green!40!black] (-4.5,-5.25) rectangle (-1.5,-1.5);
    \draw [step=0.75, very thin, color=gray] (-4.5,-5.25) grid (-1.5,-1.5);
    
    \draw (-2.5,-7) node
    {{\color{black}\scriptsize{$\mathcal{B}_{\ell}\in\mathbb{R}^{5 \times 4 \times p}$}}};

    \draw (.5, -3.5) node
    {{\color{black}\Large{$,$}}};   
    
    \draw (-6.5, -3) node {{\color{black}$R_{\ell, t} = \mu_{\ell} (s_t) \; + $}};

    \draw (-6.5,-7) node {{\color{black}\scriptsize{$R_{\ell, t}\in\mathbb{R}$}}};

    \draw (-5, -3) node [rotate = 90] {$\widehat{\hspace{5cm}}$};

    \draw (6.5, -3) node [rotate = 270] {{$\widehat{\hspace{5cm}}$}};

    \draw (3,-5.5,0) node[rotate = 0] {{\color{black}$\underbrace{\hspace{3cm}}$}};

    \draw (3,-5.8,0) node {\scriptsize{\color{black}\scriptsize{weekly data}}};

    \draw [very thin] (1.2,-3.4,0) node[rotate = 270] {{\color{black}$\underbrace{\hspace{3.5cm}}$}};

    \draw (1,-3.4,0) node[rotate = 90] {\scriptsize{\color{black}\scriptsize{regressors}}};

    \draw (7.8,-3) node {{\color{black}\large{$+\; \sigma_{\ell}(s_t) \epsilon_{\ell, t}$}}};

    \draw [->] (1.5,-1) -- (2.5,-.15);
    
    \draw (1.6,-.1) node {{\color{black}\scriptsize{lags}}};

    \draw (1.9,-4.9) node {\fontsize{5.5pt}{2pt}\selectfont $\text{VI}^{(4)}_{t-\frac{1}{4}}$};
    \draw (2.65,-4.9) node {\fontsize{5.5pt}{2pt}\selectfont$\text{VI}^{(4)}_{t-\frac{1}{2}}$};
    \draw (3.4,-4.9) node {\fontsize{5.5pt}{2pt}\selectfont$\text{VI}^{(4)}_{t-\frac{3}{4}}$};
    \draw (4.15,-4.9) node {\fontsize{5.5pt}{2pt}\selectfont$\text{VI}^{(4)}_{t-1}$};

    \draw (1.9,-4.15) node {\fontsize{5.5pt}{2pt}\selectfont$\text{IR}^{(4)}_{t-\frac{1}{4}}$};
    \draw (2.65,-4.15) node {\fontsize{5.5pt}{2pt}\selectfont$\text{IR}^{(4)}_{t-\frac{1}{2}}$};
    \draw (3.4,-4.15) node {\fontsize{5.5pt}{2pt}\selectfont$\text{IR}^{(4)}_{t-\frac{3}{4}}$};
    \draw (4.15,-4.15) node {\fontsize{5.5pt}{2pt}\selectfont$\text{IR}^{(4)}_{t-1}$};
    \draw (4.9,-4.15) node {\fontsize{5.5pt}{2pt}\selectfont$\text{VI}^{(4)}_{t-2}$};

    \draw (1.9,-3.4) node {\fontsize{5.5pt}{2pt}\selectfont$\text{ER}^{(4)}_{t-\frac{1}{4}}$};
    \draw (2.65,-3.4) node {\fontsize{5.5pt}{2pt}\selectfont$\text{ER}^{(4)}_{t-\frac{1}{2}}$};
    \draw (3.4,-3.4) node {\fontsize{5.5pt}{2pt}\selectfont$\text{ER}^{(4)}_{t-\frac{3}{4}}$};
    \draw (4.15,-3.4) node {\fontsize{5.5pt}{2pt}\selectfont$\text{ER}^{(4)}_{t-1}$};
    \draw (4.9,-3.4) node {\fontsize{5.5pt}{2pt}\selectfont$\text{IR}^{(4)}_{t-2}$};
    \draw (5.65,-3.4) node {\fontsize{5.5pt}{2pt}\selectfont$\text{VI}^{(4)}_{t-3}$};

    \draw (1.9,-2.65) node {\fontsize{5.5pt}{2pt}\selectfont$\text{BV}^{(4)}_{t-\frac{1}{4}}$};
    \draw (2.65,-2.65) node {\fontsize{5.5pt}{2pt}\selectfont$\text{BV}^{(4)}_{t-\frac{1}{2}}$};
    \draw (3.4,-2.65) node {\fontsize{5.5pt}{2pt}\selectfont$\text{BV}^{(4)}_{t-\frac{3}{4}}$};
    \draw (4.15,-2.65) node {\fontsize{5.5pt}{2pt}\selectfont$\text{BV}^{(4)}_{t-1}$};
    \draw (4.9,-2.65) node {\fontsize{5.5pt}{2pt}\selectfont$\text{ER}^{(4)}_{t-2}$};   
    \draw (5.65,-2.65) node {\fontsize{5.5pt}{2pt}\selectfont$\text{IR}^{(4)}_{t-3}$}; 

    \draw (1.9,-1.9) node {\fontsize{5.5pt}{2pt}\selectfont$\text{GV}^{(4)}_{t-\frac{1}{4}}$};
    \draw (2.65,-1.9) node {\fontsize{5.5pt}{2pt}\selectfont$\text{GV}^{(4)}_{t-\frac{1}{2}}$};
    \draw (3.4,-1.9) node {\fontsize{5.5pt}{2pt}\selectfont$\text{GV}^{(4)}_{t-\frac{3}{4}}$};
    \draw (4.15,-1.9) node {\fontsize{5.5pt}{2pt}\selectfont$\text{GV}^{(4)}_{t-1}$};
    \draw (4.9,-1.9) node {\fontsize{5.5pt}{2pt}\selectfont$\text{BV}^{(4)}_{t-2}$}; 
    \draw (5.65,-1.9) node {\fontsize{5.5pt}{2pt}\selectfont$\text{ER}^{(4)}_{t-3}$}; 

    \draw (2.65,-1.15) node {\fontsize{5.5pt}{2pt}\selectfont$\text{GV}^{(4)}_{t-\frac{5}{4}}$};
    \draw (3.4,-1.15) node {\fontsize{5.5pt}{2pt}\selectfont$\text{GV}^{(4)}_{t-\frac{3}{2}}$};
    \draw (4.15,-1.15) node {\fontsize{5.5pt}{2pt}\selectfont$\text{GV}^{(4)}_{t-\frac{7}{4}}$};
    \draw (4.9,-1.15) node {\fontsize{5.5pt}{2pt}\selectfont$\text{GV}^{(4)}_{t-2}$};
    \draw (5.65,-1.15) node {\fontsize{5.5pt}{2pt}\selectfont$\text{BV}^{(4)}_{t-3}$};

    \draw (3.4,-.4) node {\fontsize{5.5pt}{2pt}\selectfont $\text{GV}^{(4)}_{t-\frac{9}{4}}$};
    \draw (4.15,-.4) node {\fontsize{5.5pt}{2pt}\selectfont$\text{GV}^{(4)}_{t-\frac{5}{2}}$};
    \draw (4.9,-.4) node {\fontsize{5.5pt}{2pt}\selectfont$\text{GV}^{(4)}_{t-\frac{11}{4}}$};
    \draw (5.65,-.4) node {\fontsize{5.5pt}{2pt}\selectfont$\text{GV}^{(4)}_{t-3}$};

    \draw (-4.1,-4.9) node {\fontsize{7.3pt} {1pt}\selectfont $\mathcal{B}_{\ell, 511}$};
    
    \draw (-3.35,-4.9) node {\fontsize{7.3pt}{1pt}\selectfont $\mathcal{B}_{\ell, 521}$};

    \draw (-2.6,-4.9) node {\fontsize{7.3pt}{1pt}\selectfont $\mathcal{B}_{\ell, 531}$};

    \draw (-1.85,-4.9) node {\fontsize{7.3pt}{1pt}\selectfont $\mathcal{B}_{\ell, 541}$};

     \draw (-4.1,-4.15) node {\fontsize{7.3pt}{1pt}\selectfont$\mathcal{B}_{\ell, 411}$};

     \draw (-3.35,-4.15) node {\fontsize{7.3pt}{1pt}\selectfont$\mathcal{B}_{\ell, 421}$};

     \draw (-2.6,-4.15) node {\fontsize{7.3pt}{1pt}\selectfont$\mathcal{B}_{\ell, 431}$};

     \draw (-1.85,-4.15) node {\fontsize{7.3pt}{1pt}\selectfont$\mathcal{B}_{\ell, 441}$};

     \draw (-1.1,-4.15) node {\fontsize{7.3pt}{1pt}\selectfont$\mathcal{B}_{\ell, 542}$};

     \draw (-4.1,-3.4) node {\fontsize{7.3pt}{1pt}\selectfont$\mathcal{B}_{\ell, 311}$};

     \draw (-3.35,-3.4) node {\fontsize{7.3pt}{1pt}\selectfont$\mathcal{B}_{\ell, 321}$};

     \draw (-2.6,-3.4) node {\fontsize{7.3pt}{1pt}\selectfont$\mathcal{B}_{\ell, 331}$};

    \draw (-1.85,-3.4) node {\fontsize{7.3pt}{1pt}\selectfont$\mathcal{B}_{\ell, 341}$};

    \draw (-1.1,-3.4) node {\fontsize{7.3pt}{1pt}\selectfont$\mathcal{B}_{\ell, 442}$};

    \draw (-0.35,-3.4) node {\fontsize{7.3pt}{1pt}\selectfont$\mathcal{B}_{\ell, 543}$};

    \draw (-4.1,-2.65) node {\fontsize{7.3pt}{1pt}\selectfont$\mathcal{B}_{\ell, 211}$};

    \draw (-3.35,-2.65) node {\fontsize{7.3pt}{1pt}\selectfont$\mathcal{B}_{\ell, 221}$};

    \draw (-2.6,-2.65) node {\fontsize{7.3pt}{1pt}\selectfont$\mathcal{B}_{\ell, 231}$};

    \draw (-1.85,-2.65) node {\fontsize{7.3pt}{1pt}\selectfont$\mathcal{B}_{\ell, 241}$};

    \draw (-1.1,-2.65) node {\fontsize{7.3pt}{1pt}\selectfont$\mathcal{B}_{\ell, 342}$};

    \draw (-0.35,-2.65) node {\fontsize{7.3pt}{1pt}\selectfont$\mathcal{B}_{\ell, 443}$};

    \draw (-4.1,-1.9) node {\fontsize{7.3pt}{1pt}\selectfont$\mathcal{B}_{\ell, 111}$};

    \draw (-3.35,-1.9) node {\fontsize{7.3pt}{1pt}\selectfont$\mathcal{B}_{\ell, 121}$};

    \draw (-2.6,-1.9) node {\fontsize{7.3pt}{1pt}\selectfont$\mathcal{B}_{\ell, 131}$};

    \draw (-1.85,-1.9) node {\fontsize{7.3pt}{1pt}\selectfont$\mathcal{B}_{\ell, 141}$};

    \draw (-1.1,-1.9) node {\fontsize{7.3pt}{1pt}\selectfont$\mathcal{B}_{\ell, 242}$};

    \draw (-0.35,-1.9) node {\fontsize{7.3pt}{1pt}\selectfont$\mathcal{B}_{\ell, 343}$};

    \draw (-3.35,-1.15) node {\fontsize{7.3pt}{1pt}\selectfont$\mathcal{B}_{\ell, 112}$};

    \draw (-2.6,-1.15) node {\fontsize{7.3pt}{1pt}\selectfont$\mathcal{B}_{\ell, 122}$};

    \draw (-1.85,-1.15) node {\fontsize{7.3pt}{1pt}\selectfont$\mathcal{B}_{\ell, 132}$};

    \draw (-1.1,-1.15) node {\fontsize{7.3pt}{1pt}\selectfont$\mathcal{B}_{\ell, 142}$};

    \draw (-0.35,-1.15) node {\fontsize{7.3pt}{1pt}\selectfont$\mathcal{B}_{\ell, 243}$};

    \draw (-2.6,-0.4) node {\fontsize{7.3pt}{1pt}\selectfont$\mathcal{B}_{\ell, 113}$};

    \draw (-1.85,-0.4) node {\fontsize{7.3pt}{1pt}\selectfont$\mathcal{B}_{\ell, 123}$};

    \draw (-1.1,-0.4) node {\fontsize{7.3pt}{1pt}\selectfont$\mathcal{B}_{\ell, 133}$};

    \draw (-0.35,-0.4) node {\fontsize{7.3pt}{1pt}\selectfont$\mathcal{B}_{\ell, 143}$};

\end{tikzpicture}
    \caption{Graphic Representation of Tensor Regression for Macro Application}
    \label{app2}
\end{figure}

We arrange the different regressors along the rows (first mode) of the tensor covariates, data points sampled weekly for four weeks along the columns (second mode), and the weekly data points of the previous four weeks are stacked along the third axis (third mode) for $p$ lags. A graphical representation of the specification of the tensor regression model is shown in Fig. \ref{app2}, 
where $\mu_{\ell}({s_t})$ and $\sigma_{\ell}(s_t)$ are regime-specific intercepts and standard deviations, $\mathcal{B}_{\ell}$ and $\mathcal{X}_t$ are, respectively, tensor coefficients and covariates of sizes $\mathbb{R}^{5 \times 4 \times p}$, and $p$ represents the number of lags we wish to consider for the covariates.

Figure \ref{fig:insamp_mac1} of the Supplement shows the in-sample fitting of Least Squares and Linear LASSO (left column) and the in-sample fitting of MSMETR (right column). Notably, Least Squares and linear LASSO fail to capture the volatility changes in the market return. In contrast, the MSMETR can identify the most relevant episodes of market disruptions at both aggregate and disaggregate levels. 
For the aggregate analysis, when S\&P 500 is used as the dependent variable, MSMETR can identify the biggest disruption in the financial market in recent years, the 2008 global financial crisis.

\begin{figure}[t]
    \centering
    \begin{tabular}{cc}
        \includegraphics[width=.43\linewidth]{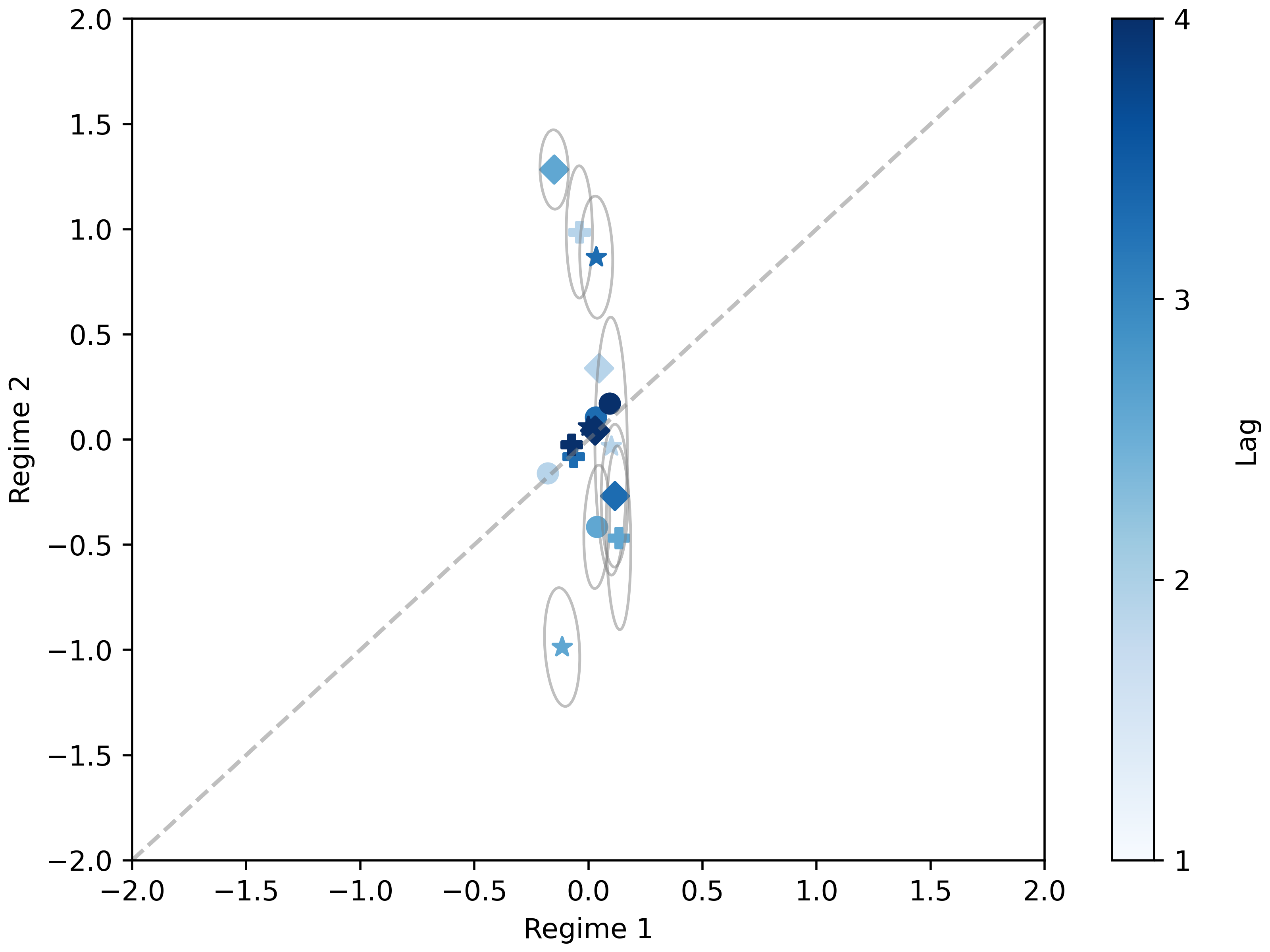} & \includegraphics[width=.43\linewidth]{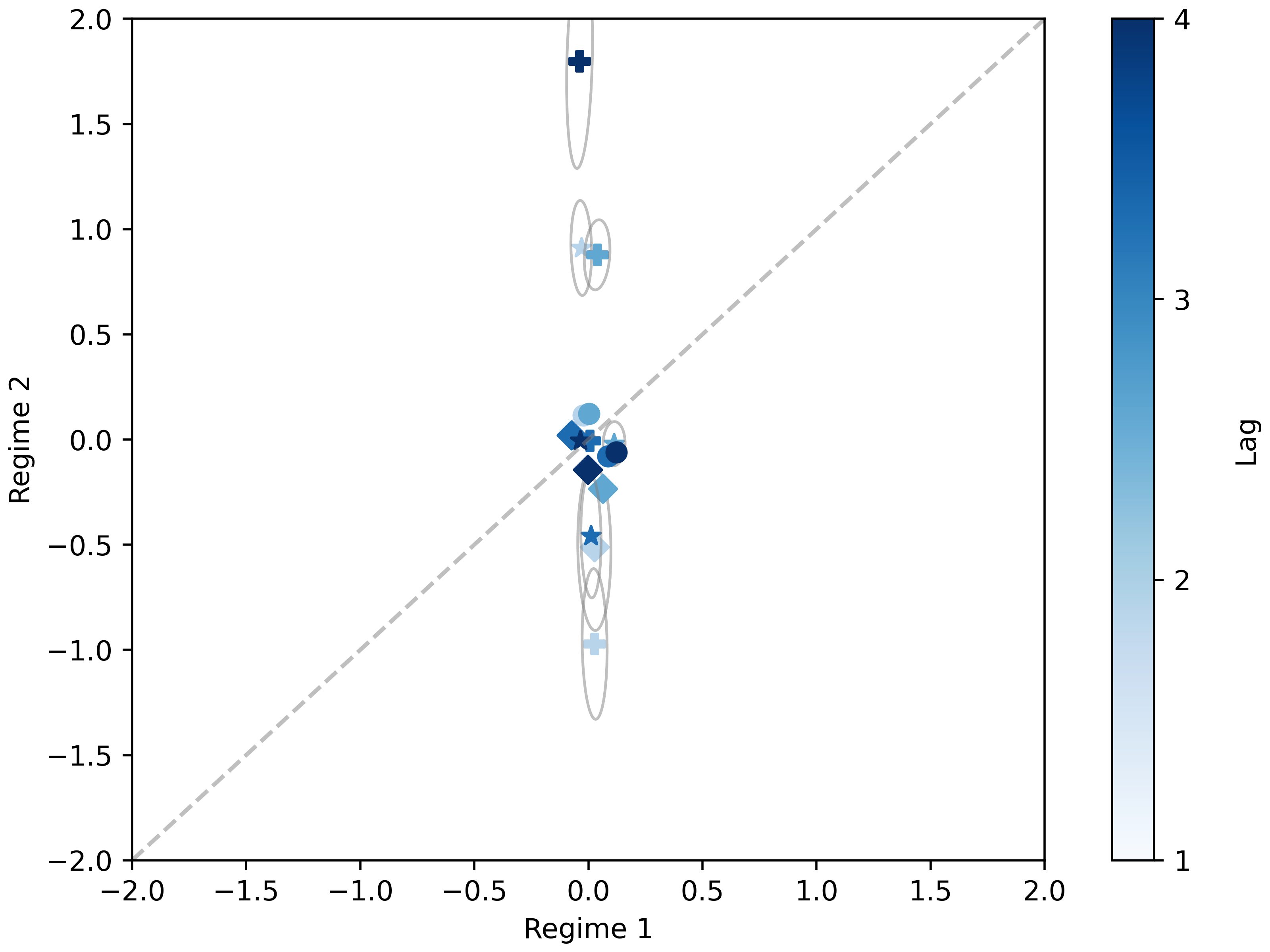}\\
        \includegraphics[width=.43\linewidth]{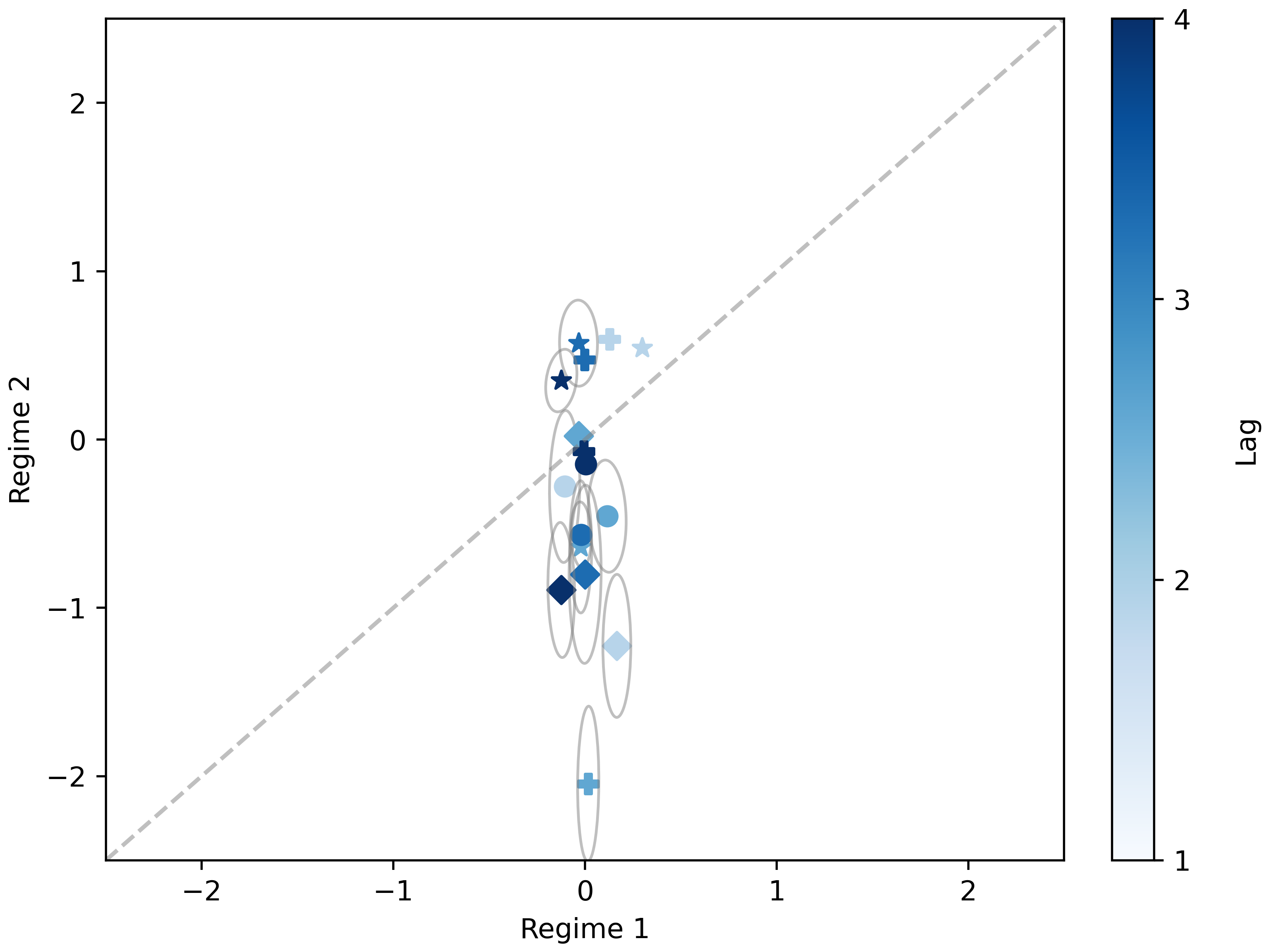} & \includegraphics[width=.43\linewidth]{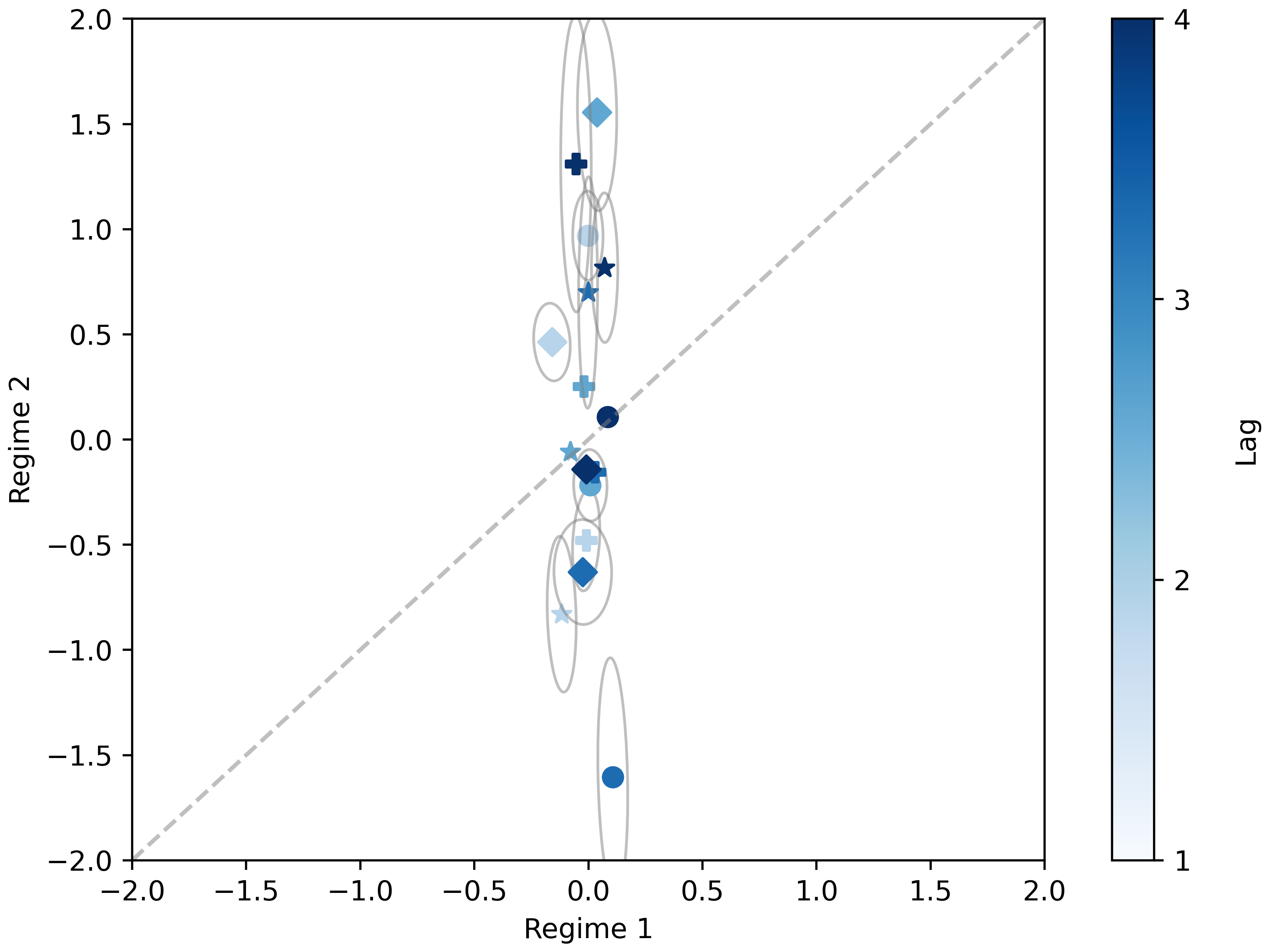}
    \end{tabular}
    \caption{\small The scatter plot shows the effects of Good Oil Volatility (left) and Bad Oil Volatility (right) on the financial sector (top) and energy sector (bottom). Different symbols represent different weeks, with $\bullet$: $t-(1+4(p-1))/4$,  ~\ding{58}: $t-2(1+4(p-1))/4$, \ding{117}: $t-3(1+4(p-1))/4$ and $\bigstar: t-4(1+4(p-1))/4$ for $p = \{1,2,3, 4\}$. Different blue shades represent  lags, from order 1 (lighter) to order 4 (darker).}
    \label{fig:coef_macro}
\end{figure}

For the disaggregated S\&P 500 analysis (bottom plot and Figure \ref{fig:insamp_mac1} of the Supplement), when sector indices are used as dependent variables, MSMETR can identify more episodes of market disruptions, including the 1997 Asia financial crisis, 2001 9/11 terrorist attack and 2002 corporate scandals and dot-com bubble together with the 2008 global financial crisis. The fact that MSMETR can capture more structural changes at a disaggregated level can be largely attributed to the heterogeneity between different sectors. Thus, MSMETR with multiple equations can also be an effective data integration tool.

Figure \ref{fig:coef_macro} shows the effects of GV and BV on financial and energy sector log-returns (see also Figure \ref{fig:coef_macro1} of the Supplement). We use different symbols to represent the weekly data sampled at different weeks for different lags $p= \{1,2,3,4\}$, with $\bullet$: $t-(1+4(p-1))/4$,  ~\ding{58}: $t-2(1+4(p-1))/4$, \ding{117}: $t-3(1+4(p-1))/4$ and $\bigstar: t-4(1+4(p-1))/4$. And we use lighter (darker) blue to represent lower (higher) lag $p$. Coefficients with $90\%$ HPD regions (grey ellipses) indicate large asymmetric effects.

For both aggregate and disaggregate analyses, GV and BV show more pronounced effects in the high-volatility regime ($s_t=2$) than in the low-volatility regime ($s_t=1$). This confirms the hypothesis of the financialization of the oil market \citep{xiao2022good,XIAO2023106969}. The HPD regions are more concentrated along the horizontal axis, most likely due to the smaller number of observations in regime $2$ compared to regime $1$. Regarding the long-term effects, GV has a larger impact on the markets at lower lags, while BV has a larger effect at higher lags. Similar asymmetries in the long and short-term impact have been documented within a univariate quantile regression framework by \cite{XIAO2023106969}.

We report the MSE, MAE for the in-sample fitting and the out-of-sample forecasting with prediction horizons of $1$-month and $5$-month in the lower panel of Table \ref{tab:stats fin} of the Supplement. Overall, tensor regression offers competing performances with LS and LASSO, and MSMETR performs strictly better in terms of in-sample fitting and short-term forecasting.

\section{Conclusion} \label{concl}
In this paper, we propose a new multiple-equation Markov Switching Tensor Regression Model to work with high dimensional data where a common hidden Markov chain process introduces dependencies between equations and allows for latent regime changes and dynamic coefficients. A low-rank representation of the tensor coefficient is used to achieve dimensionality reduction. A hierarchical prior distribution is imposed to introduce further shrinkage effects in the regression model with many regressors. Multiple prior stages allow smoothing of the effects of the low-rank representation (soft PARAFAC decomposition). We developed an MCMC sampler based on Random Partial Scan Gibbs and a back-fitting strategy. We show that the Markov chain generated by the proposed sampler is stationary and converges to the target distribution. The validity and efficiency of the sampler are demonstrated using simulations with different settings. We also tested our MSMETR with two real-world applications, where MSMETR outperforms the competing algorithms in both in-sample fitting and out-of-sample forecasting. Moreover, MSMETR provides more insight into the possible structural changes in the parameters by identifying regimes with regime-specific intercepts and variances, which are prevalent in time series data. The multiple-equation MSMETR can also capture the heterogeneity in the data at aggregate and disaggregate levels to exploit more information in the data.

The model we proposed is ready to be used for tensor regression with order 2 or 3. It can be applied to many other applications where regression on high-dimensional data is needed, and overparametrization or overfitting issues must be considered.

\subsection*{Funding}

The first author was supported by the MUR - PRIN project `\textit{Discrete random structures for Bayesian learning and prediction}' under g.a. n. 2022CLTYP4 and the Next Generation EU - `\textit{GRINS - Growing Resilient, INclusive and Sustainable}' project (PE0000018), National Recovery and Resilience Plan (NRRP) - PE9 - Mission 4. The second author was supported by NSERC of Canada discovery grants RGPIN-2018-249547  and  RGPIN-2024-04506.

\appendix

\renewcommand{\thesection}{A}
\renewcommand{\theequation}{A.\arabic{equation}}
\renewcommand{\thefigure}{A.\arabic{figure}}
\renewcommand{\thetable}{A.\arabic{table}}
\setcounter{table}{0}
\setcounter{figure}{0}
\setcounter{equation}{0}

\section{Proofs of the Results} \label{app:proofs}
\subsection{Proof of Proposition \ref{prop:Hyper}}
In the following, we drop the state subscript index $k$ and equation subscript index $\ell$ for simplicity. The variance of the coefficient entries of soft PARAFAC can be written as:
\begin{align}
    \mathbb{V}\left(B_{ij}\right) &= \mathbb{E}\left\{\mathbb{V}\left(B_{ij}\vert S, Z, W, \tau\right)\right\} = \mathbb{E}\left\{\text{Var}\left(\sum_{d=1}^D\prod_{m=1}^M\beta_{m,(ij)}^{(d)}\vert S,Z,W,\tau\right)\right\}\notag \\
    &=\mathbb{E}\left\{\sum_{d=1}^D\prod_{m=1}^M\text{Var}\left(\beta_{m,(ij)}^{(d)}\vert S,Z,W,\tau\right)\right\}=\mathbb{E}\left\{\sum_{d=1}^D\prod_{m=1}^M\tau\zeta^{(d)}\left(\sigma_m^2 + w_{m,(ij)}^{(d)}\right)\right\} \notag\\
    &=\mathbb{E}\left\{\tau^M\right\}\mathbb{E}\left\{\sum_{d=1}^D\left(\zeta^{(d)}\right)^M\right\}\mathbb{E}\left\{\prod_{m=1}^M\left(\sigma_m^2 + w_{m,(ij)}^{(d)}\right)\right\} \notag\\
    &=\frac{\Gamma(a_{\tau}+M)}{\Gamma(a_{\tau})b_{\tau}^M}D\prod_{r=0}^{M-1}\frac{\alpha/D+r}{\alpha+r}\left(\frac{a_{\sigma}}{b_{\sigma}}+\frac{2b_{\lambda}^2}{(a_{\lambda}-1)(a_{\lambda}-2)}\right)^M \label{eq.svar}
\end{align}

For the variance of the coefficient entries of hard PARAFAC, $\sigma_m^2=0$, thus
\begin{align}
    \mathbb{V}^{\text{hard}}\left(B_{ij}\right) = \frac{\Gamma(a_{\tau}+M)}{\Gamma(a_{\tau})b_{\tau}^M}D\prod_{r=0}^{M-1}\frac{\alpha/D+r}{\alpha+r}\left(\frac{2b_{\lambda}^2}{(a_{\lambda}-1)(a_{\lambda}-2)}\right)^M \label{eq.hvar}
\end{align}

It's not hard to notice that $\frac{a_{\sigma}}{b_{\sigma}}$ is the quantity that drives the additional variability of the soft PARAFAC by comparing Equations (\ref{eq.svar}) and (\ref{eq.hvar}). The goal is to set $\mathbb{V}(B_{ij}) = V^*$ and $AV = AV^*$. By exploiting $\mathbb{V}(B_{ij}) / \mathbb{V}^{\text{hard}}(B_{ij}) = (1-AV^*)^{-1}$ and setting $M=2$, we have

\begin{align}
    \frac{\mathbb{V}(B_{ij})}{\mathbb{V}^{\text{hard}}(B_{ij})} &=  \frac{\left(\frac{a_{\sigma}}{b_{\sigma}}+\frac{2b_{\lambda}^2}{(a_{\lambda}-1)(a_{\lambda}-2)}\right)^2}{\left(\frac{2b_{\lambda}^2}{(a_{\lambda}-1)(a_{\lambda}-2)}\right)^2} = \left(\frac{a_{\sigma}}{b_{\sigma}}\frac{(a_{\lambda}-1)(a_{\lambda}-2)}{2b_{\lambda}^2}+1\right)^2 = \left(1-AV^*\right)^{-1} \notag
\end{align}
Solving the above equation for $\frac{a_{\sigma}}{b_{\sigma}}$ and given $\frac{a_{\sigma}}{b_{\sigma}}$ is positive,

\begin{align}
    \frac{a_{\sigma}}{b_{\sigma}} = \left(\frac{1}{\sqrt{1-AV^*}}-1\right)\frac{2b_{\lambda}^2}{(a_{\lambda}-1)(a_{\lambda}-2)} \label{eq.asig1}
\end{align}
By setting $\mathbb{V}(B_{ij}) = V^*$ we have,
\begin{align}
    \frac{2b_{\lambda}^2}{(a_{\lambda}-1)(a_{\lambda}-2)} = \frac{b_{\tau}}{a_{\tau}}\sqrt{\frac{a_{\tau}V^*}{(a_{\tau}+1)C}} - \frac{a_{\sigma}}{b_{\sigma}}\label{eq.asig2}
\end{align}
Combing Equations (\ref{eq.asig1}) and (\ref{eq.asig2}),
\begin{align}
    \frac{a_{\sigma}}{b_{\sigma}} = \frac{b_{\tau}}{a_{\tau}}\sqrt{\frac{a_{\tau}V^*}{(a_{\tau}+1)C}}\left(1-\sqrt{1-AV^*}\right)\notag
\end{align}

\subsection{Proof of Proposition \ref{prop:Backfitting}}
For the likelihood part, we drop the state subscript index $k$ and equation subscript index $\ell$ for simplicity. We have 

\begin{eqnarray}
&&<B,X_t> = <\sum_{d=1}^{D}B_1^{(d)}\circ \cdots \circ B_{M}^{(d)},X_t>\notag\\
&&=<B_{m}^{(d)},B_1^{(d)}\circ\cdots\circ B_{m-1}^{(d)}\circ B_{m+1}^{(d)} \circ \cdots \circ B_{M}^{(d)}\circ X_t>+\sum_{d'\neq d}<B_1^{(d')}\circ \cdots \circ B_{M}^{(d')},X_t>\notag\\
&&=\sum_{j_{m}=1}^{p_{m}}<B_{m,\tilde{j}_{m}}^{(d)},(B_1^{(d)}\circ\cdots\circ B_{m-1}^{(d)}\circ B_{m+1}^{(d)} \circ \cdots \circ B_M^{(d)}\circ X_t)_{\tilde{j}_{m}}>+R^{(d)}_{t}\notag\\
&&={\boldsymbol{\beta}_{m,j_{m}}^{(d)}}'\text{vec}(B_1^{(d)}\circ\cdots\circ B_{m-1}^{(d)}\circ B_{m+1}^{(d)} \circ \cdots \circ B_M^{(d)}\circ X_t)_{\tilde{j}_{m}}+R^{(d)}_{j_m,m,t}+R^{(d)}_{t}\notag\\
&&={\boldsymbol{\beta}_{m,j_{m}}^{(d)}}' \Psi^{(d)}_{j_m,m,t}+R^{(d)}_{j_m,m,t}+R^{(d)}_{t}\notag
\end{eqnarray}
where
\begin{eqnarray}
R^{(d)}_{t}&&=\sum_{d'\neq d}<B_1^{(d')}\circ \cdots \circ B_M^{(d')},X_t> \notag\\
R^{(d)}_{j_m,m,t}&&=\sum_{j_{m}'\neq j_{m}}<B_{m,\tilde{j}_{m}'}^{(d)},(B_1^{(d)}\circ\cdots\circ B_{m-1}^{(d)}\circ B_{m+1}^{(d)} \circ \cdots \circ B_M^{(d)}\circ X_t)_{\tilde{j}_{m}'}>\notag\\
&&= \sum_{j_{m}'\neq j_{m}}<(B_{1}^{(d)}\circ \cdots \circ B_M^{(d)})_{\tilde{j}_{m}'}, (X_t)_{\tilde{j}_{m}'}>\notag\\ 
&&=<(B_{1}^{(d)}\circ \cdots \circ B_M^{(d)})_{-j_{m}}, (X_t)_{-j_{m}}>\notag\\
\Psi_{j_m, m, t}^{(d)} &&= \text{vec}(B_1^{(d)}\circ\cdots\circ B_{m-1}^{(d)}\circ B_{m+1}^{(d)} \circ \cdots \circ B_M^{(d)}\circ X_t)_{\tilde{j}_{m}} \notag
\end{eqnarray}

\subsection{Proof of Proposition \ref{prop:randGibbs}}
To simplify notation, we prove the result for $|I|$=3, but the result holds in general. Suppose $2$ components with indexes set $\boldsymbol{d}=\{d_1, d_2, d_3\}$ and $\lvert \boldsymbol{d} \rvert =3$ are randomly selected to be updated, and the order in which the $\lvert \boldsymbol{d} \rvert$ components are updated is also random. 

Assume that the set of indexes is selected at random and the order in which the components are updated is also random. Given that the other components outside of the index set $\boldsymbol{d}$ remain unchanged, we have the transition function
\begin{align*}
    K\left(\theta^{(i+1)} \vert \theta^{(i)}\right) = &\xi_1 \pi\left(\theta^{(i+1)}_{d_1}\vert \theta^{(i)}_{d_2}, \theta^{(i)}_{d_3},\theta^{(i)}_{-\boldsymbol{d}} \right)\pi\left(\theta^{(i+1)}_{d_2}\vert \theta^{(i+1)}_{d_1},\theta^{(i)}_{d_3},\theta^{(i)}_{-\boldsymbol{d}}\right)\pi\left(\theta^{(i+1)}_{d_3}\vert \theta^{(i+1)}_{d_1},\theta^{(i+1)}_{d_2},\theta^{(i)}_{-\boldsymbol{d}}\right) + \\
    &\xi_2 \pi\left(\theta^{(i+1)}_{d_1}\vert \theta^{(i)}_{d_2}, \theta^{(i)}_{d_3},\theta^{(i)}_{-\boldsymbol{d}} \right)\pi\left(\theta^{(i+1)}_{d_3}\vert \theta^{(i+1)}_{d_1},\theta^{(i)}_{d_2},\theta^{(i)}_{-\boldsymbol{d}}\right)\pi\left(\theta^{(i+1)}_{d_2}\vert \theta^{(i+1)}_{d_1},\theta^{(i+1)}_{d_3},\theta^{(i)}_{-\boldsymbol{d}}\right) + \\
    &\xi_3 \pi\left(\theta^{(i+1)}_{d_2}\vert \theta^{(i)}_{d_1}, \theta^{(i)}_{d_3},\theta^{(i)}_{-\boldsymbol{d}} \right)\pi\left(\theta^{(i+1)}_{d_1}\vert \theta^{(i+1)}_{d_2},\theta^{(i)}_{d_3},\theta^{(i)}_{-\boldsymbol{d}}\right)\pi\left(\theta^{(i+1)}_{d_3}\vert \theta^{(i+1)}_{d_1},\theta^{(i+1)}_{d_2},\theta^{(i)}_{-\boldsymbol{d}}\right) + \\
    &\xi_4 \pi\left(\theta^{(i+1)}_{d_2}\vert \theta^{(i)}_{d_1}, \theta^{(i)}_{d_3},\theta^{(i)}_{-\boldsymbol{d}} \right)\pi\left(\theta^{(i+1)}_{d_3}\vert \theta^{(i+1)}_{d_2},\theta^{(i)}_{d_1},\theta^{(i)}_{-\boldsymbol{d}}\right)\pi\left(\theta^{(i+1)}_{d_1}\vert \theta^{(i+1)}_{d_2},\theta^{(i+1)}_{d_3},\theta^{(i)}_{-\boldsymbol{d}}\right) +\\
    &\xi_5 \pi\left(\theta^{(i+1)}_{d_3}\vert \theta^{(i)}_{d_1}, \theta^{(i)}_{d_2},\theta^{(i)}_{-\boldsymbol{d}} \right)\pi\left(\theta^{(i+1)}_{d_1}\vert \theta^{(i+1)}_{d_3},\theta^{(i)}_{d_2},\theta^{(i)}_{-\boldsymbol{d}}\right)\pi\left(\theta^{(i+1)}_{d_2}\vert \theta^{(i+1)}_{d_1},\theta^{(i+1)}_{d_3},\theta^{(i)}_{-\boldsymbol{d}}\right) + \\
    &\xi_6 \pi\left(\theta^{(i+1)}_{d_3}\vert \theta^{(i)}_{d_1}, \theta^{(i)}_{d_2},\theta^{(i)}_{-\boldsymbol{d}} \right)\pi\left(\theta^{(i+1)}_{d_2}\vert \theta^{(i+1)}_{d_3},\theta^{(i)}_{d_1},\theta^{(i)}_{-\boldsymbol{d}}\right)\pi\left(\theta^{(i+1)}_{d_1}\vert \theta^{(i+1)}_{d_2},\theta^{(i+1)}_{d_3},\theta^{(i)}_{-\boldsymbol{d}}\right)
\end{align*}
where $\xi_1 \sim \xi_6$ are the probabilities of all possible orders in which $d_j$ could be updated, and $\sum_{i=1}^6 \xi_i = 1$. To save space, we ignore $\theta^{(i)}_{-\boldsymbol{d}}$ from the equations for the rest of the proof. From the detailed balance condition, we have $ K\left(\theta^{(i+1)} \vert \theta^{(i)}\right)\pi\left(\theta^{(i)}\right) =  K\left(\theta^{(i)} \vert \theta^{(i+1)}\right)\pi\left(\theta^{(i+1)}\right)$ which implies:
\begin{align*}
    \Bigl\{&\xi_1 \pi\left(\theta^{(i+1)}_{d_1}\vert \theta^{(i)}_{d_2}, \theta^{(i)}_{d_3} \right)\pi\left(\theta^{(i+1)}_{d_2}\vert \theta^{(i+1)}_{d_1},\theta^{(i)}_{d_3}\right)\pi\left(\theta^{(i+1)}_{d_3}\vert \theta^{(i+1)}_{d_1},\theta^{(i+1)}_{d_2}\right) + \\
    &\xi_2 \pi\left(\theta^{(i+1)}_{d_1}\vert \theta^{(i)}_{d_2}, \theta^{(i)}_{d_3} \right)\pi\left(\theta^{(i+1)}_{d_3}\vert \theta^{(i+1)}_{d_1},\theta^{(i)}_{d_2}\right)\pi\left(\theta^{(i+1)}_{d_2}\vert \theta^{(i+1)}_{d_1},\theta^{(i+1)}_{d_3}\right) + \\
    &\xi_3 \pi\left(\theta^{(i+1)}_{d_2}\vert \theta^{(i)}_{d_1}, \theta^{(i)}_{d_3} \right)\pi\left(\theta^{(i+1)}_{d_1}\vert \theta^{(i+1)}_{d_2},\theta^{(i)}_{d_3}\right)\pi\left(\theta^{(i+1)}_{d_3}\vert \theta^{(i+1)}_{d_1},\theta^{(i+1)}_{d_2}\right) +\\
    &\xi_4 \pi\left(\theta^{(i+1)}_{d_2}\vert \theta^{(i)}_{d_1}, \theta^{(i)}_{d_3} \right)\pi\left(\theta^{(i+1)}_{d_3}\vert \theta^{(i+1)}_{d_2},\theta^{(i)}_{d_1}\right)\pi\left(\theta^{(i+1)}_{d_1}\vert \theta^{(i+1)}_{d_2},\theta^{(i+1)}_{d_3}\right) +\\
    &\xi_5 \pi\left(\theta^{(i+1)}_{d_3}\vert \theta^{(i)}_{d_1},\theta^{(i)}_{d_2}\right)\pi\left(\theta^{(i+1)}_{d_1}\vert \theta^{(i+1)}_{d_3},\theta^{(i)}_{d_2}\right)\pi\left(\theta^{(i+1)}_{d_2}\vert \theta^{(i+1)}_{d_1}, \theta^{(i+1)}_{d_3}\right) +\\
    &\xi_6 \pi\left(\theta^{(i+1)}_{d_3}\vert \theta^{(i)}_{d_1}, \theta^{(i)}_{d_2}\right)\pi\left(\theta^{(i+1)}_{d_2}\vert \theta^{(i+1)}_{d_3},\theta^{(i)}_{d_1}\right)\pi\left(\theta^{(i+1)}_{d_1}\vert \theta^{(i+1)}_{d_2},\theta^{(i+1)}_{d_3}\right)\Bigr\}\pi\left(\theta^{(i)}\right)\\
    =\Bigl\{&\xi_1 \pi\left(\theta^{(i)}_{d_1}\vert \theta^{(i+1)}_{d_2}, \theta^{(i+1)}_{d_3} \right)\pi\left(\theta^{(i)}_{d_2}\vert \theta^{(i)}_{d_1},\theta^{(i+1)}_{d_3}\right)\pi\left(\theta^{(i)}_{d_3}\vert \theta^{(i)}_{d_1},\theta^{(i)}_{d_2}\right) + \\
    &\xi_2 \pi\left(\theta^{(i)}_{d_1}\vert \theta^{(i+1)}_{d_2}, \theta^{(i+1)}_{d_3} \right)\pi\left(\theta^{(i)}_{d_3}\vert \theta^{(i)}_{d_1},\theta^{(i+1)}_{d_2}\right)\pi\left(\theta^{(i)}_{d_2}\vert \theta^{(i)}_{d_1},\theta^{(i)}_{d_3}\right) + \\
    &\xi_3 \pi\left(\theta^{(i)}_{d_2}\vert \theta^{(i+1)}_{d_1}, \theta^{(i+1)}_{d_3} \right)\pi\left(\theta^{(i)}_{d_1}\vert \theta^{(i)}_{d_2},\theta^{(i+1)}_{d_3}\right)\pi\left(\theta^{(i)}_{d_3}\vert \theta^{(i)}_{d_1},\theta^{(i)}_{d_2}\right) +\\
    &\xi_4 \pi\left(\theta^{(i)}_{d_2}\vert \theta^{(i+1)}_{d_1}, \theta^{(i+1)}_{d_3} \right)\pi\left(\theta^{(i)}_{d_3}\vert \theta^{(i)}_{d_2},\theta^{(i+1)}_{d_1}\right)\pi\left(\theta^{(i)}_{d_1}\vert \theta^{(i)}_{d_2},\theta^{(i)}_{d_3}\right) +\\
    &\xi_5 \pi\left(\theta^{(i)}_{d_3}\vert \theta^{(i+1)}_{d_1},\theta^{(i+1)}_{d_2}\right)\pi\left(\theta^{(i)}_{d_1}\vert \theta^{(i)}_{d_3},\theta^{(i+1)}_{d_2}\right)\pi\left(\theta^{(i)}_{d_2}\vert \theta^{(i)}_{d_1}, \theta^{(i)}_{d_3}\right) +\\
    &\xi_6 \pi\left(\theta^{(i)}_{d_3}\vert \theta^{(i+1)}_{d_1}, \theta^{(i+1)}_{d_2}\right)\pi\left(\theta^{(i)}_{d_2}\vert \theta^{(i)}_{d_3},\theta^{(i+1)}_{d_1}\right)\pi\left(\theta^{(i)}_{d_1}\vert \theta^{(i)}_{d_2},\theta^{(i)}_{d_3}\right)\Bigr\}\pi\left(\theta^{(i+1)}\right)
\end{align*}
where $\theta^{(i+1)} = \left(\theta^{(i+1)}_{d_1}, \theta^{(i+1)}_{d_2}, \theta^{(i+1)}_{d_3}\right)$, this is equivalent of writing
\begin{align}
    \Biggl\{&\xi_1 \frac{\pi\left(\theta^{(i+1)}_{d_1}, \theta^{(i)}_{d_2}, \theta^{(i)}_{d_3}\right)}{\pi\left(\theta^{(i)}_{d_2}, \theta^{(i)}_{d_3}\right)}\frac{\pi\left(\theta^{(i+1)}_{d_1}, \theta^{(i+1)}_{d_2}, \theta^{(i)}_{d_3}\right)}{\pi\left(\theta^{(i+1)}_{d_1}, \theta^{(i)}_{d_3}\right)}\frac{\pi\left(\theta^{(i+1)}_{d_1}, \theta^{(i+1)}_{d_2}, \theta^{(i+1)}_{d_3}\right)}{\pi\left(\theta^{(i+1)}_{d_1}, \theta^{(i+1)}_{d_2}\right)} + \notag\\
    &\xi_2 \frac{\pi\left(\theta^{(i+1)}_{d_1}, \theta^{(i)}_{d_2}, \theta^{(i)}_{d_3}\right)}{\pi\left(\theta^{(i)}_{d_2}, \theta^{(i)}_{d_3}\right)}\frac{\pi\left(\theta^{(i+1)}_{d_1}, \theta^{(i)}_{d_2}, \theta^{(i+1)}_{d_3}\right)}{\pi\left(\theta^{(i+1)}_{d_1}, \theta^{(i)}_{d_2}\right)}\frac{\pi\left(\theta^{(i+1)}_{d_1}, \theta^{(i+1)}_{d_2}, \theta^{(i+1)}_{d_3}\right)}{\pi\left(\theta^{(i+1)}_{d_1}, \theta^{(i+1)}_{d_3}\right)} + \notag \\
    & \xi_3 \frac{\pi\left(\theta^{(i)}_{d_1}, \theta^{(i+1)}_{d_2}, \theta^{(i)}_{d_3}\right)}{\pi\left(\theta^{(i)}_{d_1}, \theta^{(i)}_{d_3}\right)}\frac{\pi\left(\theta^{(i+1)}_{d_1}, \theta^{(i+1)}_{d_2}, \theta^{(i)}_{d_3}\right)}{\pi\left(\theta^{(i+1)}_{d_2}, \theta^{(i)}_{d_3}\right)}\frac{\pi\left(\theta^{(i+1)}_{d_1}, \theta^{(i+1)}_{d_2}, \theta^{(i+1)}_{d_3}\right)}{\pi\left(\theta^{(i+1)}_{d_1}, \theta^{(i+1)}_{d_2}\right)} + \notag \\
    & \xi_4 \frac{\pi\left(\theta^{(i)}_{d_1}, \theta^{(i+1)}_{d_2}, \theta^{(i)}_{d_3}\right)}{\pi\left(\theta^{(i)}_{d_1}, \theta^{(i)}_{d_3}\right)}\frac{\pi\left(\theta^{(i)}_{d_1}, \theta^{(i+1)}_{d_2}, \theta^{(i+1)}_{d_3}\right)}{\pi\left(\theta^{(i)}_{d_1}, \theta^{(i+1)}_{d_2}\right)}\frac{\pi\left(\theta^{(i+1)}_{d_1}, \theta^{(i+1)}_{d_2}, \theta^{(i+1)}_{d_3}\right)}{\pi\left(\theta^{(i+1)}_{d_2}, \theta^{(i+1)}_{d_3}\right)} + \notag \\
    & \xi_5 \frac{\pi\left(\theta^{(i)}_{d_1}, \theta^{(i)}_{d_2}, \theta^{(i+1)}_{d_3}\right)}{\pi\left(\theta^{(i)}_{d_1}, \theta^{(i)}_{d_2}\right)}\frac{\pi\left(\theta^{(i+1)}_{d_1}, \theta^{(i)}_{d_2}, \theta^{(i+1)}_{d_3}\right)}{\pi\left(\theta^{(i)}_{d_2}, \theta^{(i+1)}_{d_3}\right)}\frac{\pi\left(\theta^{(i+1)}_{d_1}, \theta^{(i+1)}_{d_2}, \theta^{(i+1)}_{d_3}\right)}{\pi\left(\theta^{(i+1)}_{d_1}, \theta^{(i+1)}_{d_3}\right)} + \notag \\
    &\xi_6 \frac{\pi\left(\theta^{(i)}_{d_1}, \theta^{(i)}_{d_2}, \theta^{(i+1)}_{d_3}\right)}{\pi\left(\theta^{(i)}_{d_1}, \theta^{(i)}_{d_2}\right)}\frac{\pi\left(\theta^{(i)}_{d_1}, \theta^{(i+1)}_{d_2}, \theta^{(i+1)}_{d_3}\right)}{\pi\left(\theta^{(i)}_{d_1}, \theta^{(i+1)}_{d_3}\right)}\frac{\pi\left(\theta^{(i+1)}_{d_1}, \theta^{(i+1)}_{d_2}, \theta^{(i+1)}_{d_3}\right)}{\pi\left(\theta^{(i+1)}_{d_2}, \theta^{(i+1)}_{d_3}\right)}\Biggr\}\pi\left(\theta^{(i)}\right) \notag\\
    =\Biggl\{&\xi_1 \frac{\pi\left(\theta^{(i)}_{d_1}, \theta^{(i+1)}_{d_2}, \theta^{(i+1)}_{d_3}\right)}{\pi\left(\theta^{(i+1)}_{d_2}, \theta^{(i+1)}_{d_3}\right)}\frac{\pi\left(\theta^{(i)}_{d_1}, \theta^{(i)}_{d_2}, \theta^{(i+1)}_{d_3}\right)}{\pi\left(\theta^{(i)}_{d_1}, \theta^{(i+1)}_{d_3}\right)}\frac{\pi\left(\theta^{(i)}_{d_1}, \theta^{(i)}_{d_2}, \theta^{(i)}_{d_3}\right)}{\pi\left(\theta^{(i)}_{d_1}, \theta^{(i)}_{d_2}\right)} + \notag\\
    &\xi_2 \frac{\pi\left(\theta^{(i)}_{d_1}, \theta^{(i+1)}_{d_2}, \theta^{(i+1)}_{d_3}\right)}{\pi\left(\theta^{(i+1)}_{d_2}, \theta^{(i+1)}_{d_3}\right)}\frac{\pi\left(\theta^{(i)}_{d_1}, \theta^{(i+1)}_{d_2}, \theta^{(i)}_{d_3}\right)}{\pi\left(\theta^{(i)}_{d_1}, \theta^{(i+1)}_{d_2}\right)}\frac{\pi\left(\theta^{(i)}_{d_1}, \theta^{(i)}_{d_2}, \theta^{(i)}_{d_3}\right)}{\pi\left(\theta^{(i)}_{d_1}, \theta^{(i)}_{d_3}\right)} + \notag \\
    & \xi_3 \frac{\pi\left(\theta^{(i+1)}_{d_1}, \theta^{(i)}_{d_2}, \theta^{(i+1)}_{d_3}\right)}{\pi\left(\theta^{(i+1)}_{d_1}, \theta^{(i+1)}_{d_3}\right)}\frac{\pi\left(\theta^{(i)}_{d_1}, \theta^{(i)}_{d_2}, \theta^{(i+1)}_{d_3}\right)}{\pi\left(\theta^{(i)}_{d_2}, \theta^{(i+1)}_{d_3}\right)}\frac{\pi\left(\theta^{(i)}_{d_1}, \theta^{(i)}_{d_2}, \theta^{(i)}_{d_3}\right)}{\pi\left(\theta^{(i)}_{d_1}, \theta^{(i)}_{d_2}\right)} + \notag \\
    & \xi_4 \frac{\pi\left(\theta^{(i+1)}_{d_1}, \theta^{(i)}_{d_2}, \theta^{(i+1)}_{d_3}\right)}{\pi\left(\theta^{(i+1)}_{d_1}, \theta^{(i+1)}_{d_3}\right)}\frac{\pi\left(\theta^{(i+1)}_{d_1}, \theta^{(i)}_{d_2}, \theta^{(i)}_{d_3}\right)}{\pi\left(\theta^{(i+1)}_{d_1}, \theta^{(i)}_{d_2}\right)}\frac{\pi\left(\theta^{(i)}_{d_1}, \theta^{(i)}_{d_2}, \theta^{(i)}_{d_3}\right)}{\pi\left(\theta^{(i)}_{d_2}, \theta^{(i)}_{d_3}\right)} + \notag \\
    & \xi_5 \frac{\pi\left(\theta^{(i+1)}_{d_1}, \theta^{(i+1)}_{d_2}, \theta^{(i)}_{d_3}\right)}{\pi\left(\theta^{(i+1)}_{d_1}, \theta^{(i+1)}_{d_2}\right)}\frac{\pi\left(\theta^{(i)}_{d_1}, \theta^{(i+1)}_{d_2}, \theta^{(i)}_{d_3}\right)}{\pi\left(\theta^{(i+1)}_{d_2}, \theta^{(i)}_{d_3}\right)}\frac{\pi\left(\theta^{(i)}_{d_1}, \theta^{(i)}_{d_2}, \theta^{(i)}_{d_3}\right)}{\pi\left(\theta^{(i)}_{d_1}, \theta^{(i)}_{d_3}\right)} + \notag \\
    &\xi_6 \frac{\pi\left(\theta^{(i+1)}_{d_1}, \theta^{(i+1)}_{d_2}, \theta^{(i)}_{d_3}\right)}{\pi\left(\theta^{(i+1)}_{d_1}, \theta^{(i+1)}_{d_2}\right)}\frac{\pi\left(\theta^{(i+1)}_{d_1}, \theta^{(i)}_{d_2}, \theta^{(i)}_{d_3}\right)}{\pi\left(\theta^{(i+1)}_{d_1}, \theta^{(i)}_{d_3}\right)}\frac{\pi\left(\theta^{(i)}_{d_1}, \theta^{(i)}_{d_2}, \theta^{(i)}_{d_3}\right)}{\pi\left(\theta^{(i)}_{d_2}, \theta^{(i)}_{d_3}\right)}\Biggr\}\pi\left(\theta^{(i+1)}\right)\label{eq.pp3}
\end{align}
where $\pi\left(\theta^{(i)}\right) = \pi\left(\theta^{(i)}_{i_1}, \theta^{(i)}_{i_2}, \theta^{(i)}_{d_3}\right)$ and $\pi\left(\theta^{(i+1)}\right) = \pi\left(\theta^{(i+1)}_{i_1}, \theta^{(i+1)}_{i_2}, \theta^{(i+1)}_{d_3}\right)$. The equality of Equation \ref{eq.pp3} holds, if $\xi_1 = \xi_2 =\xi_3 = \xi_4=\xi_5 = \xi_6$. This proof can be easily generalized to an arbitrary randomly selected number of components as long as $\boldsymbol{d} \subset \{1,\ldots,D\}$.


\renewcommand{\thesection}{B}
\renewcommand{\theequation}{B.\arabic{equation}}
\renewcommand{\thefigure}{B.\arabic{figure}}
\renewcommand{\thetable}{B.\arabic{table}}
\setcounter{table}{0}
\setcounter{figure}{0}
\setcounter{equation}{0}

\section{Full conditional derivations} \label{app: deri}
\subsection{Full conditional distribution of the hidden state variables}
A multi-move sampling is applied to sample from the joint posterior distribution of the hidden state variables. We apply forward filtering and backward sampling  \citep{Fruhwirth2006}. Let us introduce the set of allocation variables $\boldsymbol{\xi}_{t}=(\xi_{1t},\ldots,\xi_{Kt})$, with $\xi_{kt}=\mathbf{I}(s_t=k)$. By means of dynamic factorization, the full conditional distribution of the hidden state is
\begin{equation}
    p(s_{1},\ldots,s_t\vert \mathbf{y}, B,\boldsymbol{\gamma},\boldsymbol{w},\boldsymbol{\zeta},\boldsymbol{\tau},\mathbf{p})\propto \prod_{t=1}^{T}\left(\prod_{\ell=1}^{N}p(y_{\ell,t}\vert B_{\ell}(s_{t}),\sigma^2_{\ell}(s_t) )\right)\prod_{k=1}^{K}\prod_{l=1}^{K}p_{lk}^{\xi_{kt}\xi_{l,t-1}}
\end{equation}
where $B=(B_1,\ldots,B_K)$, $\boldsymbol{\gamma}=(\boldsymbol{\gamma}_1, \ldots, \boldsymbol{\gamma}_K)$, $\boldsymbol{w}=(\boldsymbol{w}_1, \ldots, \boldsymbol{w}_K)$, $\boldsymbol{\zeta}=(\boldsymbol{\zeta}_1, \ldots, \boldsymbol{\zeta}_K)$, $\boldsymbol{\tau}=(\tau_1,\ldots,\tau_K)'$.

\subsection{Full conditional distribution of the transition probability}
\begin{align}
    p\left((p_{1i}, \ldots, p_{Ki}) \vert S, \mathbf{y}\right) &\propto \prod_{t=1}^T\prod_{l=1}^K p_{li}^{\xi_{lt}\xi_{i, t-1}}\prod_{l=1}^K p_{li}^{\nu_l - 1} 
    \propto \prod_{l=1}^K p_{il}^{\sum_{t=1}^T\xi_{lt}\xi_{i, t-1} + \nu_l - 1}  
\end{align}
which is proportional to Dirichlet distribution $\mathcal{D}ir(\Bar{\nu}_1,\ldots,\Bar{\nu}_K)$, where
$\Bar{\nu}_l = \nu_l + \sum_{t=1}^T\xi_{lt}\xi_{i, t-1}$.

\subsection{Full conditional distribution of the state-specific paratemters}
Given the conditional independence assumption, we drop the state subscript index $k$ and equation subscript index $\ell$ for simplicity in the following. From the prior for $B_{m}^{(d)}$, we have $\boldsymbol{\beta}_{m,j_{m}}^{(d)} \sim \mathcal{N}_{q_{m}}(\gamma_{m,j_{m}}^{(d)}\boldsymbol{\iota}_{q_{m}},\tau\sigma^{2}_{m}\zeta^{(d)}I_{q_{m}})$.
 
The posterior of the unknowns of the model is given by
\begin{eqnarray}
p(\boldsymbol{\beta}_{m,j_{m}}^{(d)}, \gamma_{m,j_{m}}^{(d)}, \sigma_{m}^2, w_{m,j_{m}}^{(d)},\lambda_{m}^{(d)},\tau,\zeta^{(d)} \mid \mathbf{y}, B_1,\ldots,B_T)
\end{eqnarray}
We adopt the MCMC procedure based on the Gibbs sampling algorithm to sample the unknowns from 3 blocks.

\subsubsection*{Block 1: Sampling $\zeta^{(d)}$ and $\tau$ from $p(\zeta^{(d)},\tau\mid B,\boldsymbol{\gamma},\boldsymbol{w})$}

We first sample $\boldsymbol{\zeta}$ from the joint posterior by integrating out $\tau$

\begin{eqnarray}
&&p\left(\boldsymbol{\zeta}\mid B,\boldsymbol{\gamma},\boldsymbol{w}\right)\propto \pi\left(\boldsymbol{\zeta}\right)p\left(B,\boldsymbol{\gamma}\mid\boldsymbol{w},\zeta^{(d)}\right)= \pi\left(\boldsymbol{\zeta}\right) \int_{\tau}p\left(B\mid\boldsymbol{\gamma},\boldsymbol{w},\tau\right)p\left(\boldsymbol{\gamma}\mid\boldsymbol{w},\tau\right)\pi(\tau)d\tau \notag\\
&&= \prod_{d=1}^D {\zeta^{(d)}}^{\frac{\alpha}{D}-1} \int_{\tau} \left(\prod_{d=1}^D \prod_{m=1}^M \prod_{j_{m}=1}^{p_{m}} \left(\tau \zeta^{(d)} \sigma_{m}^2\right)^{-\frac{q_{m}}{2}} \right.\notag\\
&&\cdot \exp \left\{-\frac{1}{2}\left(\boldsymbol{\beta}_{m,j_{m}}^{(d)}-\gamma_{m,j_{m}}^{(d)}\boldsymbol{\iota}_{q_{m}}\right)'\left(\tau \zeta^{(d)} \sigma_{m}^2\right)^{-1}\left(\boldsymbol{\beta}_{m,j_{m}}^{(d)}-\gamma_{m,j_{m}}^{(d)}\boldsymbol{\iota}_{q_{m}}\right)\right\}\notag\\
&&\left.\cdot \left(\tau \zeta^{(d)} w_{m,j_{m}}^{(d)}\right)^{-\frac{1}{2}} \exp \left\{-\frac{1}{2}\frac{{\gamma_{m,j_{m}}^{(d)}}^2}{\tau \zeta^{(d)} w_{m,j_{m}}^{(d)}}\right\}\right)\tau^{a_{\tau}-1}e^{-b_{\tau}\tau}d\tau \notag
\end{eqnarray} 

\begin{eqnarray}&&\propto \prod_{d=1}^D {\zeta^{(d)}}^{\frac{\alpha}{D}-1} \int_{\tau}  \left(\prod_{d=1}^D \left(\tau \zeta^{(d)}\right)^{-\frac{\sum_{m=1}^Mp_{m}(q_{m}+1)}{2}}\right.\notag\\
&&\left.\cdot\exp\left\{-\frac{1}{2\tau\zeta^{(d)}}\sum_{m=1}^M\sum_{j_{m}=1}^{p_{m}}\left(\frac{1}{\sigma_{m}^2}\left(\boldsymbol{\beta}_{m,j_{m}}^{(d)}-\gamma_{m,j_{m}}^{(d)}\boldsymbol{\iota}_{q_{m}}\right)' \left(\boldsymbol{\beta}_{m,j_{m}}^{(d)}-\gamma_{m,j_{m}}^{(d)}\boldsymbol{\iota}_{q_{m}}\right)+\frac{1}{w_{m,j_{m}}^{(d)}}{\gamma_{m,j_{m}}^{(d)}}^2\right)\right\}\right)\notag \\
&& \cdot \tau^{a_{\tau}-1}e^{-b_{\tau}\tau}d\tau \notag
\end{eqnarray} 

Let us define $C_d = \sum_{m=1}^M\sum_{j_{m}=1}^{p_{m}}\left(\frac{1}{\sigma_{m}^2}\left(\boldsymbol{\beta}_{m,j_{m}}^{(d)}-\gamma_{m,j_{m}}^{(d)}\boldsymbol{\iota}_{q_{m}}\right)' \left(\boldsymbol{\beta}_{m,j_{m}}^{(d)}-\gamma_{m,j_{m}}^{(d)}\boldsymbol{\iota}_{q_{m}}\right)+\frac{1}{w_{m,j_{m}}^{(d)}}{\gamma_{m,j_{m}}^{(d)}}^2\right)$, $I_0 = \sum_{m=1}^Mp_{m}(q_{m}+1) = M\prod_{m=1}^M p_m + \sum_{m=1}^M p_m$

\begin{eqnarray}
p\left(\boldsymbol{\zeta}\mid B,\boldsymbol{\gamma},\boldsymbol{w}\right)&&\propto \prod_{d=1}^D {\zeta^{(d)}}^{\frac{\alpha}{D}-\frac{I_0}{2}-1} \int_{\tau} \tau^{a_{\tau}-\frac{DI_0}{2}-1}\exp\left\{-b_{\tau}\tau-\frac{\sum_{d=1}^D C_d}{2\tau\zeta^{(d)}}\right\}d\tau \notag
\end{eqnarray}

By definition, $\sum_{d=1}^D \zeta^{(d)} =1$ which allows us to write $\sum_{d=1}^D(b_{\tau}\tau\zeta^{(d)})=b_{\tau}\tau\sum_{d=1}^D \zeta^{(d)}=b_{\tau}\tau$ \citep{billio2023tensor}, moreover, by letting $a_{\tau}=\alpha$ (\cite{guhaniyogi2017bayesian})

\begin{eqnarray}
p\left(\boldsymbol{\zeta}\mid B,\boldsymbol{\gamma},\boldsymbol{w}\right) \propto \int_{\tau}\left(\prod_{d=1}^D{\zeta^{(d)}}^{\frac{\alpha}{D}-\frac{I_0}{2}-1}\right)\tau^{(\alpha-\frac{DI_0}{2})-1}\exp\left\{-\frac{1}{2}\sum_{d=1}^D \left(\frac{C_d}{\tau\zeta^{(d)}}+2b_{\tau}\tau\zeta^{(d)}\right)\right\}d\tau\label{eq.zeta}
\end{eqnarray}
We recognize that equation (\ref{eq.zeta}) is the kernel of a Generalized Inverse Gaussian for $\phi^{(d)} = \tau\zeta^{(d)}$

\begin{eqnarray}
\phi^{(d)} \sim \text{GiG } \left(\frac{\alpha}{D}-\frac{I_0}{2}, 2b_{\tau}, C_d\right) \notag
\end{eqnarray}

We then obtain $\zeta^{(d)}$ by renormalizing: $\zeta^{(d)} = \phi^{(d)}/\sum_{l=1}^D \phi^{(d)}$ \citep{billio2023tensor}.

The full conditional for $\tau$ can be derived as follows
\begin{eqnarray}
&&p\left(\tau\mid B,\boldsymbol{\gamma},\boldsymbol{w},\boldsymbol{\zeta}\right) \propto \pi(\tau)p\left(B, \boldsymbol{\gamma}\mid\zeta^{(d)},\tau,\boldsymbol{w}\right)=\pi(\tau)p\left(B|\boldsymbol{\gamma},\boldsymbol{w},\zeta^{(d)},\tau\right)p\left(\boldsymbol{\gamma}|\boldsymbol{w},\zeta^{(d)},\tau\right)\notag\\
&& = \tau^{a_{\tau}-1}e^{-b_{\tau}\tau} \prod_{d=1}^D \prod_{m=1}^M \prod_{j_{m}=1}^{p_{m}} \left(\tau \zeta^{(d)} \sigma_{m}^2\right)^{-\frac{q_{m}}{2}} \notag\\
&&\cdot \exp \left\{-\frac{1}{2}\left(\boldsymbol{\beta}_{m,j_{m}}^{(d)}-\gamma_{m,j_{m}}^{(d)}\boldsymbol{\iota}_{q_{m}}\right)'\left(\tau \zeta^{(d)} \sigma_{m}^2\right)^{-1}\left(\boldsymbol{\beta}_{m,j_{m}}^{(d)}-\gamma_{m,j_{m}}^{(d)}\boldsymbol{\iota}_{q_{m}}\right)\right\}\notag\\
&&\cdot \left(\tau \zeta^{(d)} w_{m,j_{m}}^{(d)}\right)^{-\frac{1}{2}} \exp \left\{-\frac{1}{2}\frac{{\gamma_{m,j_{m}}^{(d)}}^2}{\tau \zeta^{(d)} w_{m,j_{m}}^{(d)}}\right\}\notag\\
&&\propto \tau^{a_{\tau}-\frac{DI_0}{2}-1}e^{-b_{\tau}\tau}\notag\\
&&\cdot \prod_{d=1}^D\exp\left\{-\frac{1}{2\tau\zeta^{(d)}}\sum_{m=1}^M\sum_{j_{m}=1}^{p_{m}}\left(\frac{1}{\sigma_{m}^2}\left(\boldsymbol{\beta}_{m,j_{m}}^{(d)}-\gamma_{m,j_{m}}^{(d)}\boldsymbol{\iota}_{q_{m}}\right)' \left(\boldsymbol{\beta}_{m,j_{m}}^{(d)}-\gamma_{m,j_{m}}^{(d)}\boldsymbol{\iota}_{q_{m}}\right)+\frac{1}{w_{m,j_{m}}^{(d)}}{\gamma_{m,j_{m}}^{(d)}}^2\right)\right\}\notag\\
&&=\tau^{a_{\tau}-\frac{DI_0}{2}-1}\exp\left\{-\frac{1}{2}\left(\sum_{d=1}^D \frac{C_d}{\tau\zeta^{(d)}}+2b_{\tau}\tau\right)\right\}\notag
\end{eqnarray}

Therefore, the full conditional of $\tau$ is also a Generalized Inverse Gaussian
\begin{eqnarray}
p\left(\tau\mid B,\boldsymbol{\gamma},\boldsymbol{w},\boldsymbol{\zeta}\right) \propto \text{GiG } \left(a_{\tau}-\frac{DI_0}{2}, 2b_{\tau}, \sum_{d=1}^D \frac{C_d}{\zeta^{(d)}}\right) \notag
\end{eqnarray}

\subsubsection*{Block 2: Sampling $\lambda_{m}^{(d)}$ and $w_{m,j_{m}}^{(d)}$ from $p(\lambda_{m}^{(d)}, w_{m,j_{m}}^{(d)}|\gamma_{m,j_{m}}^{(d)}, \tau, \zeta^{(d)})$}

Notice that by the construction of the prior distributions, $\gamma_{m,j_{m}}^{(d)}$ follows a double exponential distribution given $\lambda_{m}^{(d)}$, $\tau$, $\zeta^{(d)}$
\begin{eqnarray}
\gamma_{m,j_{m}}^{(d)} \sim \text{DE }\left(0,\frac{\sqrt{\tau\zeta^{(d)}}}{\lambda_{m}^{(d)}}\right)\notag
\end{eqnarray}

The full conditional of $\lambda_{m}^{(d)}$ can be written as
\begin{eqnarray}
&&p\left(\lambda_{m}^{(d)}\mid\gamma_{m,j_{m}}^{(d)}, \tau, \zeta^{(d)}\right)\propto \pi(\lambda_{m}^{(d)}) p\left(\gamma_{m,j_{m}}^{(d)}\mid\lambda_{m}^{(d)},\tau, \zeta^{(d)}\right)\notag\\
&&\propto\left(\tau\zeta^{(d)}\right)^{-\frac{p_{m}}{2}}\left(\lambda_{m}^{(d)}\right)^{a_{\lambda}+p_{m}-1}\exp\left\{-\left(\frac{\sum_{j_{m}=1}^{p_{m}}\left|\gamma_{m,j_{m}}^{(d)}\right|}{\sqrt{\tau\zeta^{(d)}}}+b_{\lambda}\right)\lambda_{m}^{(d)}\right\}\notag\\
&& \propto \mathcal{G}a \left(a_{\lambda}+p_{m}, \frac{\sum_{j_{m}=1}^{p_{m}}\left|\gamma_{m,j_{m}}^{(d)}\right|}{\sqrt{\tau\zeta^{(d)}}}+b_{\lambda}\right)\notag
\end{eqnarray}

The full conditional for $w_{m,j_{m}}^{(d)}$ is
\begin{eqnarray}
&&p\left(w_{m,j_{m}}^{(d)}\mid\gamma_{m,j_{m}}^{(d)},\lambda_{m}^{(d)}, \tau, \zeta^{(d)}\right) \propto \pi\left(w_{m,j_{m}}^{(d)}\right)p\left(\gamma_{m,j_{m}}^{(d)}\mid\lambda_{m}^{(d)}, \tau, \zeta^{(d)}, w_{m,j_{m}}^{(d)}\right)\notag\\
&&\propto {w_{m,j_{m}}^{(d)}}^{\frac{1}{2}-1}\exp \left\{-\frac{1}{2}\left({\lambda_{m}^{(d)}}^2w_{m,j_{m}}^{(d)}+\frac{{\gamma_{m,j_{m}}^{(d)}}^2}{\tau\zeta^{(d)}w_{m,j_{m}}^{(d)}}\right)\right\}\notag\\
&&\propto \text{GiG }\left(\frac{1}{2}, {\lambda_{m}^{(d)}}^2, \frac{{\gamma_{m,j_{m}}^{(d)}}^2}{\tau\zeta^{(d)}}\right)\notag
\end{eqnarray}

\subsubsection*{Block 3: Sampling $\boldsymbol{\beta}_{m,j_{m}}^{(d)}, \gamma_{m,j_{m}}^{(d)}, \sigma_{m}^2, \mu, \sigma^2$ from $p\left(\boldsymbol{\beta}_{m,j_{m}}^{(d)}, \gamma_{m,j_{m}}^{(d)}, \sigma_{m}^2, \mu, \sigma^2 \mid \mathbf{y}, B_1,\ldots,B_T\right)$}

We derive the full conditional of $\boldsymbol{\beta}_{m,j_{m}}^{(d)}$ in a way such that it only depends on observed data by integrating out $\gamma_{m,j_{m}}^{(d)}$. The total number of $\boldsymbol{\beta}_{m,j_{m}}^{(d)}$ we need to sample is $D\sum_{m=1}^Mp_{m}$.

\begin{eqnarray}
&&p\left(\boldsymbol{\beta}_{m,j_{m}}^{(d)}\mid \mathbf{y},B_1,\ldots,B_T, \sigma_{m}^2\right) \propto p\left(\mathbf{y}\mid\boldsymbol{\beta}_{m,j_{m}}^{(d)}, \sigma_{m}^2, B_1,\ldots,B_T\right)\int_{\gamma}p\left(\boldsymbol{\beta}_{m,j_{m}}^{(d)}\mid\gamma_{m,j_{m}}^{(d)},\sigma_{m}^2\right)p\left(\gamma_{m,j_{m}}^{(d)}\right)d\gamma\notag\\
&&\propto \prod_{t\in\mathcal{T}} \exp\left\{-\frac{1}{2}\frac{\left(y_t- \mu -\left<B,B_t\right>\right)^2}{\sigma^2}\right\}\int_{\gamma}p\left(\boldsymbol{\beta}_{m,j_{m}}^{(d)}\mid\gamma_{m,j_{m}}^{(d)},\sigma_{m}^2\right)p\left(\gamma_{m,j_{m}}^{(d)}\right)d\gamma\notag
\end{eqnarray}
where $\mathcal{T}\subset\{1,\ldots, T\}$ contains all the indexes of the observations such that the latent variable $s_t$ takes a given value in $\{1,\ldots, K\}$. Thanks to the result in Proposition \ref{prop:Backfitting}, the
terms at the exponent in the likelihood write as:
\begin{eqnarray}
&&\left(y_t- \mu - {\boldsymbol{\beta}_{m,j_{m}}^{(d)}}'\Psi_{j_m, m, t}^{(d)}-R_{j_m, m, t}^{(d)}-R_{t}^{(d)}\right)^2=\notag\\
&&\left(y_t- \mu - R_{j_m, m, t}^{(d)}-R_{t}^{(d)}\right)^2+{\boldsymbol{\beta}_{m,j_{m}}^{(d)'}}\Psi_{j_m, m, t}^{(d)}\Psi_{j_m, m, t}^{(d)'}\boldsymbol{\beta}_{m,j_{m}}^{(d)}-2{\boldsymbol{\beta}_{m,j_{m}}^{(d)}}'\Psi_{j_m, m, t}^{(d)}\left(y_t-\mu-R_{j_m, m, t}^{(d)}-R_{t}^{(d)}\right)\notag
\end{eqnarray}

Define $\tilde{y}_t = y_t-\mu-R_{j_m, m, t}^{(d)}-R_{t}^{(d)}$, then the likelihood can be written as
\begin{eqnarray}
p\left(\mathbf{y}\mid\boldsymbol{\beta}_{m,j_{m}}^{(d)},\sigma_{m}^2, B_1,\ldots,B_T\right)
\propto \exp \left\{-\frac{1}{2\sigma^2}\sum_{t\in\mathcal{T}}\left({\boldsymbol{\beta}_{m,j_{m}}^{(d)}}'\Psi_{j_m, m, t}^{(d)}\Psi_{j_m, m, t}^{(d)'}\boldsymbol{\beta}_{m,j_{m}}^{(d)}-2{\boldsymbol{\beta}_{m,j_{m}}^{(d)}}'\Psi_{j_m, m, t}^{(d)}\tilde{y}_t\right)\right\}\notag
\end{eqnarray}

For the integration part, given $p\left(\boldsymbol{\beta}_{m,j_{m}}^{(d)} \mid \gamma_{m,j_{m}}^{(d)}\right)$ and $p\left(\gamma_{m,j_{m}}^{(d)}\right)$ are normal then the marginal distribution is normal with mean
$\mathbb{E}\left(\boldsymbol{\beta}_{m,j_{m}}^{(d)}\right)  = \mathbb{E}\left(\mathbb{E}\left(\boldsymbol{\beta}_{m,j_{m}}^{(d)} \mid \gamma_{m,j_{m}}^{(d)}\right)\right) = \boldsymbol{0}$, and variance  $\mathbb{V}\left(\boldsymbol{\beta}_{m,j_{m}}^{(d)}\right)  = \mathbb{V}\left(\mathbb{E}\left(\boldsymbol{\beta}_{m,j_{m}}^{(d)} \mid \gamma_{m,j_{m}}^{(d)}\right)\right) + \mathbb{E}\left(\mathbb{V}\left(\boldsymbol{\beta}_{m,j_{m}}^{(d)} \mid \gamma_{m,j_{m}}^{(d)}\right)\right) = \left(\tau \zeta^{(d)} w_{m,j_{m}}^{(d)} + \tau \zeta^{(d)} \sigma_{m}^2\right)I_{q_{m}}$.
\medskip

Let $\xi = \tau\zeta^{(d)}\left(w_{m,j_{m}}^{(d)}+\sigma_{m}^2\right)$, then the full conditional of $\boldsymbol{\beta}_{m,j_{m}}^{(d)}$ can be written as
\begin{eqnarray}
&&p\left(\boldsymbol{\beta}_{m,j_{m}}^{(d)}\mid \mathbf{y},B_1,\ldots,B_T, \sigma_{m}^2\right)\notag\\ 
&&\propto\exp \left\{-\frac{1}{2}\left[{\boldsymbol{\beta}_{m,j_{m}}^{(d)}}'\left(\sum_{t\in\mathcal{T}}\frac{\Psi_{j_m, m, t}^{(d)}\Psi_{j_m, m, t}^{(d)'}}{\sigma^2} + \frac{1}{\xi}I_{q_{m}}\right)\boldsymbol{\beta}_{m,j_{m}}^{(d)}-2{\boldsymbol{\beta}_{m,j_{m}}^{(d)}}'\sum_{t\in\mathcal{T}}\frac{\Psi_{j_m, m, t}^{(d)}\tilde{y}_t}{\sigma^2}\right]\right\} \notag\\
&&\propto \mathcal{N}\left(\left(\sum_{t\in\mathcal{T}}\frac{\Psi_{j_m, m, t}^{(d)}\Psi_{j_m, m, t}^{(d)'}}{\sigma^2} + \frac{1}{\xi}I_{q_{m}}\right)^{-1}\sum_{t\in\mathcal{T}}\frac{\Psi_{j_m, m, t}^{(d)}\tilde{y}_t}{\sigma^2}, \left(\sum_{t\in\mathcal{T}}\frac{\Psi_{j_m, m, t}^{(d)}\Psi_{j_m, m, t}^{(d)'}}{\sigma^2} + \frac{1}{\xi}I_{q_{m}}\right)^{-1}\right) \notag
\end{eqnarray}

The full conditional for $\gamma_{m,j_{m}}^{(d)}$ given $\boldsymbol{\beta}_{m,j_{m}}^{(d)}$ can be written as
\begin{eqnarray}
&&p\left(\gamma_{m,j_{m}}^{(d)}\mid\boldsymbol{\beta}_{m,j_{m}}^{(d)}\right) \propto p\left(\boldsymbol{\beta}_{m,j_{m}}^{(d)}|\gamma_{m,j_{m}}^{(d)}\right) \pi\left(\gamma_{m,j_{m}}^{(d)}\right) \notag\\
    && \propto \exp\left\{-\frac{1}{2\tau\zeta^{(d)}}\left[\frac{q_{m}w_{m,j_{m}}^{(d)}+\sigma_{m}^2}{w_{m,j_{m}}^{(d)}\sigma_{m}^2}\left(\gamma_{m,j_{m}}^{(d)}-\frac{w_{m,j_{m}}^{(d)}}{q_{m}w_{m,j_{m}}^{(d)}+\sigma_{m}^2}{\boldsymbol{\beta}_{m,j_{m}}^{(d)}}'\boldsymbol{\iota}_{q_{m}}\right)^2\right]\right\}\notag\\
    && \propto \mathcal{N}\left(\frac{w_{m,j_{m}}^{(d)}}{q_{m}w_{m,j_{m}}^{(d)}+\sigma_{m}^2}{\boldsymbol{\beta}_{m,j_{m}}^{(d)}}'\boldsymbol{\iota}_{q_{m}}, \frac{\tau\zeta^{(d)}w_{m,j_{m}}^{(d)}\sigma_{m}^2}{q_{m}w_{m,j_{m}}^{(d)}+\sigma_{m}^2}\right)\notag
\end{eqnarray}

The full conditional for $\sigma_{m}^2$ can be written as
\begin{eqnarray}
&&p\left(\sigma_{m}^2\mid\boldsymbol{\beta}_{m,j_{m}}^{(d)}\right) \propto \prod_{d=1}^D\prod_{j_{m}=1}^{p_{m}} p\left(\boldsymbol{\beta}_{m,j_{m}}^{(d)}\mid\sigma_{m}^2\right)\pi\left(\sigma_{m}^2\right) \notag \\
    && = \left(\sigma_{m}^2\right)^{a_{\sigma_m}-D\frac{p_{m}q_{m}}{2}-1} \exp \left\{-\frac{1}{2}\left(\frac{\sum_{d=1}^D\sum_{j_{m}=1}^{p_{m}}\left(\boldsymbol{\beta}_{m,j_{m}}^{(d)}-\gamma_{m,j_{m}}^{(d)}\boldsymbol{\iota}_{q_{m}}\right)' \left(\boldsymbol{\beta}_{m,j_{m}}^{(d)}-\gamma_{m,j_{m}}^{(d)}\boldsymbol{\iota}_{q_{m}}\right)}{\tau \zeta^{(d)} \sigma_{m}^2}+2b_{\sigma_m}\sigma_{m}^2\right)\right\}\notag \\
    && \propto \text{GiG }\left(a_{\sigma_m}-D\frac{p_{m}q_{m}}{2}, 2b_{\sigma_m}, \frac{\sum_{d=1}^D\sum_{j_{m}=1}^{p_{m}}\left(\boldsymbol{\beta}_{m,j_{m}}^{(d)}-\gamma_{m,j_{m}}^{(d)}\boldsymbol{\iota}_{q_{m}}\right)' \left(\boldsymbol{\beta}_{m,j_{m}}^{(d)}-\gamma_{m,j_{m}}^{(d)}\boldsymbol{\iota}_{q_{m}}\right)}{\tau \zeta^{(d)}}\right) \notag
\end{eqnarray}

The full conditional for $\sigma^2$ can be written as:
\begin{eqnarray}
    &&p\left(\sigma^2 \mid \mathbf{y}\right) \propto p\left(\mathbf{y} \mid \sigma^2\right)\pi\left(\sigma^2\right) \notag\\
    && \propto \left(\sigma^2\right)^{-\left(a_{\sigma}+\frac{T}{2}\right)-1}\exp \left\{-\frac{1}{\sigma^2}\left(\frac{1}{2}\sum_{t=1}^T\left(y_t - \left<B, X_t\right>-\mu\right)^2 +b_{\sigma}\right)\right\} \propto \mathcal{IG} \left(a_{\sigma}^*, b_{\sigma}^*\right)\notag
\end{eqnarray}
where $a_{\sigma}^* = a_{\sigma}+\frac{T}{2}$ and $b_{\sigma}^*  = \frac{1}{2}\sum_{t=1}^T\left(y_t - \left<B, X_t\right> - \mu\right)^2 +b_{\sigma}$

Let $\mu^* = \sum_{t=1}^T\left(y_t - \left<B, X_t\right>\right) {\sigma_{\mu}^*}^2$ and 
    ${\sigma_{\mu}^*}^2  = \left(\frac{T}{\sigma^2}+\frac{1}{\sigma_{\mu}^2}\right)^{-1}$, the full conditional of $\mu$ is:
\begin{align}
    & p\left(\mu \mid \mathbf{y}\right) \propto p\left(\mathbf{y} \mid \mu \right) \pi \left(\mu\right) \notag
     \propto \exp \left\{-\frac{1}{2\sigma^2}\left[T\mu^2 -2\mu\sum_{t=1}^T\left(y_t - \left<B, X_t\right>\right)\right] - \frac{1}{2}\frac{\mu^2}{\sigma_{\mu}^2}\right\} \notag\\
    & = \exp \left\{-\frac{1}{2}\left[\left(\frac{T}{\sigma^2}+\frac{1}{\sigma_{\mu}^2}\right)\mu^2 - 2\mu\frac{\sum_{t=1}^T\left(y_t - \left<B, X_t\right>\right)}{\sigma^2}\right]\right\}
     \propto \mathcal{N}\left(\mu^*, {\sigma_{\mu}^*}^2\right) \notag
\end{align}

\bibliographystyle{apalike}
\bibliography{ref}

\clearpage

\pagenumbering{arabic}
\begin{center}
	\Large{Markov Switching Multiple-equation Tensor Regressions}
\\\large{Online supplementary material}
\end{center}
	\vspace{-5pt}
	
	\vspace{0.5cm}

This supplement consists of two Appendices C and D.  Appendix C contains simulation results, while Appendix D includes more details on the empirical applications.

\clearpage

\newpage

\renewcommand{\thesection}{C}
\renewcommand{\theequation}{C.\arabic{equation}}
\renewcommand{\thefigure}{C.\arabic{figure}}
\renewcommand{\thetable}{C.\arabic{table}}
\setcounter{table}{0}
\setcounter{figure}{0}
\setcounter{equation}{0}
\section{Simulation results} \label{sec: sim_resu}
\subsection{Simple tensor regression}
For the simple tensor regression, we designed four different experimental settings. In each setting, the dimension of the true coefficients is $20\times 20$, and the true rank value and sparsity level are different following the scenarios one can find in real-data applications from many fields, including neuroscience, image processing and network analysis. 
\begin{itemize}
    \item[$(\mathcal{S}_1)$] In the first setting, the true coefficient is a diagonal matrix representing a scenario of full rank and a high level of sparsity. It can be interpreted as the adjacency matrix of a graph without edges between distinct nodes (empty graph) configuration, where edges between a node and itself (self-loops) are allowed. See \cite{newman2018networks} for an introduction to networks.
    \item[$(\mathcal{S}_2)$] In the second setting, the cross pattern in the coefficient values returns a symmetric persymmetric matrix \citep{computations1989gene} with a lower level of sparsity and a lower true rank. It can be interpreted as the adjacency matrix of a graph with a hub, a node with a number of edges that greatly exceeds the average. 
    \item[$(\mathcal{S}_3)$] The third setting is similar to the second and differs only in the configuration of the coefficient matrix, which contains a circular pattern. It can be interpreted as the adjacency matrix of a graph where the nodes are either disconnected from the other nodes or connected, directly or indirectly, to all the other nodes (connected component).
    \item[$(\mathcal{S}_4)$] In the last setting, the true coefficient matrix is symmetric, not persymmetric. It can be interpreted as the adjacency matrix of a core-periphery graph. A core-periphery structure is a common feature in social networks, where one expects to find a core of nodes with a high degree of connections surrounded by a periphery of nodes with a lower degree of connections.
\end{itemize}
Since the coefficients are binary in the above settings, we also consider more challenging settings in which Gaussian noise is added to the coefficient values to remove exact zeros. These real-valued settings are denoted by $\tilde{\mathcal{S}}_1$ to $\tilde{\mathcal{S}}_4$. For each setting, two types of data generating processes (DGP) are examined:  i) with covariates randomly drawn \textit{i.i.d} from a standard normal distribution; ii) with covariates following an AR(1) process. The number of observations generated from each DGP is $400$. We check model performances and sampling efficiency under three rank orders: $D = \{3, 5, 7\}$. 

\begin{figure}[p]
    \centering
    \includegraphics[width=.8\textwidth]{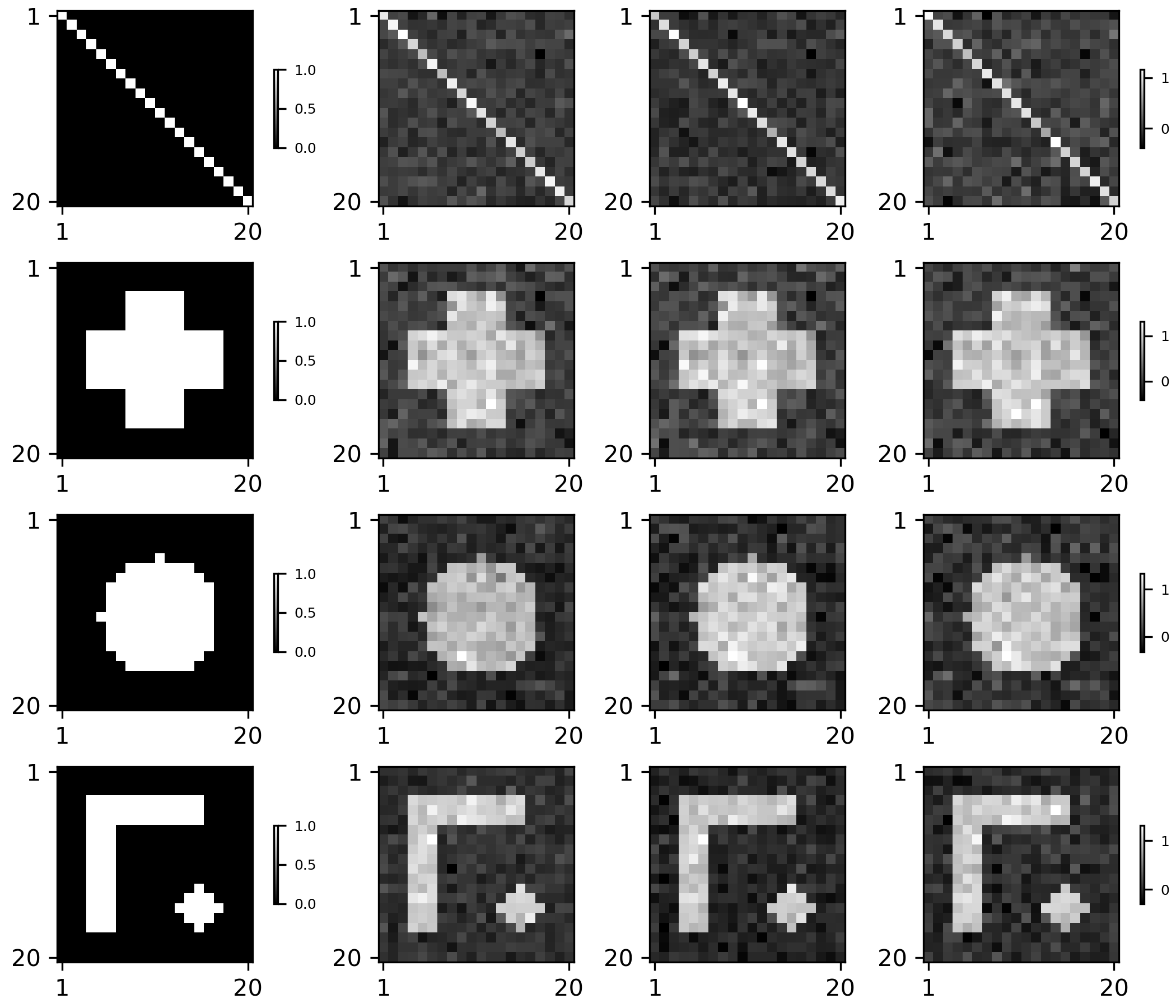}
    \caption{First column contains the true coefficient value; columns from two to four contain the estimation results for different ranks $D=3,5,7$. The four different experimental settings $\mathcal{S}_j$, $j=1,\ldots,4$ are in the rows.}
    \label{fig:sim_main}
\end{figure}

\begin{figure}[h!]
    \centering
    \begin{tabular}{c}  
     (a) AR(1) covariates and noisy true coefficients\\
    \includegraphics[trim = 0.5cm 0cm 0cm 1cm, clip, width=.8\textwidth]{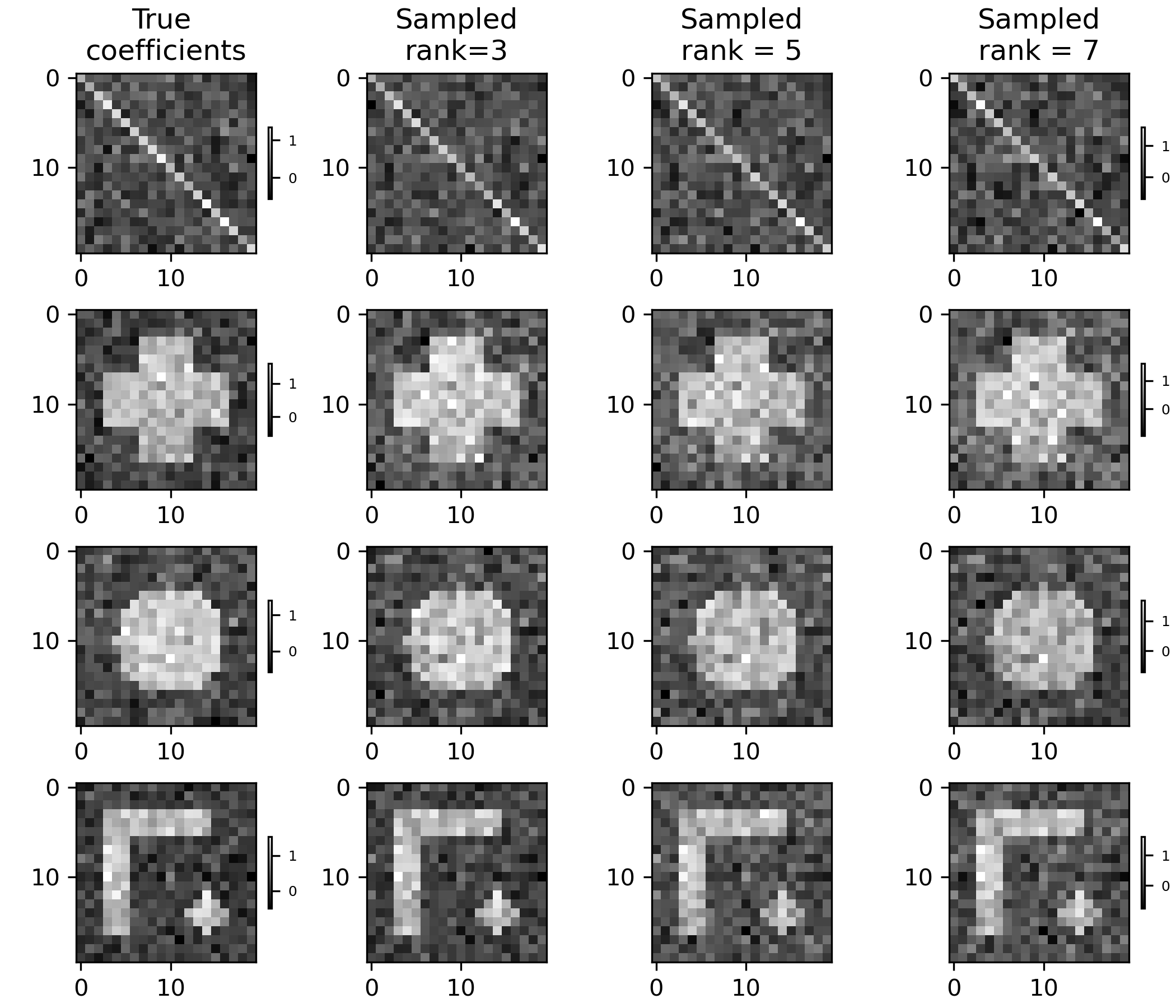}\\
     (b) IID covariates and noisy true coefficients\\
\includegraphics[trim = 0.5cm 0cm 0cm 1cm, clip, width=.8\textwidth]{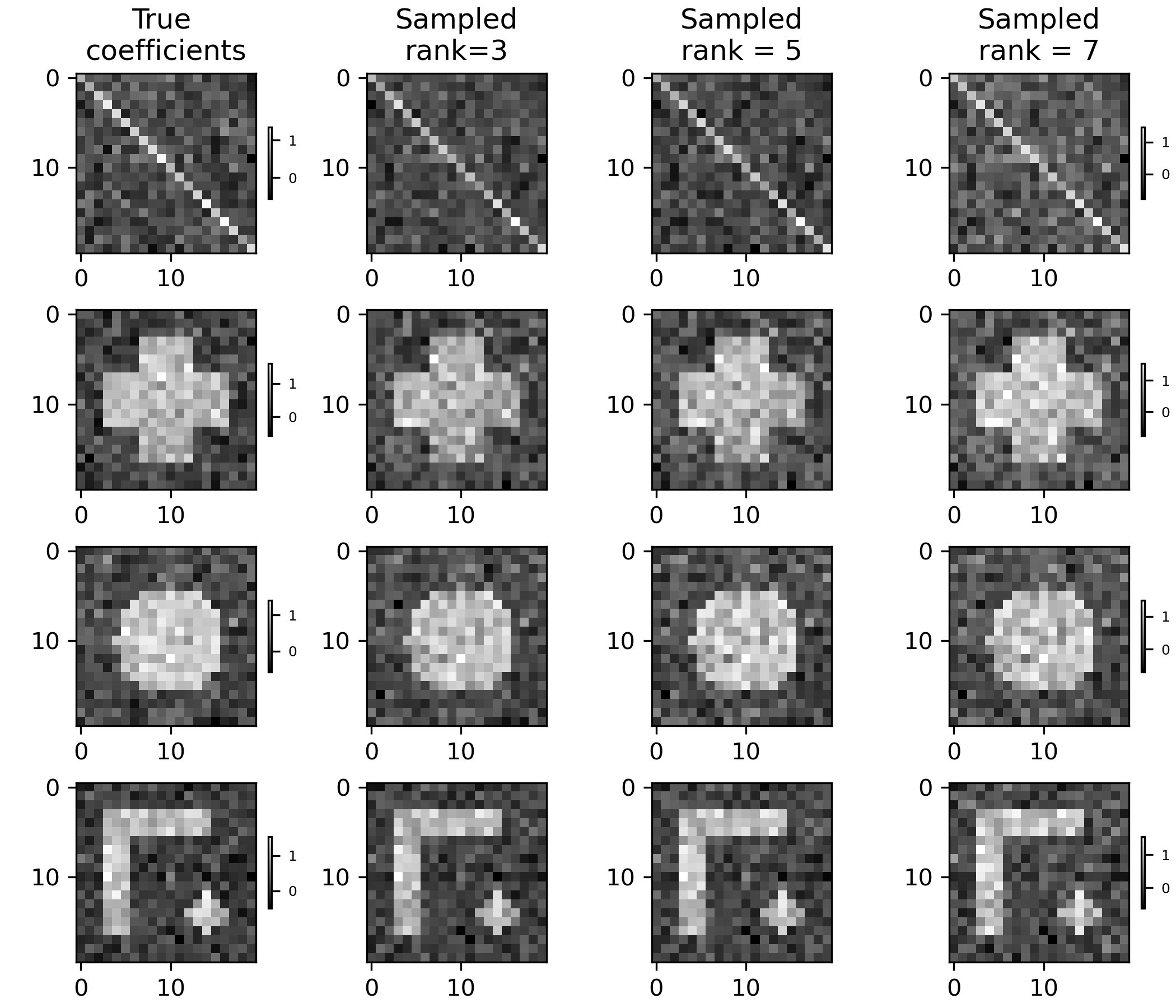}
\end{tabular}
    \caption{In the first column, the true coefficient value, in the columns from two to four, the estimation results for different ranks $D=3,5,7$. The four different experimental settings $\mathcal{S}_j$, $j=1,\ldots,4$ are in the rows.}
    \label{fig:sim_main_noisy}
\end{figure}

We ran the proposed Gibbs sampler for $10,000$ iterations.  Figure \ref{fig:sim_main} shows the simulation results for the four settings with exact zeros in the true coefficients. Further illustrations of the simulation results for each setting are given Figure \ref{fig:sim_main_noisy}. The panels show the results with AR(1) setting and for the noisy coefficients setting, respectively. In all scenarios, the algorithm can recover the true structure of the coefficients. All the ranks applied in the algorithm led to good estimations, and higher ranks, such as $5$ or $7$, do not offer significant improvements.
\begin{table}[t]
\centering
\begin{tabular}{c|ccc|ccc}
 &\multicolumn{3}{c|}{$IID$ covariates} &\multicolumn{3}{c}{$AR(1)$ covariates}\\
\hline
Setting& $D=3$ & $D=5$ & $D=7$ & $D=3$ & $D=5$ & $D=7$\\
\hline
\multicolumn{7}{c}{(a) Simple regression with binary coefficients}\\
\hline
$\mathcal{S}_1$&$0.0149$ &$0.0251$ & $0.0313$&$0.0144$ &$0.0230$ & $0.0431$ \\
$\mathcal{S}_2$&$0.0498$ &$0.0735$ & $0.0801$&$0.0703$ &$0.0822$ & $0.1061$ \\
$\mathcal{S}_3$&$0.0321$ &$0.0570$ & $0.0890$&$0.0666$ &$0.0491$ & $0.0830$ \\
$\mathcal{S}_4$&$0.0255$ &$0.0312$ & $0.0550$&$0.0349$ &$0.0467$ & $0.0514$\\
\hline
\multicolumn{7}{c}{(b) Simple regression with real-valued coefficients}\\
\hline
$\tilde{\mathcal{S}}_1$&$0.0290$ &$0.0432$ & $0.0508$&$0.0320$ &$0.0421$ & $0.0460$ \\
$\tilde{\mathcal{S}}_2$&$0.0612$ &$0.1193$ & $0.1193$& $0.1222$&$0.0865$ & $0.0920$ \\
$\tilde{\mathcal{S}}_3$&$0.0709$ &$0.0879$ & $0.0809$& $0.0726$ &$0.0751$ & $0.0714$ \\
$\tilde{\mathcal{S}}_4$&$0.0433$ &$0.0622$ & $0.0774$&$0.0538$ &$0.0537$ & $0.0738$\\
\hline
\multicolumn{7}{c}{(c) Markov-switching regression}\\
\hline
$\mathcal{S}_1^{MS}$&$0.0059$ &$0.0069$ & $0.0087$&$0.0079$ &$0.0063$ & $0.0065$ \\
$\mathcal{S}_2^{MS}$&$0.0075$ & $0.0055$ & $0.0121$&$0.0062$ & $0.0072$ & $0.0078$\\
\hline
        \end{tabular}
        \caption{Mean square errors for the coefficients of the simple and switching regressions (panels a and b) in different settings (rows), data generating process and PARAFAC rank (columns).}
        \label{tab:my_label}
\end{table}

\subsection{MS tensor regression}
For the MSMETR, we consider two regimes with a coefficient matrix of size $12 \times 12$. The sample size is $800$, and two different experimental settings are considered.

\begin{itemize}
\item[$(\mathcal{S}^{MS}_1)$] In the first setting, the true coefficients are an anti-diagonal and diagonal matrix. They represent the setting where the true coefficients of the two regimes have similar levels of sparsity and ranks.
\item[$(\mathcal{S}^{MS}_2)$] In the second setting, we replaced the anti-diagonal matrix with a cross matrix. They represent the setting where the two regimes' true coefficients vary in sparsity and rank.
\end{itemize}
Due to regime identification and label-switching issues (\cite{Fruhwirth2006}), an identification constraint disentangles the two regimes. We study the effectiveness of alternative identification strategies. In the first setting, we assume an identification constraint on the trace of the coefficients in the two regimes: $\text{tr}(B_{1}) < \text{tr}(B_{2})$. In the second setting, use a constraint on the Frobenius norm $\rVert \cdot \rVert_F$ of the coefficients to identify the regimes, that is $\lVert B_{1} \rVert_F > \lVert B_{2} \rVert_F$.

We study the proposed MCMC algorithm's efficiency by examining the MCMC chain empirical autocorrelation function (ACF) and the mean square error (MSE) of the true and sampled coefficient entries. The number of Gibbs iterations in each experiment is $3000$ with $1500$ burn-in iterations.

\begin{figure}[p]
    \centering
    \begin{tabular}{cc}
        \includegraphics[width=.45\linewidth]{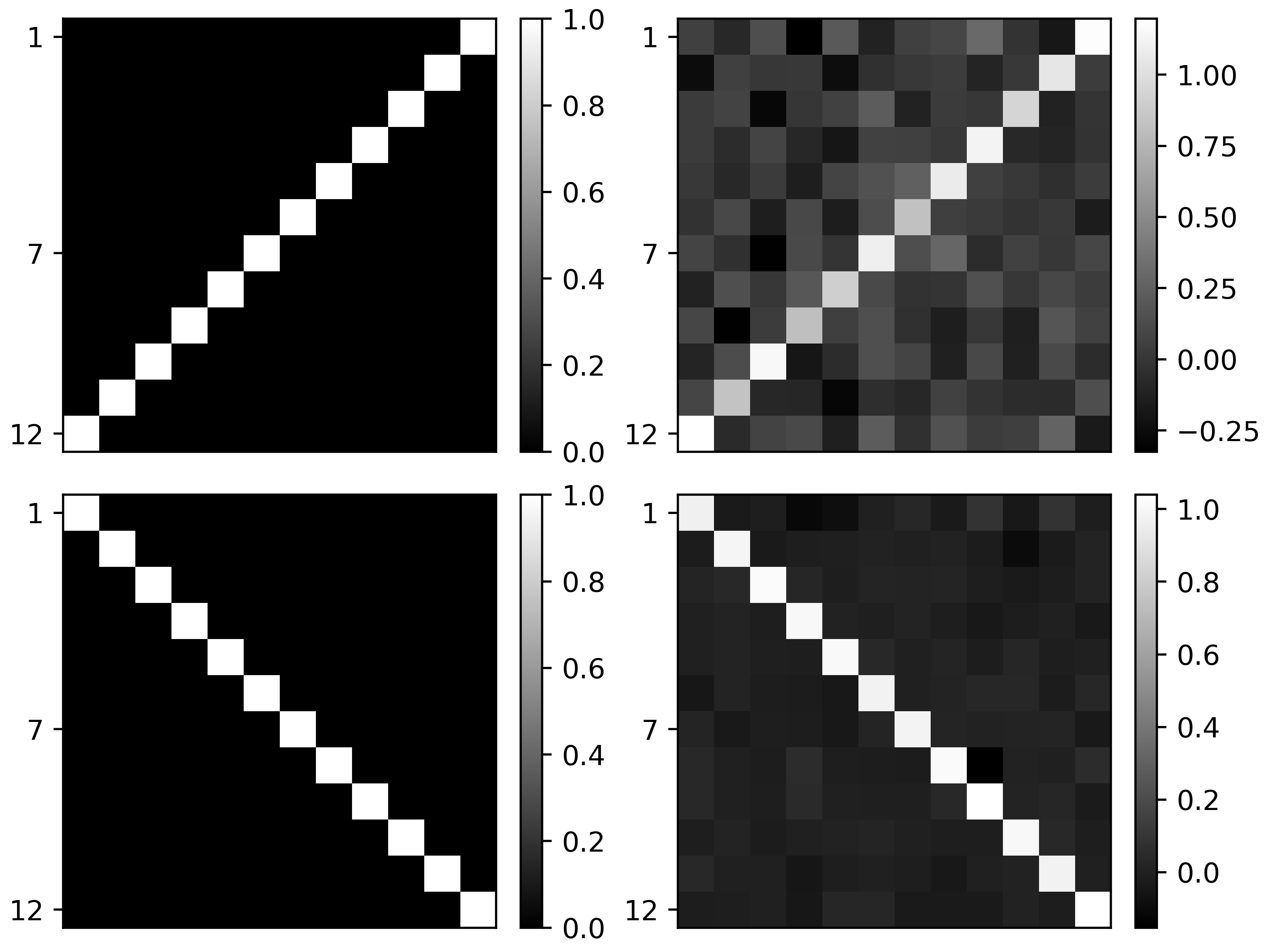}&        \includegraphics[width=.45\linewidth]{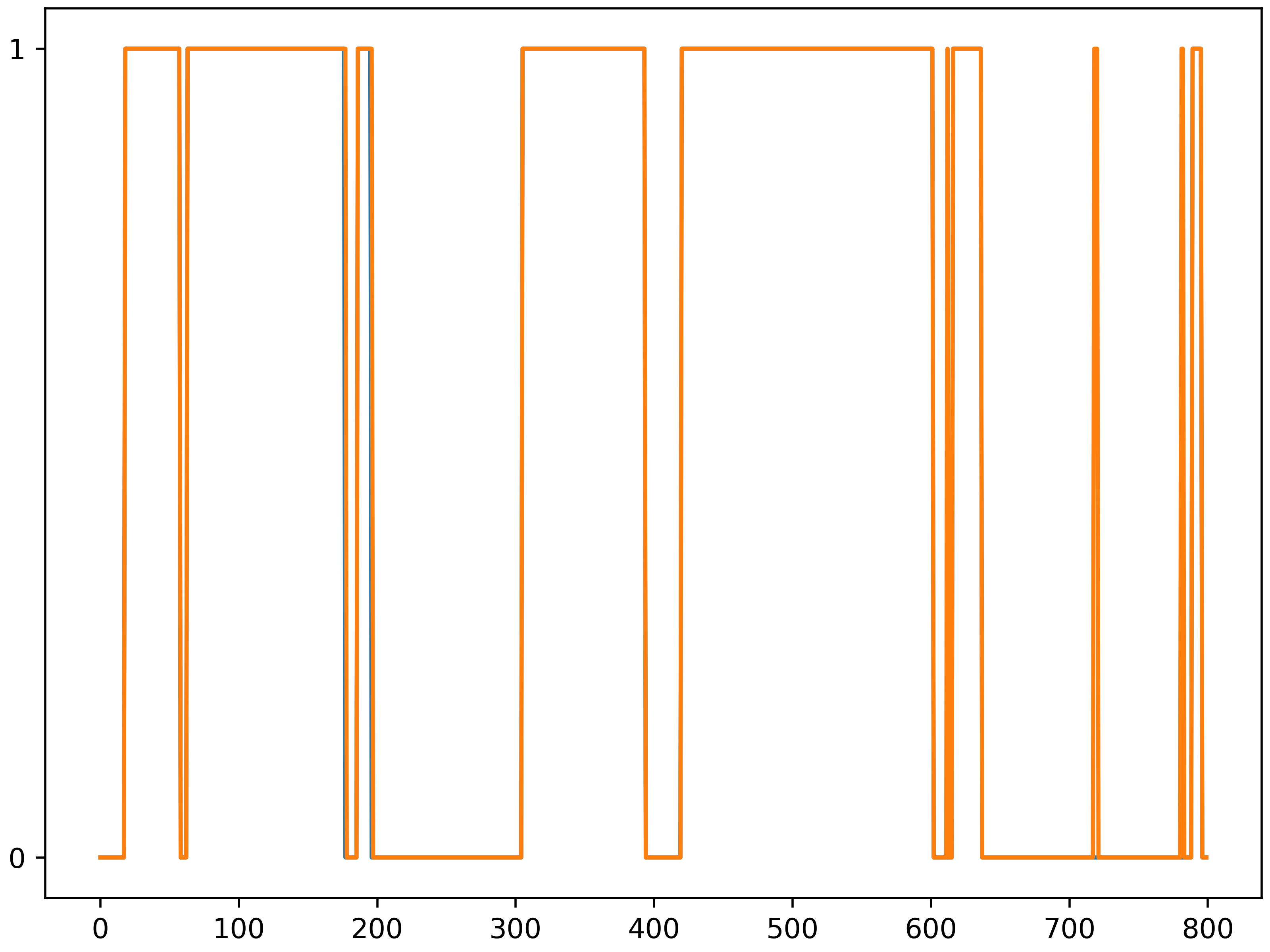}\\
        \includegraphics[width=.45\linewidth]{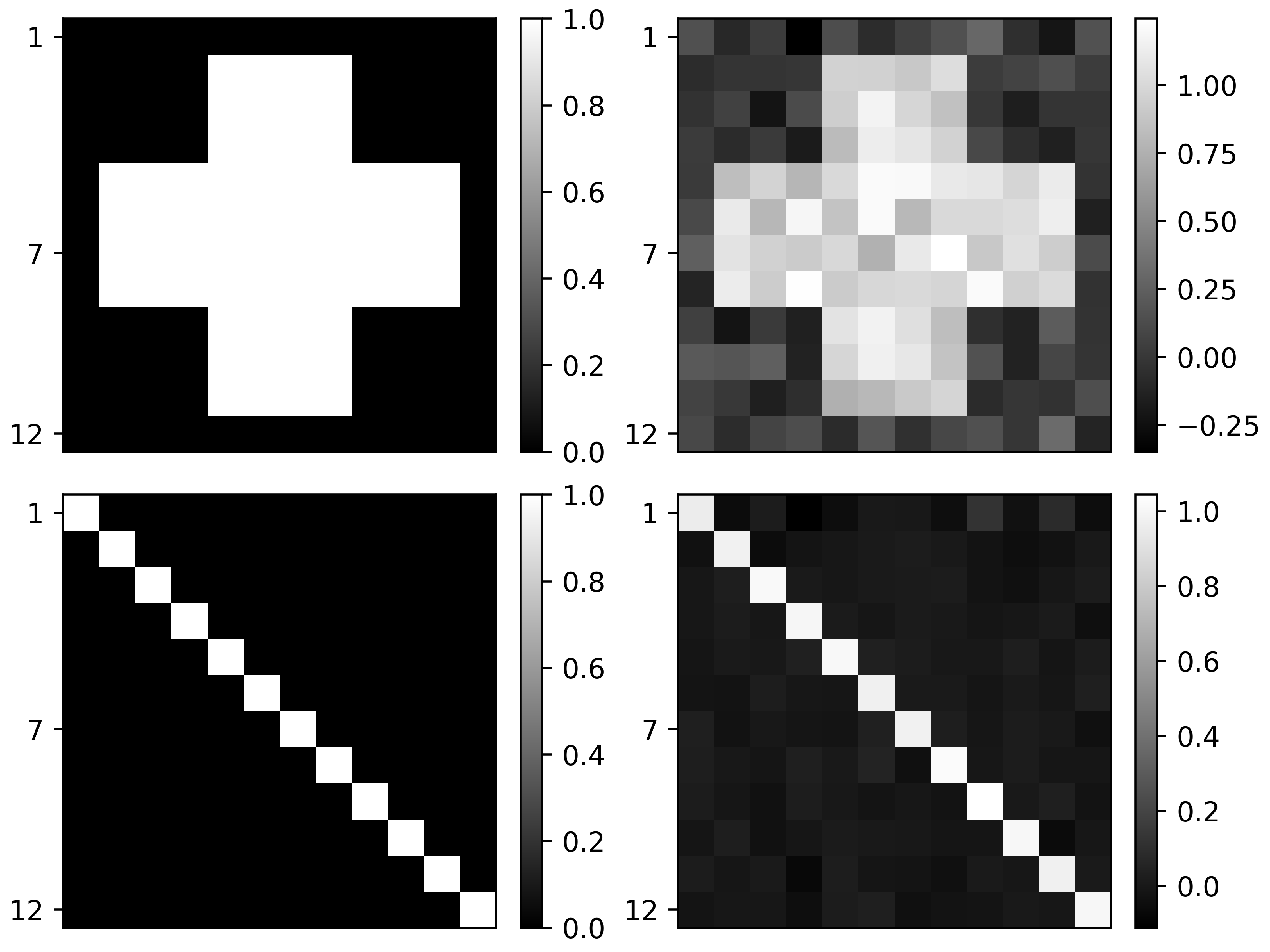}&
        \includegraphics[width=.45\linewidth]{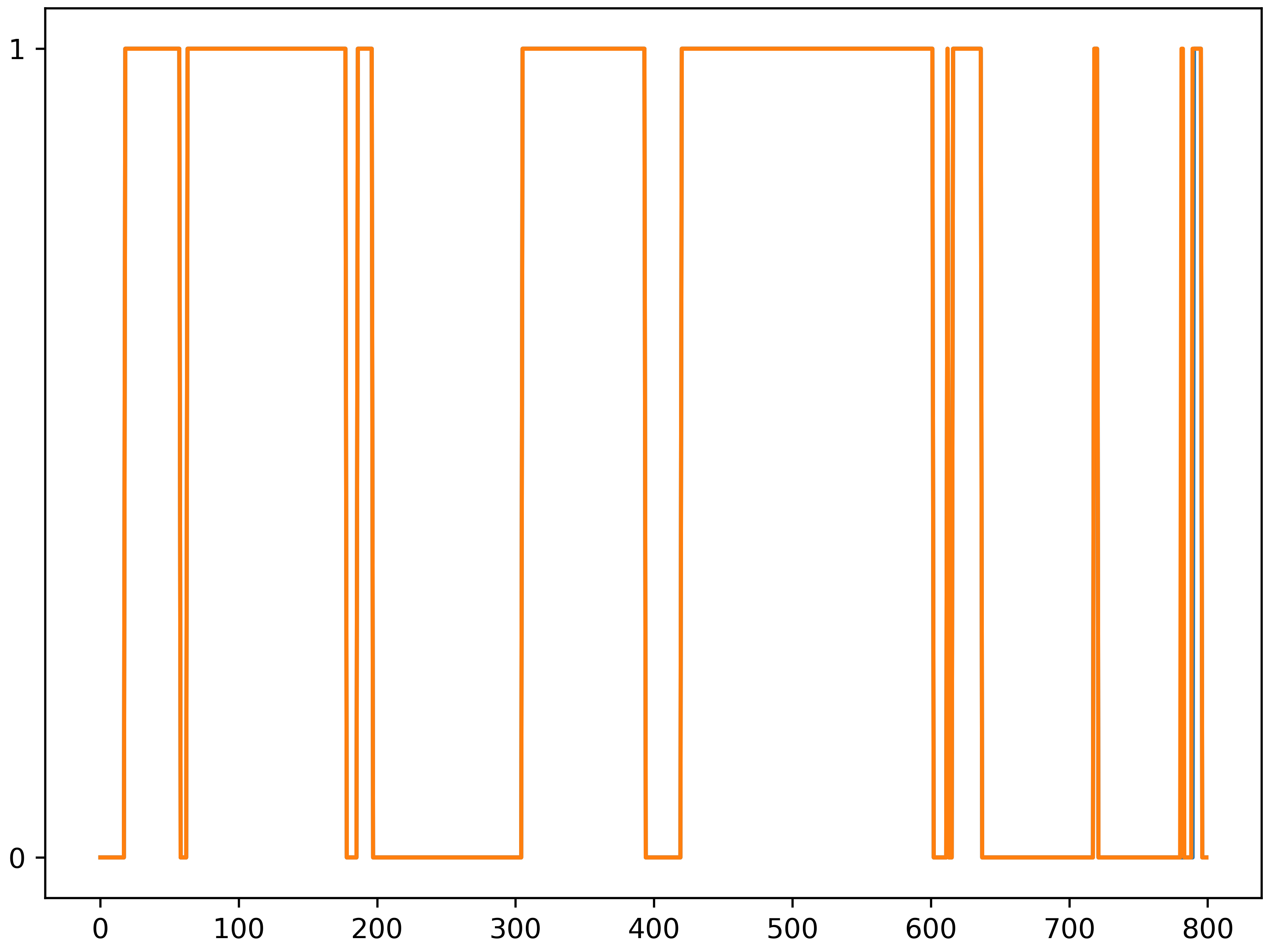}
        \end{tabular}
    \caption{\small Markov-switching model with Diagonal and Anti-diagonal coefficients (first row) and with Cross and Diagonal coefficients (second row). Left plots: true (left column) and estimated (right column) coefficients. Right plots: true (orange) and estimated (blue) hidden states.}
    \label{fig:ms_diag}
\end{figure}

 The regression model and the MCMC algorithm provide reasonably accurate estimates of the true coefficient values, and the regimes are successfully identified in the two different experimental settings (see Figure \ref{fig:ms_diag} for an illustration). The MSE of the actual outcome $\mathbf{y}$ and estimated outcome $\hat{\mathbf{y}}$ is documented in the third column of Table \ref{tab:conv_res}. The MSE decreases rapidly as the MCMC iterations increase, from $7.1119$ at the 10th iteration to $1.1618$ at the 100th iteration, and stabilizes around value $1.0$ after convergence (see Figure \ref{fig:mse} in Appendix). The second column of Table \ref{tab:conv_res} shows the ACF of the parameters and the hidden states in two different experimental settings. 

\begin{table}[t]
\centering
\begin{tabular}{lccccc}
\hline
\multicolumn{6}{c}{Setting $\mathcal{S}^{MS}_1$ (anti-diag / diag)}                                                                            \\ \hline
\multicolumn{1}{l|}{}                                   & ACF(1) & ACF(5) & \multicolumn{1}{c|}{ACF(10)} & MSE(10) & MSE(100) \\ \hline
\multicolumn{1}{l|}{Parameters ($\boldsymbol{\theta}$)} &\begin{tabular}[c]{@{}c@{}}$0.4085$\\ $(0.3145)$\end{tabular}&\begin{tabular}[c]{@{}c@{}}$0.3279$\\ $(0.2328)$\end{tabular}& \multicolumn{1}{c|}{\begin{tabular}[c]{@{}c@{}}$0.3158$\\ $(0.0980)$\end{tabular}} &$0.0559$&$0.0083$          \\
\multicolumn{1}{l|}{States($s_t$)}                      &\begin{tabular}[c]{@{}c@{}}$0.5624$\\ $(0.5448)$\end{tabular}&\begin{tabular}[c]{@{}c@{}}$0.5437$\\ $(0.3878)$\end{tabular}& \multicolumn{1}{c|}{\begin{tabular}[c]{@{}c@{}}$0.5333$\\ $(0.1942)$\end{tabular}} &$0.2725$&$0.0113$          \\ \hline
\multicolumn{6}{c}{Setting $\mathcal{S}^{MS}_2$ (cross / diag)}                                                                                \\ \hline
\multicolumn{1}{l|}{Parameters ($\boldsymbol{\theta}$)} &\begin{tabular}[c]{@{}c@{}}$0.5139$\\ $(0.4247)$\end{tabular}&\begin{tabular}[c]{@{}c@{}}$0.4425$\\ $(0.2819)$\end{tabular}& \multicolumn{1}{c|}{\begin{tabular}[c]{@{}c@{}}$0.4410$\\ $(0.1650)$\end{tabular}} & $0.1773$ & $0.0106$          \\
\multicolumn{1}{l|}{States($s_t$)}                      &\begin{tabular}[c]{@{}c@{}}$0.5294$\\ $(0.5153)$\end{tabular}&\begin{tabular}[c]{@{}c@{}}$0.5166$\\ $(0.3649)$\end{tabular}& \multicolumn{1}{c|}{\begin{tabular}[c]{@{}c@{}}$0.5077$\\ $(0.1831)$\end{tabular}}        &$0.3013$&$0.0050$   \\ \hline
\end{tabular}
\caption{\small  Results on convergence for the two different experimental settings. The second column reports the ACFs of the parameters and the hidden states before and after thinning. The results after thinning are reported in parentheses. The third column reports the MSE of the parameters and hidden states at the 10th and 100th Gibbs iteration.}
\label{tab:conv_res}
\end{table}

\begin{figure}[h]
    \centering
    \begin{subfigure}{.49\textwidth}
        \centering
        \includegraphics[width=.8\linewidth]{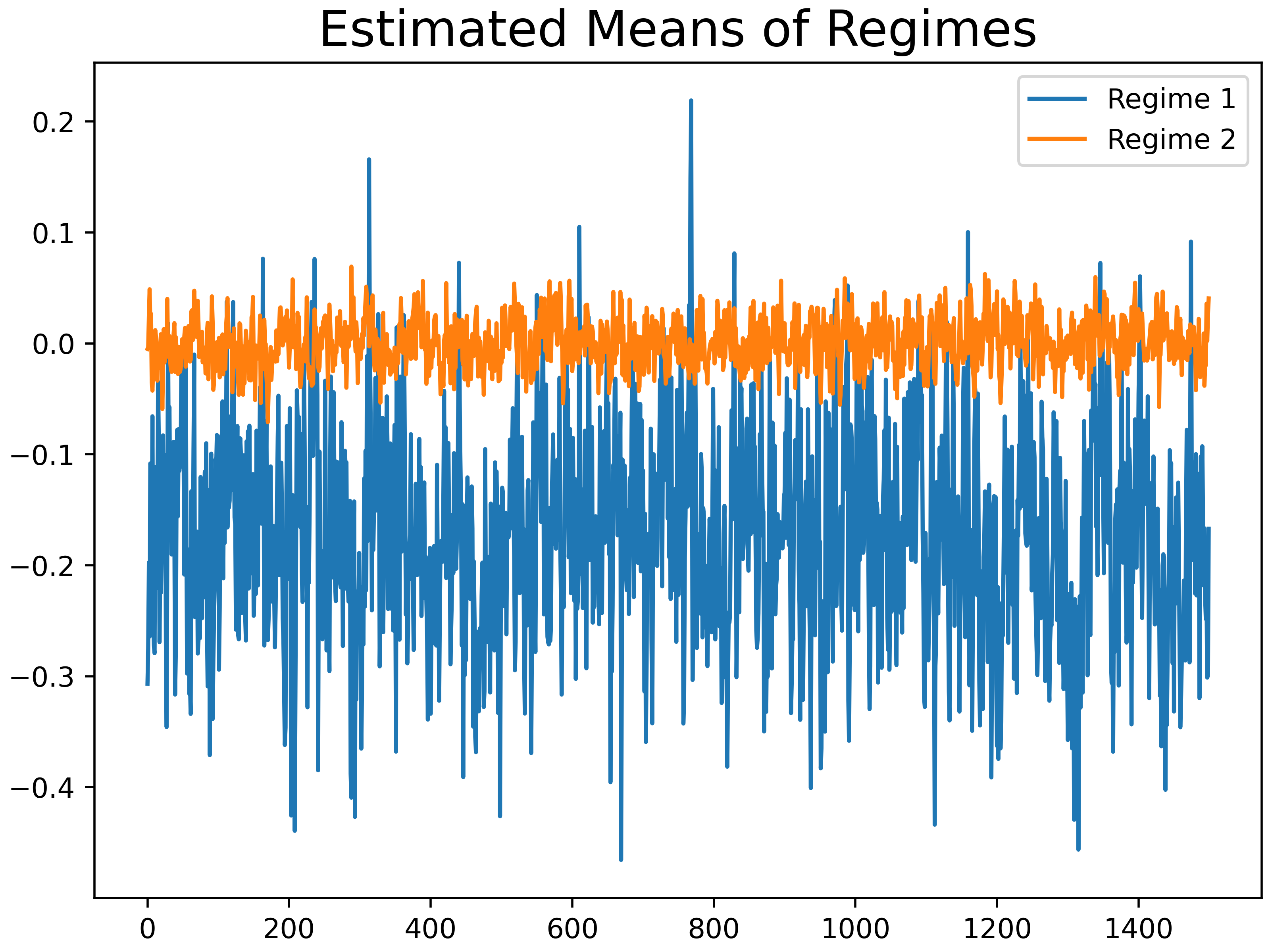}
    \end{subfigure}
    \begin{subfigure}{.49\textwidth}
        \centering
        \includegraphics[width=.8\linewidth]{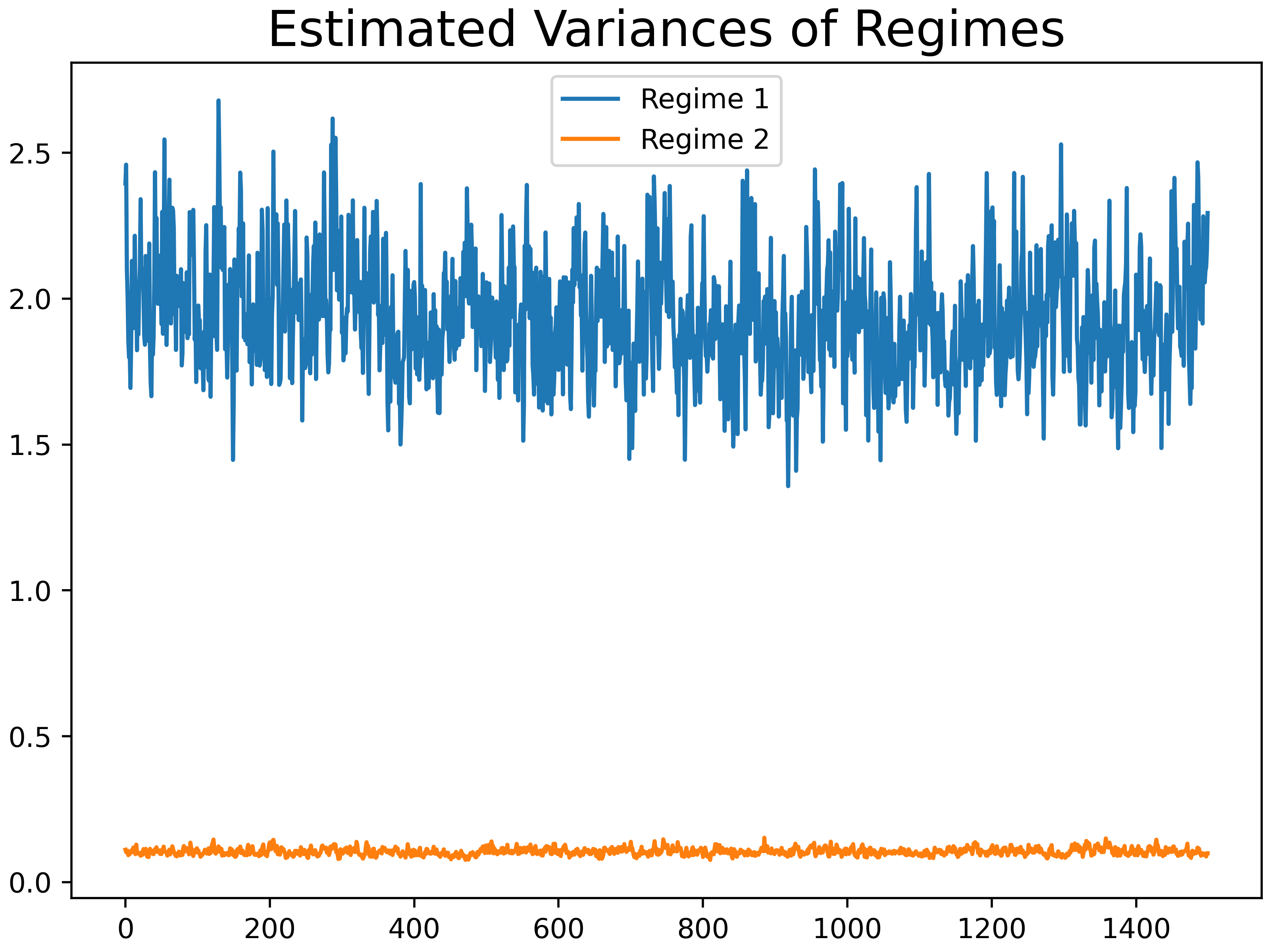}
    \end{subfigure}
    \caption{Trace plots after removing the burn-in samples for regime-specific intercepts (left) and variances (right) for the $\mathcal{S}^{MS}_2$ experimental setting. True values are $\mu_1 = \mu_2 = 0, \sigma^2_1 = 2, \sigma^2_2 = 0.1$.}
    \label{fig:mu_var}
\end{figure}

\begin{figure}[h]
    \centering
    \begin{subfigure}{.49\textwidth}
        \centering
        \includegraphics[width=.8\linewidth]{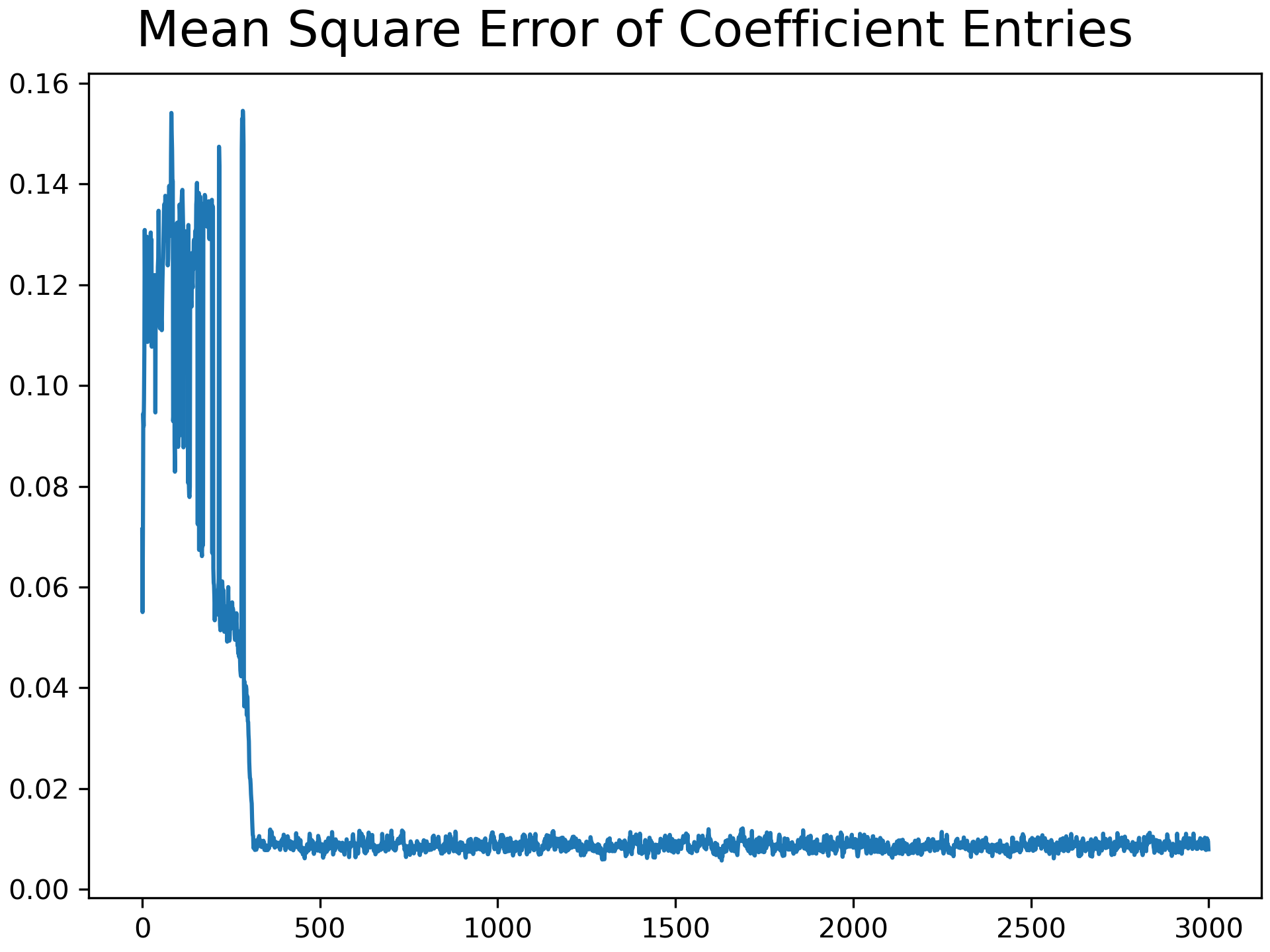}
    \end{subfigure}
    \begin{subfigure}{.49\textwidth}
        \centering
        \includegraphics[width=.8\linewidth]{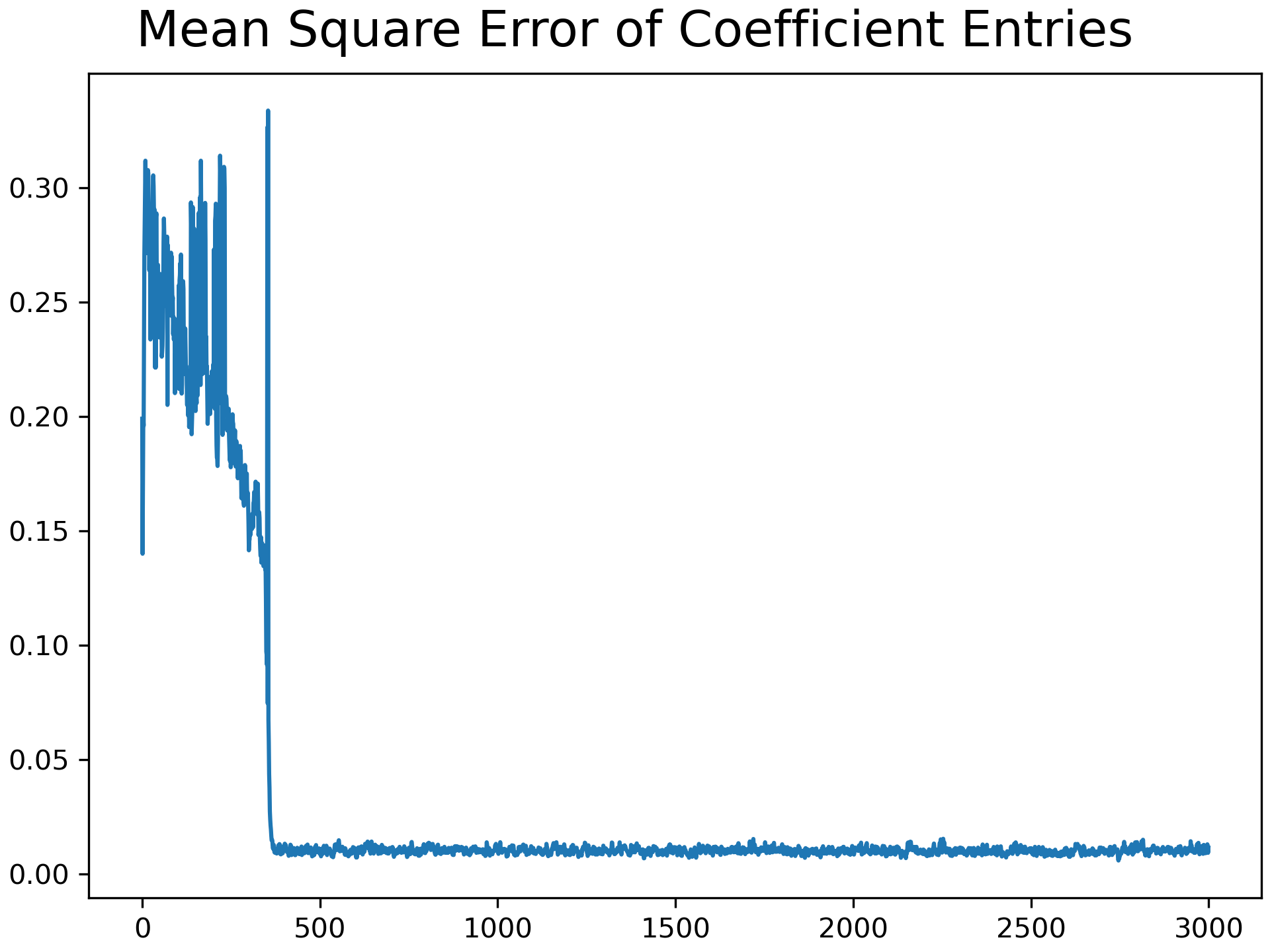}
    \end{subfigure}
    \caption{MSE of the outputs for the $\mathcal{S}^{MS}_1$ (left) and $\mathcal{S}^{MS}_2$  (right) experimental setting.}
    \label{fig:mse}
\end{figure}

\subsection{Robusteness checks}
\begin{figure}[t]
    \centering
    \includegraphics[width=.8\textwidth]{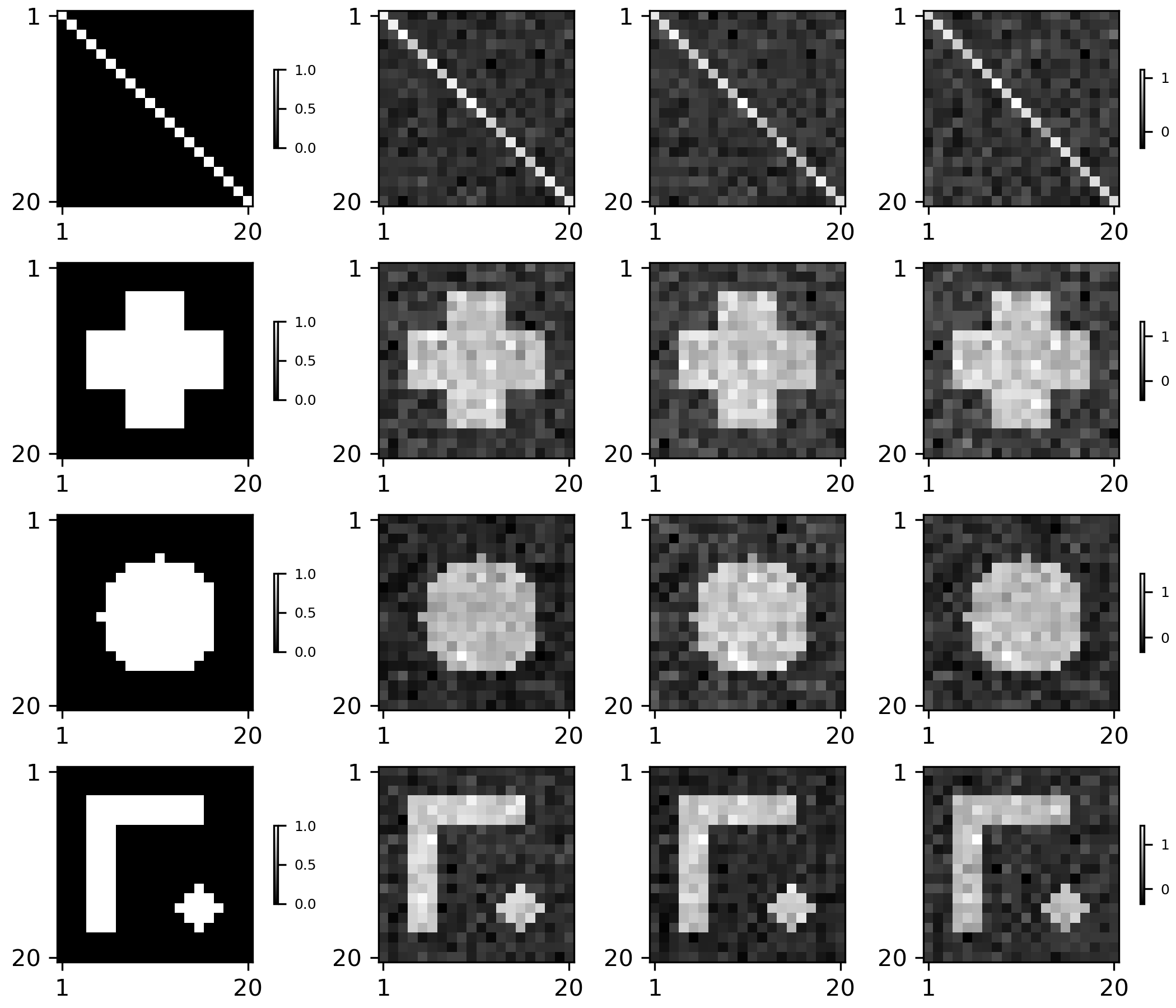}
    \caption{Robustness check subject to a different set of hyper-parameters. In the first column, the true coefficient value, the estimation results for different ranks $D=3,5,7$ in the columns from two to four. The four different experimental settings $\mathcal{S}_j$, $j=1,\ldots,4$ are in the rows.}
    \label{fig:sim_main_robust}
\end{figure}

To test further the algorithm performance with respect to the choice of hyper-parameters, we carried out robustness checks by changing the prior mean and prior variance for $\sigma_{m}^2, \tau$ and $\lambda_{m}^{(d)}$. The simulation results for the new set of hyperparameters indicated in Table \ref{tab:rob} are given in  Figure \ref{fig:sim_main_robust}.  The weight inference results are robust in the choice of hyperparameter values.

\begin{table}[h]
\centering
\caption{Values of hyper-parameters chosen for benchmark and robustness check}\label{tab:rob}
\begin{tabular}{lcc}
\hline
              & benchmark                 & robustness                \\ \hline
$\alpha$      & 1                         & 1                         \\ \hline
$a_{\sigma}$  & 0.5                       & 0.5                       \\ \hline
$b_{\sigma}$  & 8.5$\sqrt{C}$             & 2$\sqrt{C}$               \\ \hline
$a_{\tau}$    & 3                         & 3                         \\ \hline
$b_{\tau}$    & 33.75$/b_{\sigma}$ & 33.75$/b_{\sigma}$ \\ \hline
$a_{\lambda}$ & 3                         & 3                         \\ \hline
$b_{\lambda}$ & $a_{\lambda}^{1/4}$       & $a_{\lambda}^{1/2}$       \\ \hline
\end{tabular}
\end{table}

\clearpage

\newpage

\renewcommand{\thesection}{D}
\renewcommand{\theequation}{D.\arabic{equation}}
\renewcommand{\thefigure}{D.\arabic{figure}}
\renewcommand{\thetable}{D.\arabic{table}}
\setcounter{table}{0}
\setcounter{figure}{0}
\setcounter{equation}{0}

\section{Further Empirical Results}

\begin{table}[h!]
\centering
\begin{tabular}{lcc|cccc}
\hline
                & \multicolumn{2}{c|}{\multirow{2}{*}{In Sample}}    & \multicolumn{4}{c}{Out of Sample}  \\
                & \multicolumn{2}{c|}{}                              & \multicolumn{2}{c}{\textit{1-day (month) ahead}} & \multicolumn{2}{c}{\textit{5-day (month) ahead}}   \\ \cline{2-7}
                & MSE   & MAE                    & MSE          & MAE          & MSE         & MAE         \\ \hline
\multicolumn{7}{l}{\textit{Application 1: VIX and OVX on macro indicators}}                                                                                                                            \\ \hline
Least Square    & 0.3049 & 0.4266 & 0.1945 & 0.3474 & 0.3668 & 0.5211         \\
LASSO           &  0.4207  &  0.5259 &   0.5199 &  0.6363   &     0.6940 & 0.7589          \\
Tensor    &   0.3097 & 0.4324 & 0.2540 & 0.4232 &   0.3581 & 0.5182        \\
MS Tensor       &   \textbf{0.0907}    &  \textbf{0.2393}   &   \textbf{0.1409}   &    \textbf{0.3342}   &    \textbf{0.1379}     &        \textbf{0.3063}              \\ \hline
\multicolumn{7}{l}{\textit{Application 2: S\&P 500 on  oil prices (Aggregate Analysis)}}                                                                                                                   \\ \hline 
Least Square    &   0.4418    &    0.5126    &       0.2394       &      0.4893        &      0.1986       &       0.4073        \\
LASSO           &  0.4692  &  0.5264   &   0.2058   &  0.4537   &     \textbf{0.1487}        &     \textbf{0.3548}      \\
Tensor          &   0.6073    &     0.6084  &     \textbf{0.0445}   &  \textbf{0.2111}  &  0.3442 &    0.4227           \\
MS Tensor       &   \textbf{0.3901}    &  \textbf{0.4794}   &   0.5248   &    0.7244   &    0.2664    &        0.4522      \\ \hline
\multicolumn{7}{l}{\textit{Application 2: Financial Sector on  oil prices (Disaggregated S\&P 500 analysis)}}                                                                                                                   \\ \hline 
Least Square    &   0.5198   &  0.5530   & 0.3082     & 0.5551     & 0.1107     &  0.2631      \\
LASSO  & 0.5532  &  0.5650   &  0.2684   & 0.5181  &  0.1111 & 0.2555        \\
Tensor          &  0.6968   & 0.6204    &  \textbf{0.0035}    & \textbf{0.0594}   &  \textbf{0.0574}   & \textbf{0.2025}         \\
MS Tensor  &  \textbf{0.3370}   & \textbf{0.4308}    & 0.0586   & 0.2421  & 0.0878    & 0.2696        \\ \hline
\multicolumn{7}{l}{\textit{Application 2: Energy Sector on  oil prices (Disaggregated S\&P 500 analysis)}}                                                                                                                   \\ \hline 
Least Square    &  0.4587    & 0.5354    & 0.6606     & 0.8128     & 0.2879 &  0.4931       \\
LASSO  & 0.4869  & 0.5516 & 0.1756 & 0.4191  & \textbf{0.2740}  & \textbf{0.4545}        \\
Tensor          &  0.5379   & 0.5731    & 0.2414     & 0.4914   & 0.2788    & 0.4949      \\
MS Tensor       &  \textbf{0.3323}   &  \textbf{0.4362}  & \textbf{0.0090}  &  \textbf{0.0949}   &   0.3325   &  0.5153    \\ \hline
\multicolumn{7}{l}{\textit{Application 2: Other sectors on  oil prices (Disaggregated S\&P 500 analysis)}}                                                                                                                   \\ \hline 
Least Square    &   0.4389   & 0.5097    & 0.5919     & 0.7694  & 0.4054  & 0.5672        \\
LASSO  & 0.4675  & 0.5254 & 0.3940 & 0.6277  & \textbf{0.2908}  & 0.5071         \\
Tensor          &  0.4643   & 0.5219    & \textbf{0.1072}     & \textbf{0.3274}   &  0.2935   & \textbf{0.4871}      \\
MS Tensor   & \textbf{0.3064}  & \textbf{0.4105}    &   0.9498  & 0.9746   & 0.7229  & 0.7488   \\ \hline
\end{tabular}
\caption{In-sample fitting and out-of-sample forecasting performance. Results for the best-performing model are in boldface.}
\label{tab:stats fin}
\end{table}

\begin{figure}[p]
    \centering
    \begin{tabular}{cc}
        \includegraphics[width=.47\linewidth]{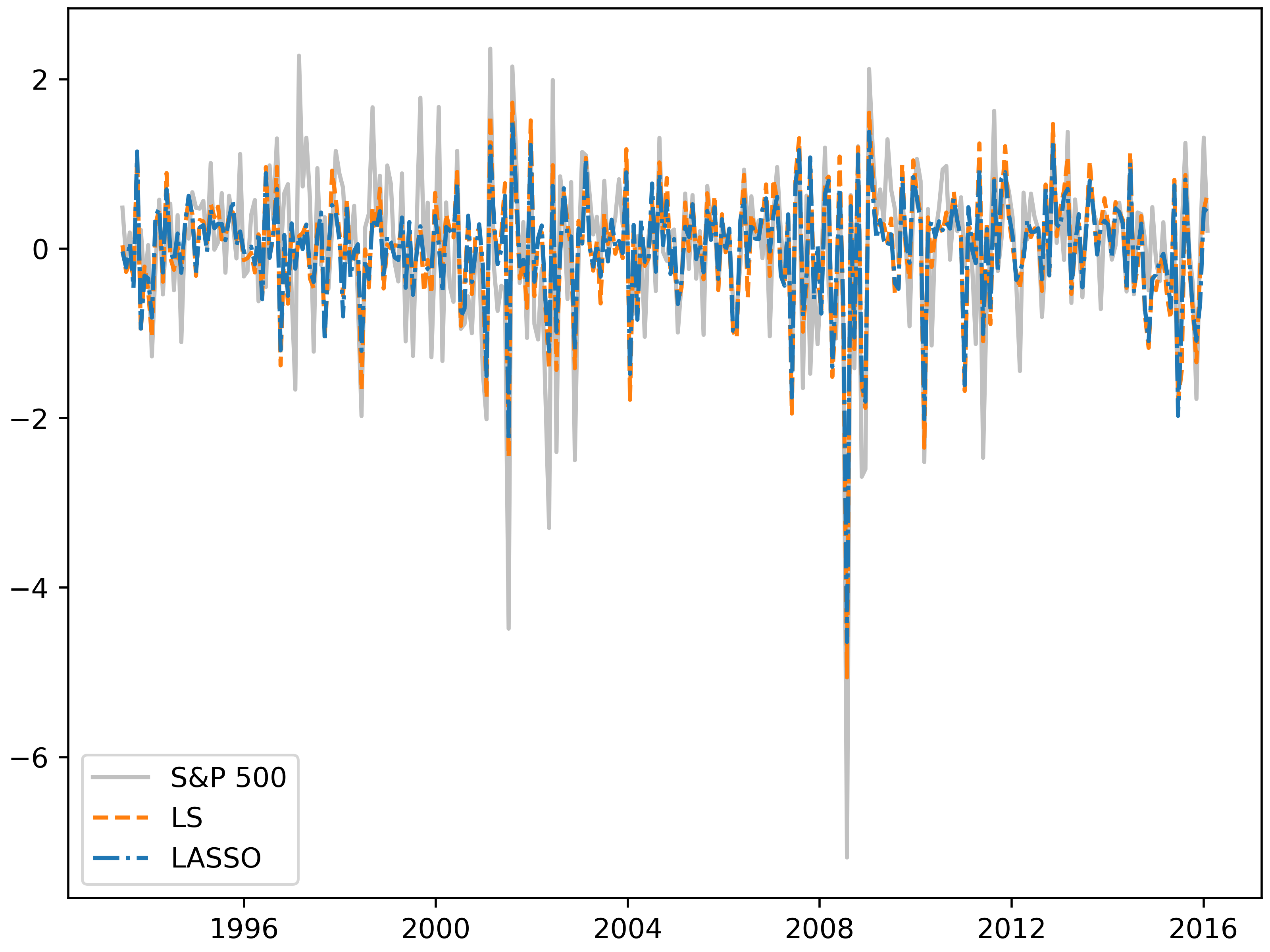} & \includegraphics[width=.47\linewidth]{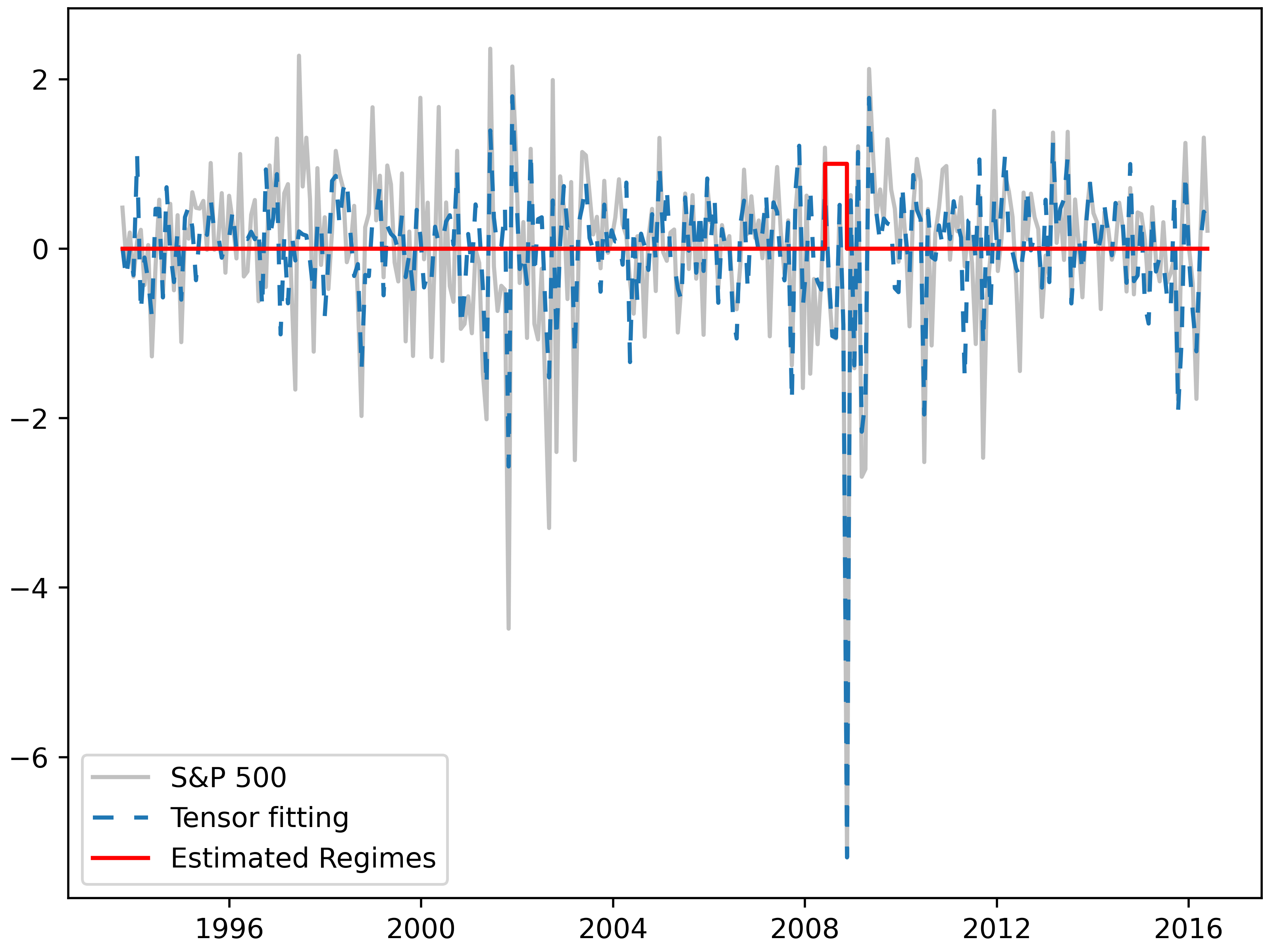} \\
        \includegraphics[width=.47\linewidth]{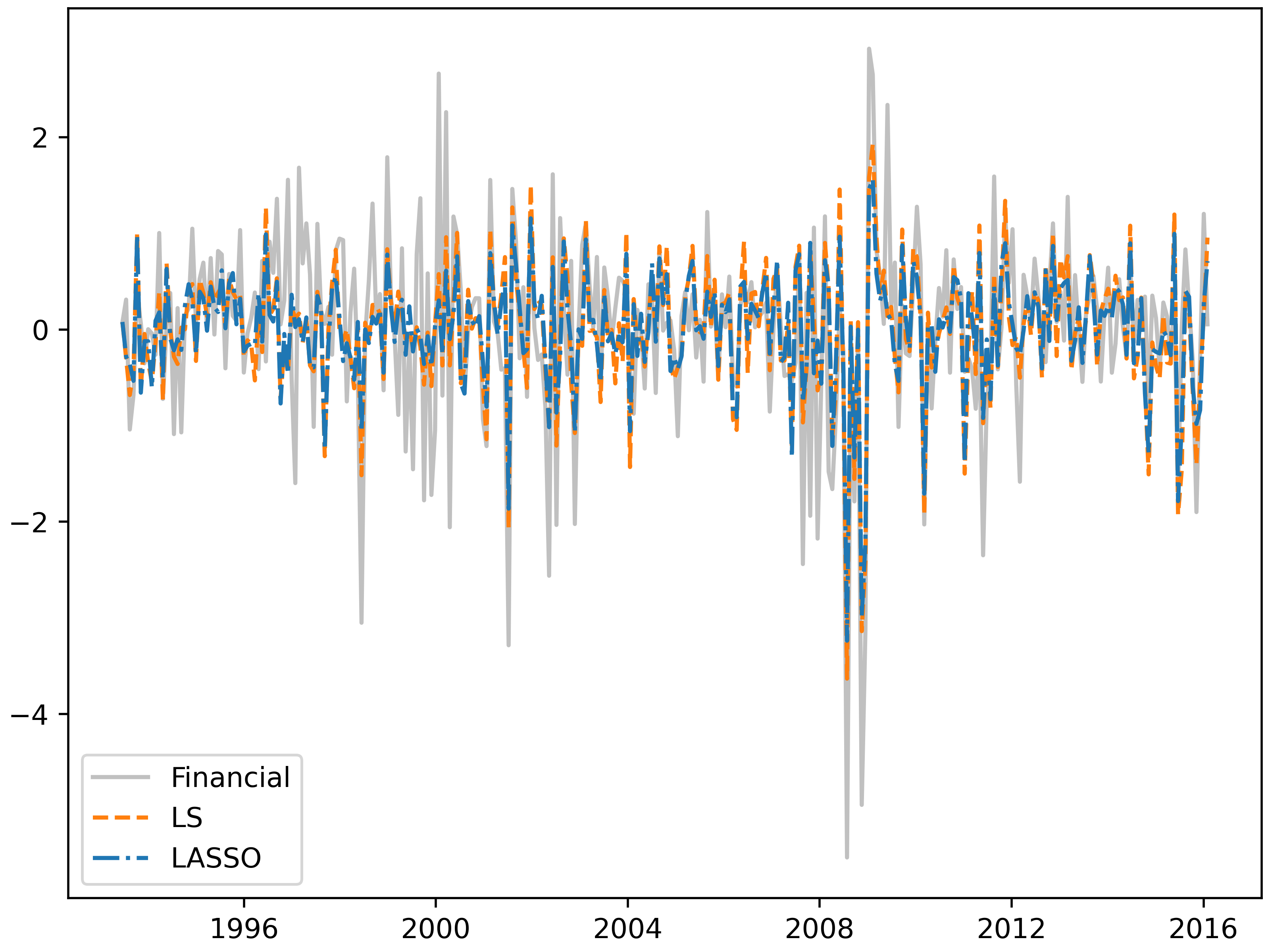} & \includegraphics[width=.47\linewidth]{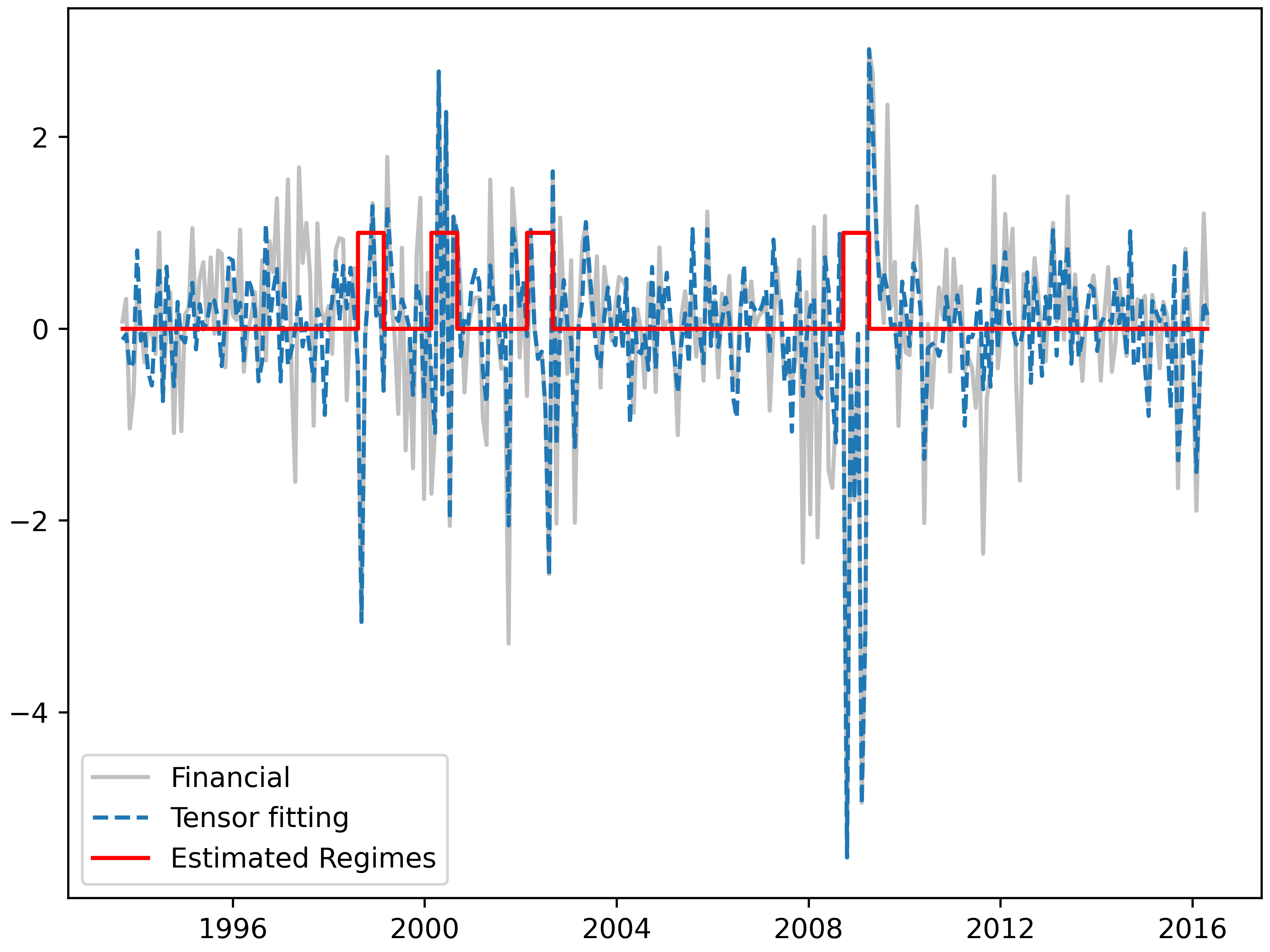}\\
        \includegraphics[width=.47\linewidth]{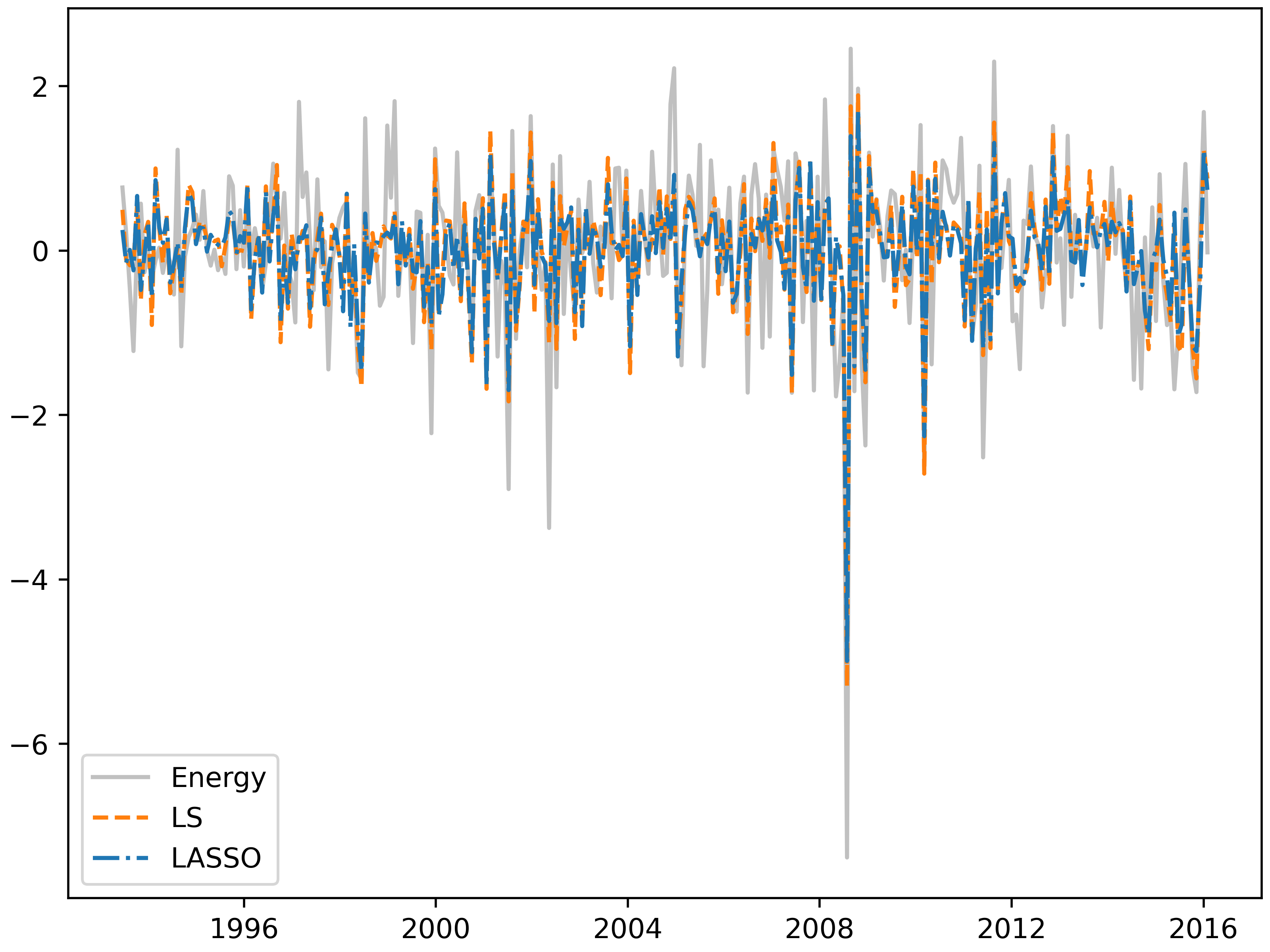} & \includegraphics[width=.47\linewidth]{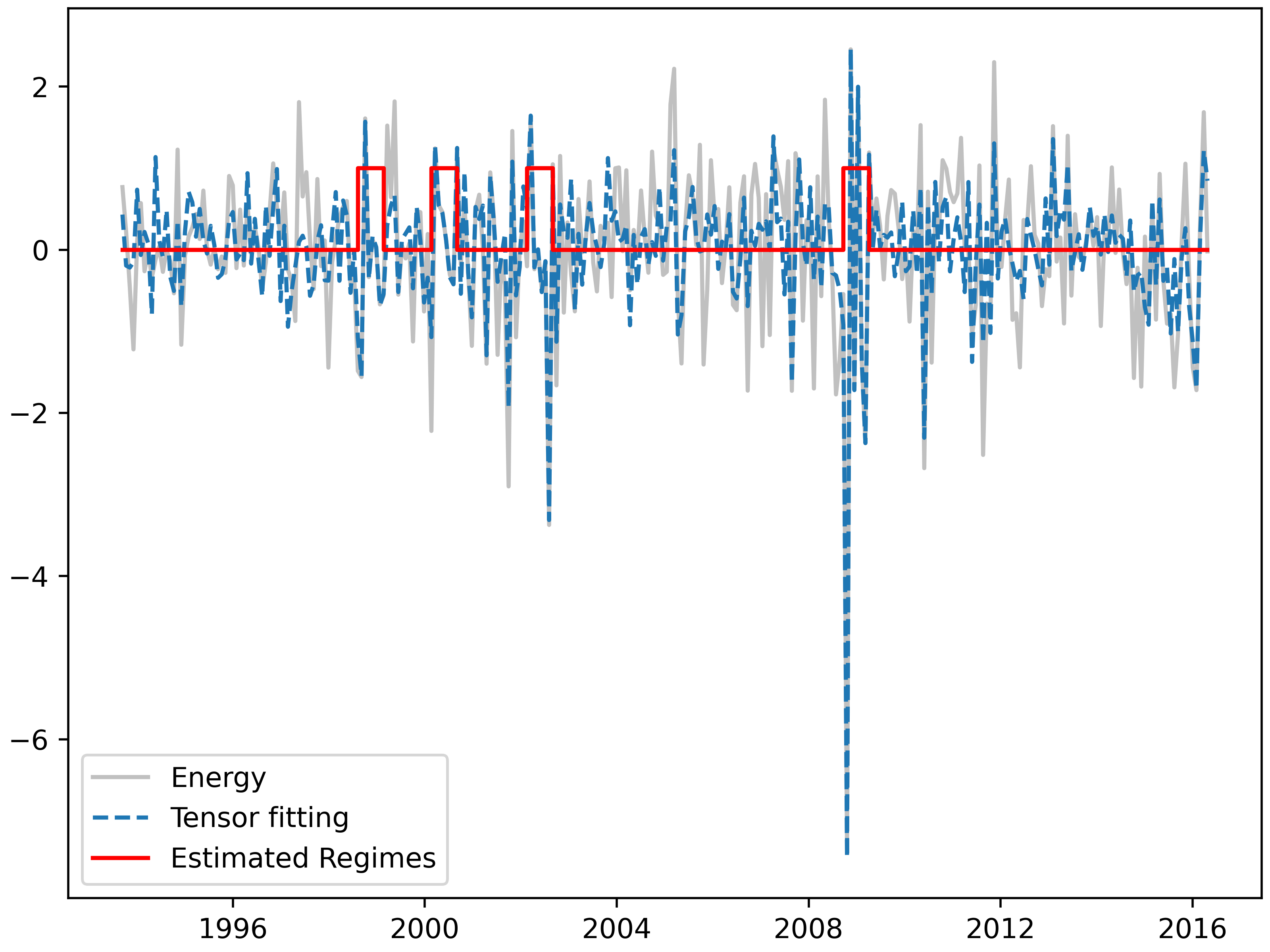}\\
        \includegraphics[width=.47\linewidth]{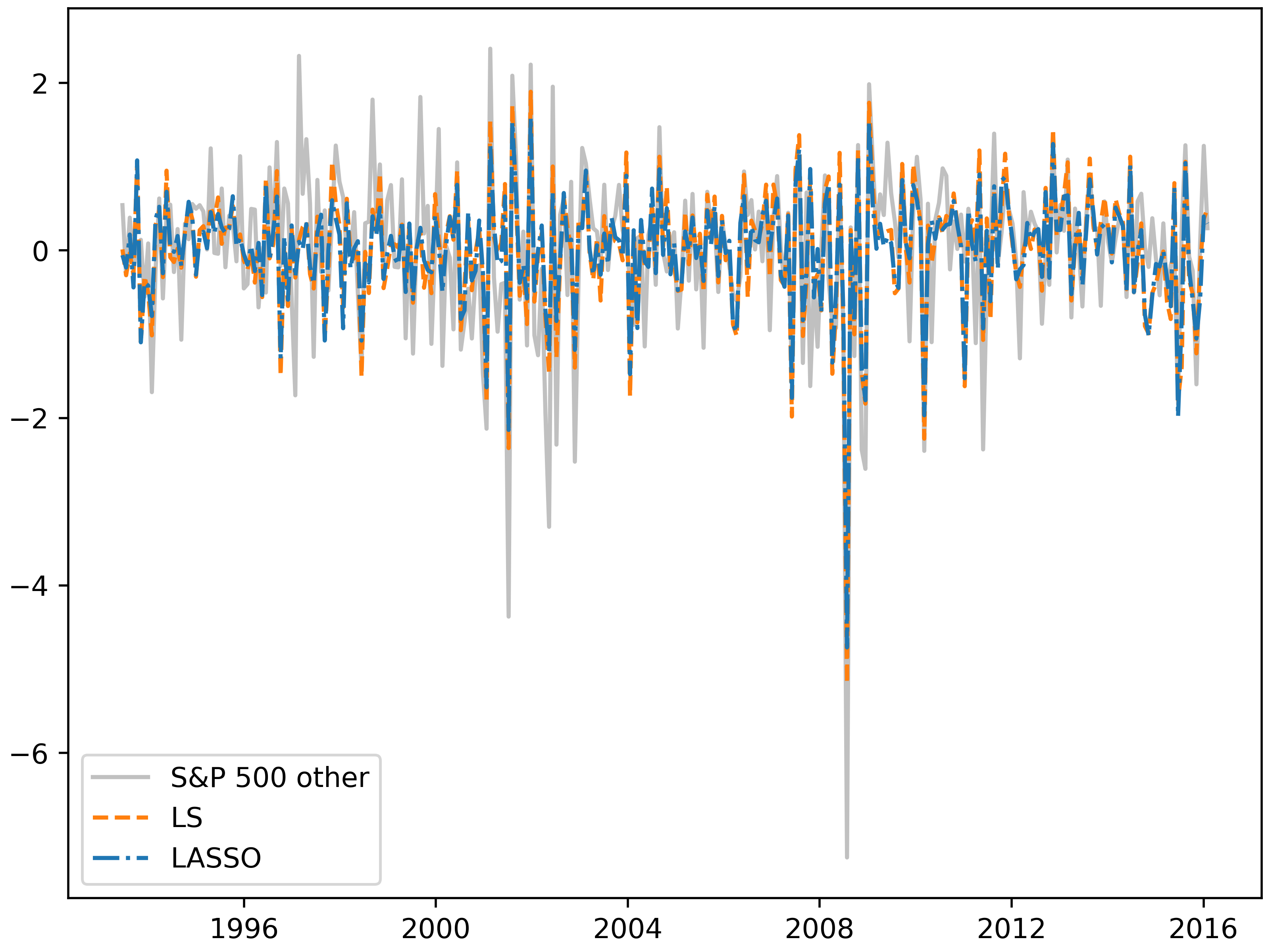} & \includegraphics[width=.47\linewidth]{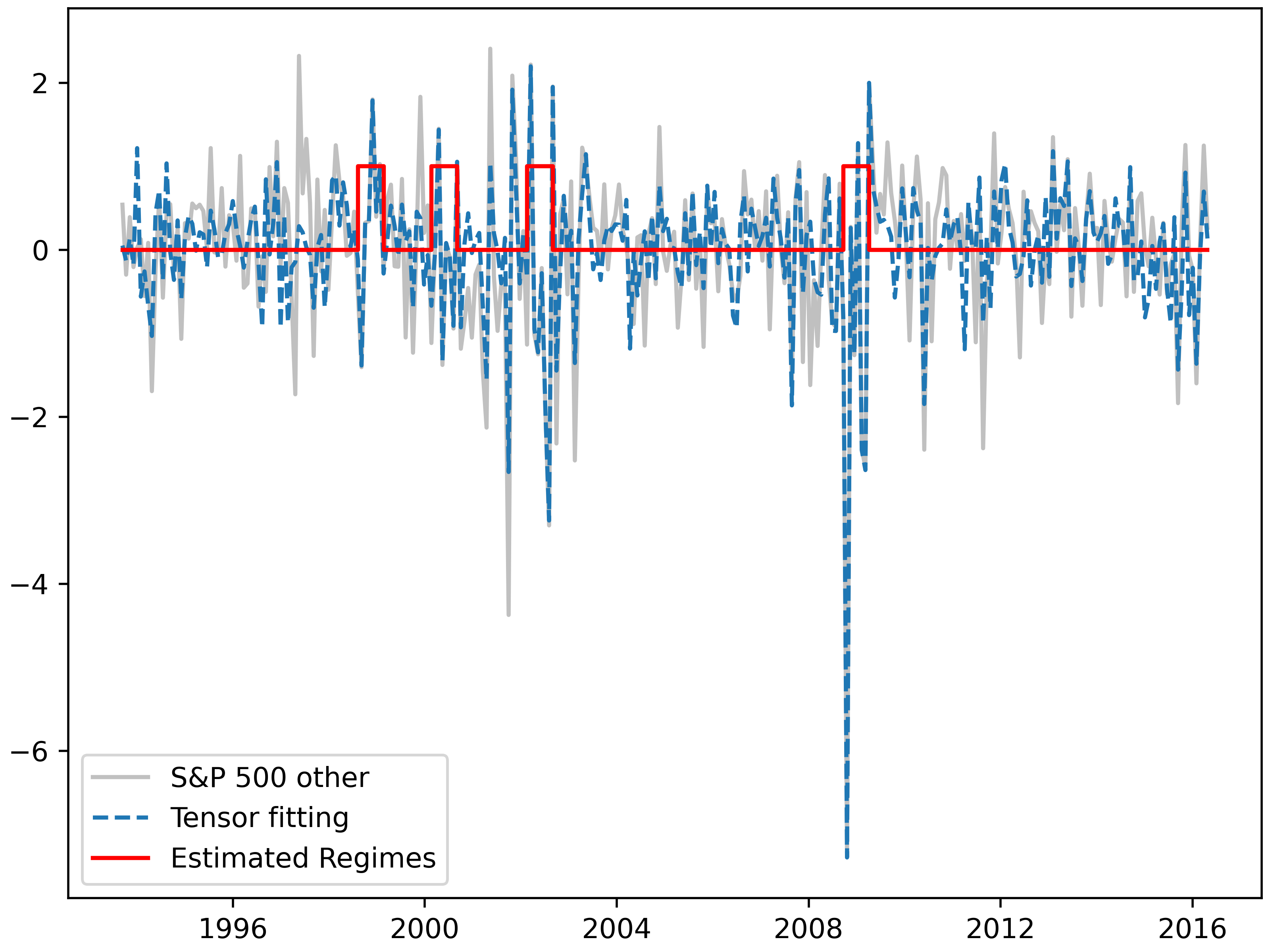}
    \end{tabular}
    \caption{\small Left: In-sample fitting results for Least Square (orange dashed line) and LASSO (blue dashed line). Right: Tensor Regression with Markov Switching (blue dashed line) and estimated hidden states (red solid line). True data is shown in solid silver line.}
    \label{fig:insamp_mac1}
\end{figure}

\begin{figure}[p]
    \centering
    \begin{tabular}{cc}
        \includegraphics[width=.47\linewidth]{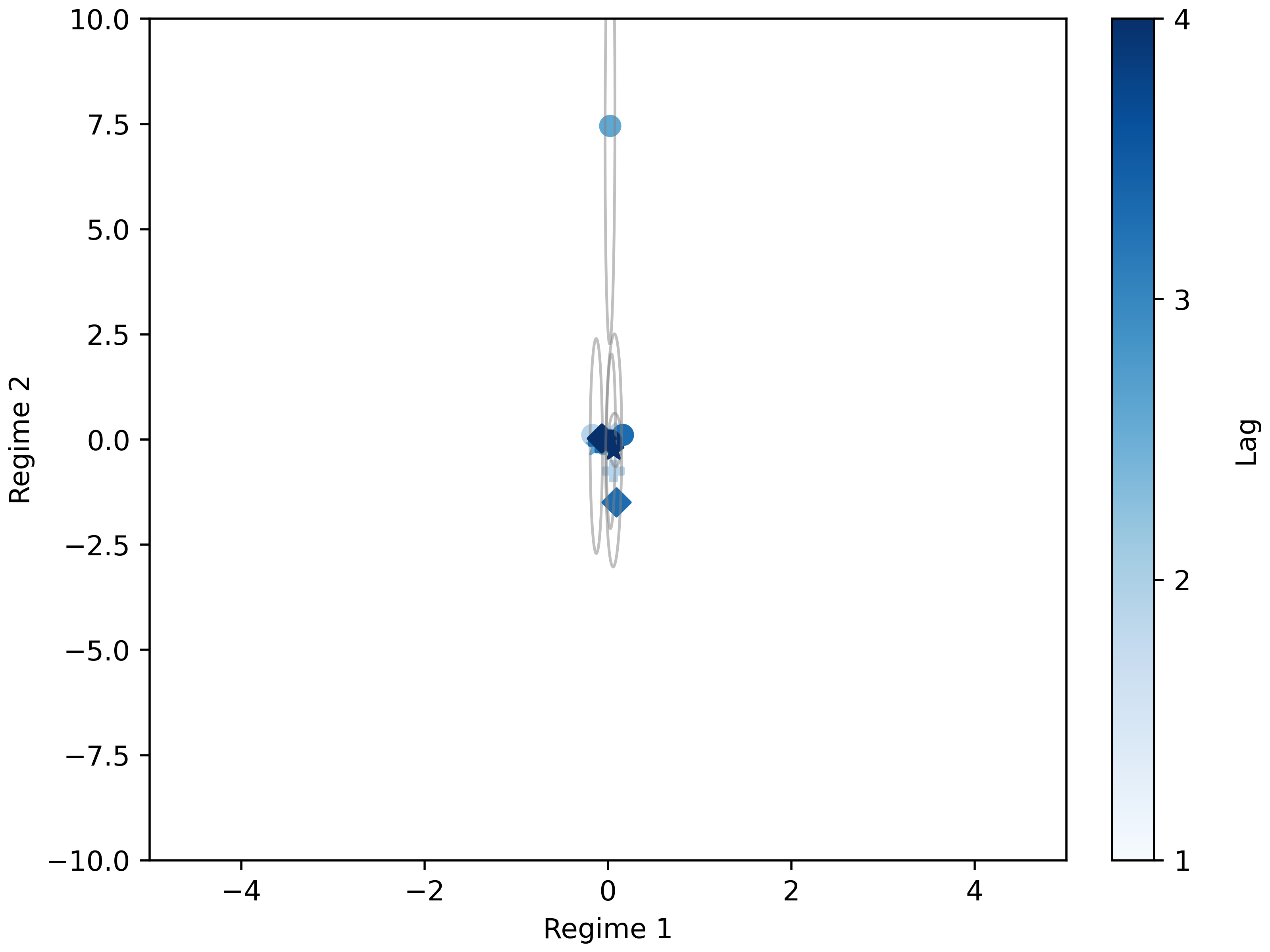} & \includegraphics[width=.47\linewidth]{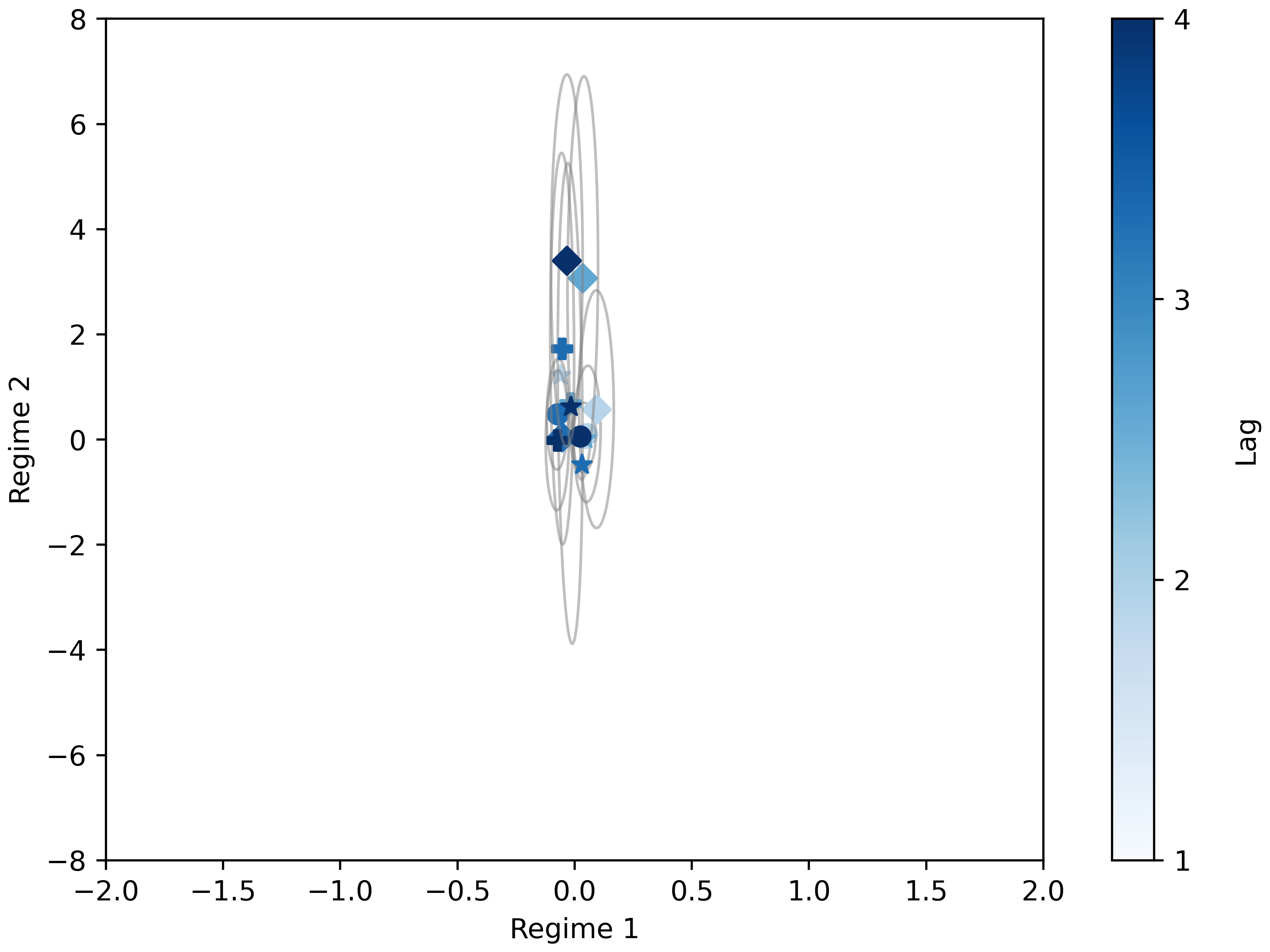} \\
        \includegraphics[width=.47\linewidth]{images/scatter_gv_fin.png} & \includegraphics[width=.47\linewidth]{images/scatter_bv_fin.png}\\
        \includegraphics[width=.47\linewidth]{images/scatter_gv_eng.png} & \includegraphics[width=.47\linewidth]{images/scatter_bv_eng.png}
    \end{tabular}
    \caption{\small The scatter plot shows the effects of Good Oil Volatility (left column) and Bad Oil Volatility (right column) on S\&P 500 (first row), financial sector (second row) and energy sector (third row). Different symbols represent the weekly data sampled at different weeks, with $\bullet$: $t-(1+4(p-1))/4$,  ~\ding{58}: $t-2(1+4(p-1))/4$, \ding{117}: $t-3(1+4(p-1))/4$ and $\bigstar: t-4(1+4(p-1))/4$ for $p = \{1,2,3, 4\}$. Different shades of blue represent different lags, from order 1 (lighter) to order 4 (darker).}
    \label{fig:coef_macro1}
\end{figure}

\clearpage

\end{document}